\documentclass{aa}
\usepackage[varg]{txfonts}
\usepackage{graphics}

\def\cc{\,{\rm cm^{-3}}}
\def\cm2{\,{\rm cm^{-2}}}
\def\pc2{\,{\rm pc^{2}}}
\def\kms{\,{\rm {km\,s^{-1}}}}
\def\thirco{\,{\rm ^{13}CO}}
\def\h2{\,{\rm H_{2}}}
\def\kkms{\,{\rm {K\,km s^{-1}}}}
\def\co{\,{\rm ^{12}CO}}

\def\ci{\hbox{{\rm [C {\scriptsize I}]}}}
\def\cii{\hbox{{\rm [C {\scriptsize II}]}}}
\def\nii{\hbox{{\rm [N {\scriptsize II}]}}}

\def\pci{\,{\rm ^{3}P_{1}-^{3}P_{0}\,[C {\scriptsize I}]}}
\def\pbci{\,{\rm ^{3}P_{2}-^{3}P_{1}\,[C {\scriptsize I}]}}

\def\coa10{$^{12}$CO($1\rightarrow 0$)}
\def\cob21{$^{12}$CO($2\rightarrow 1$)}
\def\coc32{$^{12}$CO($3\rightarrow 2$)}
\def\cod43{$^{12}$CO($4\rightarrow 3$)}
\def\coe54{$^{12}$CO($5\rightarrow 4$)}
\def\cof76{$^{12}$CO($7\rightarrow 6$)}
\def\13coc32{$^{13}$CO($3\rightarrow 2$)}

\def\hcna10{HCN($1\rightarrow 0$)}
\def\hcnb43{HCN($4\rightarrow 3$)}
 
\def\etal{et\,al.\ }
\def\aua{{\rm A\&A,} }
\def\auas{{\rm A\&AS,} }

\def\apj{{\rm ApJ,} }

\def\apjs{{\rm ApJS,} }

\def\araa{{\rm ARAA,} }
\def\mnras{{\rm MNRAS,} }

\usepackage{color}

\begin{document}

\title{Neutral carbon and CO in 76 (U)LIRGs and starburst galaxy centers}

\subtitle{A method to determine molecular gas properties in luminous galaxies}

\author{F.P. Israel\inst{1}
 \and   M.J.F. Rosenberg\inst{1}
 \and   P. van der Werf\inst{1} 
        }

\offprints{F.P. Israel}
 
  \institute{Sterrewacht Leiden, Leiden University, P.O. Box 9513,
             2300 RA Leiden, The Netherlands 
}

\authorrunning{F.P. Israel \etal }

\titlerunning{[CI] and CO in galaxies}

\date{Received ????; accepted ????}
 
\abstract{In this paper we present fluxes in the $\ci$ lines of neutral carbon at the centers of some 76 galaxies with far-infrared luminosities ranging from $10^{9}$ to $10^{12}$ L$_{\odot}$, as obtained with the Herschel Space Observatory and ground-based facilities, along with the line fluxes of the $J$=7-6, $J$=4-3, $J$=2-1 $\co$, and $J$=2-1 $\thirco$ transitions.}{With this dataset, we determine the behavior of the observed lines with respect to each other and then investigate whether they can be used to characterize the molecular ISM of the parent galaxies in simple ways and how the molecular gas properties define the model results.}{ In most starburst galaxies, the $\ci$ to $\thirco$ line flux ratio is much higher than in Galactic star-forming regions, and it is correlated to the total far-infrared luminosity.  The $\ci$ (1-0)/$\co$ (4-3), the $\ci$ (2-1)/$\co$ (7-6), and the $\ci$ (2-1)/(1-0) flux ratios are correlated, and they trace the excitation of the molecular gas.}{In the most luminous infrared galaxies (LIRGs), the interstellar medium (ISM) is fully dominated by dense ($n(\h2)=10^{4}-10^{5}\,\cc$) and moderately warm ($T_{kin}\approx30$ K) gas clouds that appear to have low [C$^{\circ}$]/[CO] and [$\thirco$]/[$\co$] abundances. In less luminous galaxies, emission from gas clouds at lower densities becomes progressively more important, and a multiple-phase analysis is required to determine consistent physical characteristics. Neither the $\co$ nor the $\ci$ velocity-integrated line fluxes are good predictors of molecular hydrogen column densities in individual galaxies.  In particular, so-called $X(\ci)$ conversion factors are not superior to $X(\co)$ factors. The methods and diagnostic diagrams outlined in this paper also provide a new and relatively straightforward means of deriving the physical characteristics of molecular gas in high-redshift galaxies up to $z$=5, which are otherwise hard to determine.} {} \keywords{Galaxies: ISM -- starburst --- statistics; ISM: molecules; Submillimeter: ISM }

\maketitle
 
\section{Introduction}

%
\begin{table*}
\caption[]{Galaxy sample}
\begin{center}
\begin{tabular}{lccccccc}
\hline
\noalign{\smallskip}
Name             &RA(2000)$^{a}$&DEC(2000)$^{a}$&V$_{LSR}^{a}$&$D^{a}$&lg $FIR^{b}$&lg $L_{FIR}$&Table$^{a}$\\ 
                 & ( h m s)    &  (d m s)  &($\kms$)& (Mpc)& (W m$^{-2}$) & (L$_{\odot}$) & \\
\noalign{\smallskip}
\hline
\noalign{\smallskip}
NGC 34            & 00 11 06.5 & -12 06 26 &  5881 &  79.3 & -12.12 & 11.21 & 2 \\
IC10              & 00 20 17.3 & +59 18 14 &  -348 &  0.85 & -11.72 & 7.62  & 5 \\
NGC 253           & 00 47 33.1 & -25 17 18 &   243 &   3.4 & -10.42 & 10.15 & 2,5 \\
NGC 278           & 00 52 04.3 & +47 33 02 &   840 &  11.3 & -11.87 &  9.76 & 5 \\
MGC+12-02-001     & 00 54 03.6 & +73 05 12 &  4706 &  66.1 & -11.77 & 12.35 & 2 \\  
IC 1623(Arp 236)  & 01 07 47.2 & -17 30 25 &  6016 &  80.7 & -11.97 & 11.37 & 2 \\  
NGC 660           & 01 43 02.4 & +13 38 42 &   910 &  12.2 & -11.46 & 10.26 & 5 \\
NGC 891           & 02 22 33.4 & +42 20 57 &   528 &   9.4 & -11.53 &  9.89 & 5 \\
Maffei 2          & 02 41 55.0 & +59 36 15 &   189 &   3.1 & -11.23 &  9.11 & 5 \\
NGC 1068/M77      & 02 42 40.7 & -00 00 48 &  1010 &  15.2 & -11.04 & 10.75 & 4.5 \\
NGC 1056          & 02 42 48.3 & +28 34 27 &  1545 &  21.4 & -12.49 & 10.00 & 4 \\
NGC 1275 (Per~A)  & 03 19 48.1 & +41 30 42 &  5264 &  70.9 & -12.47 & 10.71 & 3 \\
NGC 1365          & 03 33 36.4 & -36 08 25 &  1636 &  21.5 & -11.36 & 10.86 & 2 \\
IC 342            & 03 46 48.5 & +68 05 47 &   126 &   3.5 & -11.36 &  8.60 & 5 \\ 
NGC1482           & 03 54 38.9 & -20 30 10 &  1916 &  25.4 & -11.79 & 10.50 & 3 \\
NGC 1614          & 04 33 59.8 & -08 34 44 &  4778 &  64.2 & -11.82 & 11.33 & 2 \\
IRAS F05189-2524  & 05 21 01.4 & -25 21 45 & 12760 & 173   & -12.22 & 11.78 & 2 \\   
NGC 2146          & 06 18 37.7 & +78 21 25 &   893 &  16.7 & -11.16 & 10.53 & 2 \\
Henize 2-10       & 08 36 15.1 & -26 24 34 &   873 &  10.4 & -11.95 &  9.56 & 5 \\
NGC 2623 (Arp 243)& 08 38 24.1 & +25 45 17 &  5549 &  79.4 & -11.94 & 11.34 & 2 \\
NGC 2798          & 09 17 22.8 & +41 59 59 &  1725 &  28.6 & -11.96 & 10.43 & 3 \\
IRAS 09022-3615   & 09 04 12.7 & -36 27 01 & 17880 & 248   & -12.29 & 11.98 & 4 \\  
UGC 05101         & 09 35 51.6 & +61 21 11 &  1802 & 164   & -12.19 & 11.71 & 4 \\
NGC 3034/M 82     & 09 55 52.7 & +69 40 46 &   228 &   3.3 & -10.28 & 10.23 & 5 \\ 
NGC 3079          & 10 01 57.8 & +55 40 47 &  1260 &  20.7 & -11.60 & 10.38 & 5 \\ 
NGC 3227          & 10 23 30.6 & +19 51 54 &  1157 &  20.3 & -12.32 &  9.48 & 4 \\
NGC 3256          & 10 27 51.3 & -43 54 13 &  2804 &  37.0 & -11.34 & 11.34 & 2 \\
NGC 3521          & 11 05 48.6 & -00 02 09 &   801 &   8.0 & -11.71 &  9.57 & 3 \\
NGC 3627/M66      & 11 20 14.9 & +12 59 30 &   727 &   6.5 & -11.61 &  9.50 & 3 \\
NGC 3628          & 11 20 17.0 & +13 35 23 &   469 &   8.5 & -11.54 &  9.59 & 5 \\ 
Arp 299/IC 694/NGC 3690&11 28 31.0&+58 33 41& 3089 &  49.1 & -11.32 & 11.45 & 2 \\
ESO320-G030       & 11 53 11.7 & -39 07 49 &  3232 &  39.2 & -11.78 & 11.03 & 2 \\   
NGC 3982          & 11 56 28.1 & +55 07 31 &  1109 &  21.0 & -12.37 &  9.73 & 4 \\
NGC 4038          & 12 01 53.0 & -18 52 03 &  1642 &  23.3 & -11.65 & 10.57 & 4,5 \\
NGC 4039          & 12 01 53.5 & -18 53 10 &  1642 &  23.3 & -11.65 & 10.57 & 4 \\
NGC 4051          & 12 03 09.6 & +44 31 53 &   700 &  12.9 & -12.27 &  9.53 & 4 \\
NGC 4151          & 12 10 32.6 & +39 24 21 &   995 &  20.0 & -12.51 &  9.15 & 4 \\
NGC 4254/M99      & 12 18 49.6 & +14 24 59 &  2407 &  39.8 & -11.78 & 10.90 & 3 \\
NGC 4321/M100     & 12 22 54.8 & +15 49 19 &  1571 &  14.1 & -11.88 &  9.90 & 3 \\
NGC 4388          & 12 25 46.7 & +12 39 44 &  2524 &  41.4 & -12.24 &  9.76 & 4 \\
NGC 4418          & 12 26 54.6 & -00 52 39 &  2179 &  34.7 & -11.73 & 10.83 & 2 \\
IC 3639           & 12 40 52.8 & -36 45 21 &  3275 &  35.3 & -12.23 & 10.51 & 4 \\
NGC 4536          & 12 34 27.0 & +02 11 17 &  1808 &  30.8 & -11.81 & 10.64 & 3 \\
NGC 4631          & 12 42 08.0 & +32 32 29 &   806 &   7.6 & -11.50 &  9.74 & 3 \\
NGC 4736/M94      & 12 50 53.0 & +41 07 15 &   308 &  4.83 & -11.50 &  9.35 & 3,5 \\
Mrk 231/7Zw490    & 12 56 14.2 & +56 52 25 & 12642 & 178   & -11.8  & 12.19 & 2,5 \\
NGC 4826/M64      & 12 56 43.6 & +21 40 59 &   357 &   3.8 & -11.66 &  9.23 & 3,5 \\ 
NGC 4945          & 13 05 27.5 & -49 28 06 &   555 &   4.3 & -10.67 &  9.94 & 3,5 \\ 
IRAS 13120-5453   & 13 15 06.3 & -55 09 23 &  9222 & 134   & -11.71 & 12.01 & 2 \\   
NGC 5055/M63      & 13 15 49.3 & +42 01 45 &   484 &   8.3 & -11.66 &  9.66 & 3 \\
Arp 193/IC 883    & 13 20 35.3 & +34 08 22 &  6985 & 103   & -12.10 & 11.38 & 2 \\
NGC 5128/Cen~A    & 13 25 27.6 & -43 01 09 &   269 &   3.8 & -11.01 &  9.62 & 4,5 \\
NGC 5135          & 13 25 44.0 & -29 50 01 &  4105 &  57.7 & -12.04 & 10.97 & 2 \\
ESO 173-G015      & 13 27 23.8 & -57 29 22 &  2918 &  32.4 & -11.44 & 11.28 & 2 \\   
NGC 5194/M51      & 13 29 52.7 & +47 11 43 &   672 &   9.1 & -11.59 & 10.85 & 3,5 \\ 
NGC 5236/M83      & 13 37 00.9 & -29 51 56 &   245 &   4.0 & -11.22 &  9.34 & 4,5 \\ 
Mrk 273/1Zw71     & 13 44 42.1 & +55 53 13 & 11326 & 162   & -11.9  & 12.00 & 2 \\
Circinus          & 14 13 09.9 & -65 20 21 &   434 &   2.9 & -10.92 &  9.76 & 5 \\ 
\noalign{\smallskip}                                                  
\hline                                                                
\end{tabular}
\end{center}
\label{targetlist}
\end{table*}

\begin{table*}
\addtocounter{table}{-1}
\caption[]{Galaxy sample continued}
\begin{center}
\begin{tabular}{lccccccc}
\hline
\noalign{\smallskip}
Name             &RA(2000)$^{a}$&DEC(2000)$^{a}$&V$_{LSR}^{a}$&$D^{a}$&lg $FIR^{b}$&lg $L_{FIR}$&Table$^{a}$\\ 
                 & ( h m s)    &  (d m s)  &($\kms$)& (Mpc)& (W m$^{-2}$) & (L$_{\odot}$) & \\
\noalign{\smallskip}
\hline
\noalign{\smallskip}
CGCG 049-057      & 15 13 13.1 & +07 13 32 &  3897 &  61.9 & -11.96 & 11.01 & 2 \\   
Arp 220/IC 1127   & 15 34 57.1 & +23 30 11 &  5434 &  82.9 & -11.31 & 11.96 & 2,5 \\
NGC 6090/Mrk496   & 16 11 40.7 & +52 27 24 &  8785 & 126   & -12.47 & 11.21 & 5 \\
NGC 6240          & 16 52 58.9 & +02 24 03 &  7339 & 109   & -11.96 & 11.56 & 2 \\
IRAS F17207-0014  & 17 23 21.9 & -00 17 01 & 12834 & 183   & -11.83 & 12.17 & 2 \\   
SagA*/MilkyWay    & 17 45 40.0 & -29 00 28 &     0 & 0.0085&  ---   & 0.00  & 4 \\
IC 4687           & 18 13 39.6 & -57 43 31 &  5200 &  77.3 & -12.08 & 11.14 & 2 \\   
IRAS F18293-3413  & 18 32 41.1 & -34 11 27 &  5449 &  81.1 & -11.75 & 11.51 & 2 \\   
NGC 6946          & 20 34 52.3 & +60 09 14 &   385 &   5.5 & -11.48 & 10.30 & 3,5 \\
NGC 7130          & 21 48 19.5 & -34 57 04 &  4842 &  68.7 & -12.06 & 11.06 & 4 \\
NGC 7172          & 22 02 01.9 & -31 52 11 & 12603 &  37.6 & -12.45 & 10.08 & 4 \\
NGC 7331          & 22 37 04.0 & +34 24 56 &  816  &  14.4 & -11.78 & 10.01 & 3 \\
NGC 7469          & 23 03 15.6 & +08 52 26 &  4892 &  67.0 & -11.88 & 11.29 & 2 \\ 
NGC 7469          & 23 03 15.6 & +08 52 26 &  4892 &  67.0 & -11.88 & 11.29 & 2 \\
NGC 7552          & 23 16 10.7 & -42 35 05 &  1608 &  22.5 & -11.44 & 10.76 & 2 \\
NGC 7582          & 23 18 23.5 & -42 22 14 &  1575 &  22.0 & -11.61 & 10.49 & 5 \\
NGC 7771          & 23 51 24.8 & +20 06 42 &  4277 &  58.0 & -11.97 & 11.08 & 2 \\
Mrk 331           & 23 51 26.8 & +20 35 10 &  5541 &  74.9 & -12.09 & 11.18 & 2 \\
\noalign{\smallskip}                                                  
\hline                                                                
\end{tabular}
\end{center}
(a) The number in this column refers to the table where the respective line fluxes can be found
\label{targetlist}
\end{table*}

%
\begin{table*}
\caption[]{\ci\ and CO line fluxes from HerCULES galaxies}
\begin{center}
\begin{tabular}{lcccccccrr}
\noalign{\smallskip}
\hline
\noalign{\smallskip}
Name              & \multicolumn{6}{c}{Observed line flux}                     & Herschel & Ref$^a$ \\
                  &CO (4-3)&CO (7-6)&\ci\ (1-0)&\ci\ (2-1)&CO (2-1)&$^{13}$CO (2-1)& obsid    &  \\  
                  & 461 GHz& 807 GHz& 492 GHz& 809 GHz& 230 GHz& 220 GHz&                  &     \\
                  &\multicolumn{4}{c}{(10$^{-17}$ W m$^{-2}$)}&\multicolumn{2}{c}{(10$^{-19}$ W m$^{-2}$)} &\\
\noalign{\smallskip}
\hline
\noalign{\smallskip}
NGC 34            &   ---  &  2.54  &  0.75  &  1.25  &   10   &   ---  & 1342199253 & PS14 \\
NGC 253           &   128  &   182  &  41.5  &   114  & 2355   & 229    & 1342210847 & R14, *\\
MGC+12-02-001     &  3.07  &  2.69  &  1.05  &  2.59  &   ---  &   ---  & 1342213377 & - \\
IC 1623           &  4.25  &  3.41  &  2.40  &  3.83  &   ---  &   ---  & 1342212314 & - \\
IRAS F05189-2524  &   ---  &  0.87  &  0.25  &  0.61  &   10   &   1.5  & 1342192833 & P12, PS14\\
NGC 1365          &  41.7  &  26.2  &  19.6  &  31.8  &  385   &  34    & 1342204021 & I, * \\
NGC 1614          &  2.61  &  3.43  &  1.10  &  2.67  &   95   &   ---  & 1342192831 & A95 \\
NGC 2146          &  17.0  &  17.8  &  5.68  &  15.3  &  402   &  35    & 1342193810 & I09a * \\
NGC 2623          &  1.92  &  2.84  &  0.86  &  2.44  &   21   &   ---  & 1342219553 & P12 \\
NGC 3256          &  17.7  &  18.9  &  7.87  &  16.0  &  261   &  10    & 1342201201 & S06 \\
Arp 299C/IC 694   &  5.75  &  5.34  &  1.79  &  3.82  &   74   &   8.9  & 1342199250 & S12, A95, *\\
Arp 299B          &  5.17  &  5.38  &  1.99  &  3.88  &   77   &   7.1  & 1342199249 & S12, A95, *\\
Arp 299A/NGC 3690 &  8.71  &  12.0  &  2.44  &  7.81  &  139   &   6.9  & 1342199248 & S12, A95, *\\
ESO 320-G030      &  4.39  &  4.54  &  1.72  &  3.44  &   ---  &   ---  & 1342210861 & - \\
NGC 4418          &  1.17  &  5.17  &  1.83  &  1.08  &   ---  &   ---  & 1342187780 & - \\
Mrk 231           &   ---  &  2.46  &  0.58  &  1.34  &   24   &   0.45 & 1342210493 & P12 \\
IRAS 13120-5453   &   ---  &  6.70  &  2.35  &  5.39  &   ---  &   ---  & 1342212342 & - \\
Arp 193/IC 883    &  3.20  &  3.41  &  1.35  &  3.70  &   65   &   1.9  & 1342209853 & P12, P14 \\
NGC 5135          &  3.53  &  3.04  &  3.36  &  6.18  &   95   &   6.4  & 1342212344 & P12, PS14 \\
ESO 173-G015      &  12.1  &  17.5  &  4.90  &  14.5  &   ---  &   ---  & 1342202268 & - \\
Mrk 273           &   ---  &  2.82  &  0.51  &  1.76  &   21   &   2.6  & 1342209850 & P12 \\
CGCG 049-057      &   ---  &  2.99  &  0.96  &  1.41  &   46   &   1.7  & 1342212346 & P12 \\
Arp 220           &  6.77  &  13.4  &  2.95  &  7.42  &   86   &   4.3  & 1342190674 & P12 \\
NGC 6240          &  7.78  &  16.6  &  3.04  &  9.54  &  114   &   1.9  & 1342214831 & P14 \\
IRAS F17207-0014  &   ---  &  4.47  &  1.12  &  2.66  &   53   &   ---  & 1342192829 & P12 \\
IC 4687           &  1.67  &  1.56  &  0.91  &  1.57  &   ---  &   ---  & 1342192993 & - \\
IRAS F18293-3413  &  5.88  &  6.24  &  4.26  &  7.43  &   ---  &   ---  & 1342192830 & - \\
NGC 7469          &  3.24  &  3.74  &  2.14  &  4.63  &   73   &   5.3  & 1342199252 & P12, I09a, * \\
NGC 7552          &  12.4  &  11.6  &  5.46  &  10.8  &  208   &  22    & 1342198428 & A95 \\
NGC 7771          &  4.49  &  3.49  &  5.86  &  7.29  &   ---  &  ---   & 1342212317 & - \\
Mrk 331           &  2.45  &  2.30  &  1.34  &  2.57  &   ---  &  ---   & 1342212316 & - \\
\noalign{\smallskip}
\hline
\end{tabular}
\end{center}
\label{herculesdat}
Note: a. References to ground-based $J$=2-1 $\co$ and $^{13}$CO observations: 
A95 =  Aalto et al. (1995), I09a = Israel (2009a); 
I = Israel (JCMT, this paper); P12 = Papadopoulos et al (2012); 
P14 = Papadopoulos et al. (2014), PS14 = Pereira-Santaella et al. (2014); 
R14 = Rosenberg et al. (2014a); S06 = Sakamoto et al. (2006). S12 = Sliwa et 
al. (2012). For galaxies marked by an asterisk (*) we have determined 
$J$=2-1 $\co$ and $\thirco$ fluxes corresponding to the {\it SPIRE} aperture 
by integration over the (JCMT) map.
\end{table*}

%
\begin{table*}
\caption[]{\ci\ and CO line fluxes from Herschel archive galaxies}
\begin{center}
\begin{tabular}{lcccccccrr}
\noalign{\smallskip}
\hline
\noalign{\smallskip}
Name              & \multicolumn{6}{c}{Observed line flux}                     & Herschel & Ref$^a$ \\
                  &CO (4-3)&CO (7-6)&\ci\ (1-0)&\ci\ (2-1)&CO (2-1)&$^{13}$CO (2-1)& obsid    &  \\  
                  & 461 GHz& 807 GHz& 492 GHz& 809 GHz& 230 GHz& 220 GHz&                  &     \\
                  &\multicolumn{4}{c}{(10$^{-17}$ W m$^{-2}$)}&\multicolumn{2}{c}{(10$^{-19}$ W m$^{-2}$)} &\\
\noalign{\smallskip}
\hline
\noalign{\smallskip}
NGC 1275/Per A    & 0.86   &  0.56  & 0.48   & 1.05   &   13   &   2.0  & 1342249055 & BI98 \\
NGC 1482          &  0.52  &  0.19  & 0.32   & 0.61   &   61   &   ---  & 1342248233 & A07 \\
NGC 2798          &  1.48  &  2.65  & 0.67   & 2.11   &   ---  &   ---  & 1342252892 & \\
NGC 3521          &  0.49  &  0.09  & 0.59   & 0.50   &   ---  &   ---  & 1342247743 & \\
NGC 3627/M66      &  0.80  &  0.26  & 0.68   & 0.61   &   92   &   7.3  & 1342247604 & I \\
NGC 4254/M99      &  0.82  &  0.18  & 1.14   & 1.20   &   58   &   5.4  & 1342236997 & I \\ 
NGC 4321/M100     &  0.60  &  0.18  & 0.38   & 0.47   &  100   &   7.8  & 1342247572 & I \\
NGC 4536          &  1.71  &  1.99  & 0.23   & 2.27   &   90   &   7.4  & 1342237025 & I \\
NGC 4631          &  1.65  &  0.57  & 0.83   & 1.60   &  119   &   5.5  & 1342247553 & I09a, * \\
NGC 4736/M94      &  0.68  &  0.26  & 0.51   & 0.86   &   69   &   5.9  & 1342245851 & GP00, Ba06 \\
NGC 4826/M64      &  0.75  &  0.34  & 0.82   & 1.04   &  301   &  40    & 1342246992 & IB02, I09a, * \\
NGC 4945          & 34.9   & 90.8   & 21.5   & 91.7   & 1570   & 131    & 1342212343 & W04 \\
NGC 5055/M63      &  1.08  &  0.31  & 0.95   & 1.60   &   78   &  13    & 1342237026 & I \\
NGC 5194/M51      &  1.67  &  0.62  & 1.21   & 2.36   &  205   &  28    & 1342201202 & IB02, ITB, * \\ 
NGC 5713          &  0.57  &  0.23  & 0.26   & 0.42   &   49   &   5.6  & 1342248250 & I \\ 
Circinus          &  41.3  &  36.2  &  17.1  &  41.8  &  592   &  56    & 1342251313 & H08 \\ 
NGC 6946          &  5.07  &  5.74  & 2.20   & 5.36   &  789   &  79    & 1342243603 & IB01, IB02, * \\
NGC 7331          &  1.03  &  0.16  & 0.56   & 0.95   &   20   &   3.4  & 1342245871 & IB99, * \\
\noalign{\smallskip}
\hline
\end{tabular}
\end{center}
\label{archivedat}
Note: a. References to ground-based $J$=2-1 $\co$ and $^{13}$CO observations: 
A07 = Albrecht $\etal$ (2007); Ba06 = Bayet $\etal$ (2006); BI98 = Bridges 
$\&$ Irwin (1998); GP00 = Gerin $\&$ Phillips (2000); H08 = Hitschfeld et al. 
(2008);  I09a = Israel (2009a); I = Israel (JCMT, this paper); IB01 = Israel 
$\&$ Baas (2001); IB02 = Israel $\&$ Baas (2002); ITB = Israel, Tilanus, 
$\&$ Baas (2006); W04 = Wang $\etal$ (2004).
For galaxies marked by an asterisk (*) we have determined 
$J$=2-1 $\co$ and $\thirco$ fluxes corresponding to the {\it SPIRE} aperture 
by integration over the (JCMT) map.
\end{table*}

%
\begin{table*}
\caption[]{Published {\it Herschel-SPIRE} \ci\ and CO line fluxes}
\begin{center}
\begin{tabular}{lccccccr}
\noalign{\smallskip}
\hline
\noalign{\smallskip}
Name              & \multicolumn{6}{c}{Observed line flux}                    & Ref$^{a}$ \\
                  &CO (4-3)&CO (7-6)&\ci\ (1-0)&\ci\ (2-1)& CO(2-1)&$^{13}$CO (2-1)&  \\  
                  & 461 GHz& 807 GHz& 492 GHz  & 809 GHz  & 230 GHz& 220 GHz      &  \\
                  &\multicolumn{4}{c}{(10$^{-17}$ W m$^{-2}$)}&\multicolumn{2}{c}{(10$^{-19}$ W m$^{-2}$)}& \\
\noalign{\smallskip}
\hline
\noalign{\smallskip}  
NGC 1056          &    --  &  0.61  &    --  &  0.77  &  10.4  &  ---     & PS13 \\ 
NGC 1068/M 77     &  21.7  &  21.1  &  12.8  &  21.1  &   866  &  76      & Sp12, P12* \\
IRAS 09022-3615   &    --  &  3.33  &  1.22  &  1.52  &  ---   &  ---     & \\
UGC 05101         &    --  &  1.05  &    --  &  1.28  &   9.7  &   0.13   & PS13 \\  
NGC 3034/M 82     &  113   &  209   &  33.8  &  117   &  2070  &  142     & P10, W03* \\  
NGC 3227          &  2.90  &  1.81  &  2.27  &  4.66  &  26.5  &   6.9    & PS13, I \\ 
NGC 3982          &  1.81  &  0.47  &    --  &  1.00  &   4.3  &  ---     & PS13 \\ 
NGC 4038          &  5.47  &  1.48  &  1.00  &  1.86  &  88.2  &   3.7    & Sc14, I \\  
NGC 4039          &  2.92  &  1.85  &  1.65  &  2.41  & 109.6  &   7.7    & Sc14, I \\  
NGC 4051          &    --  &  0.73  &    --  &  0.78  &  24.9  &   1.3    & PS13, I \\ 
NGC 4151          &    --  &  0.36  &    --  &  0.84  &   3.0  &  ---     & PS13 \\ 
NGC 4388          &  2.39  &  1.31  &  1.45  &  2.67  &   5.5  &  ---     & PS13 \\ 
IC 3639           &  1.21  &  ---   &  ---   &  0.58  &  20.9  &  ---     & PS13  \\ 
NGC 5128/Cen A    &  11.3  &  4.16  &  11.5  &  30.2  &   172  &  12      & I14* \\  
NGC 5236/M 83     &  11.5  &  8.82  &  4.54  &  14.0  &   582  &  62      & K14, IB01, *\\  
NGC 7130          &  3.30  &  2.55  &  1.64  &  2.67  &  44.9  &  ---     & PS13 \\  
NGC 7172          &    --  &  ---   &  3.82  &  4.50  &  26.0  &  ---     & PS13 \\ 
NGC 7582          &  6.93  &  6.56  &  3.44  &  7.59  &   139  &  ---     & PS13, A95 \\
\noalign{\smallskip}
\hline
\noalign{\smallskip}  
MilkyWay ($\times 10^{-6}$)& 2.70  &  1.52  &  1.68  &1.68    &  0.98  &  ---     & F99 \\   
\noalign{\smallskip}
\hline
\end{tabular}
\end{center}
\label{spiredat}
Note: a. Reference to line fluxes: A95 =  Aalto et al. (1995), F99 = 
Fixsen et al. (1999) -- COBE measurement; I14 = Israel et al. (2014); I = Israel (JCMT, 
this paper); K14 = Kamenetzky $\etal$ (2014) P10 = Panuzzo et al. 
(2010); P12 = Papadopoulos et al. (2012), P14 = Papadopoulos et al 
(2014); PS13 = Pereira-Santaella et al. (2013); PS14 = 
Pereira-Santaella et al. (2014); Sc14 = Schirm et al (2014); 
Sp12 = Spinoglio et al. (2012); W03 = Ward et al. (2003).
For galaxies marked by an asterisk (*) we have determined 
$J$=2-1 $\co$ and $\thirco$ fluxes corresponding to the {\it SPIRE} aperture 
by map integration.
\end{table*}

\begin{table*}
\caption[]{Published ground-based \ci\ and CO line fluxes}
\begin{center}
\begin{tabular}{lccccccr}
\noalign{\smallskip}
\hline
\noalign{\smallskip}
Name              & \multicolumn{6}{c}{Observed line flux$^{a}$}                    & Ref$^b$ \\
                  &CO (4-3)&CO (7-6)&\ci\ (1-0)&\ci\ (2-1)&CO (2-1)&$^{13}$CO (2-1)&  \\  
                  & 461 GHz& 807 GHz& 492 GHz  & 809 GHz  & 230 GHz& 220 GHz      &  \\
                  &\multicolumn{4}{c}{(10$^{-17}$ W m$^{-2}$)}&\multicolumn{2}{c}{(10$^{-19}$ W m$^{-2}$)}& \\
\noalign{\smallskip}
\hline
\noalign{\smallskip}  
IC10              &  0.89  &  1.00  &  0.52  &  0.84  &    25  &   1.4    & GP00, Ba06 \\
NGC 253           &  101   &  ---   &  31.5  &  ---   &  1210  &  104     & IWB, IB02  \\
NGC 278           &  0.89  &  ---   &  0.54  &  ---   &  17.9  &   1.9    & IB02, I09b \\
NGC 660           &  8.39  &  ---   &  3.37  &  ---   &   154  &  11      & IB02, I09b \\
NGC 891           &  ---   &  ---   &  0.79  &  ---   &   118  &  12      & Ba06, I \\
Maffei 2          &  40.6  &  ---   &  2.17  &  ---   &   275  &  22      & IB02, IB03 \\ 
NGC 1068          &  10.8  &  ---   &  5.33  &  ---   &   247  &  17      & IB02, I09a \\
IC 342            &  20.6  &  ---   &  2.94  &  ---   &   193  &  24      & IB02, IB03 \\ 
Henize2-10        &  1.40  &  4.00  &  0.36  &  ---   &    20  &   1.2    & GP00, Ba06 \\
NGC 3079          &  11.4  &  ---   &  7.61  &  ---   &   189  &  11      & IB02, I09b, GP00 \\  
NGC 3628          &  10.9  &  ---   &  4.13  &  ---   &   169  &  13      & IB02, I09b \\
NGC 4038          &   2.6  &  0.6   &  0.94  &  ---   &   110  &  14      & GP00, Ba06 \\
NGC 4736/M94      &  ---   &  ---   &  0.57  &  ---   &    69  &   5.9    & GP00, Ba06 \\
NGC 4826/M64      &  8.49  &  ---   &  5.11  &  ---   &   126  &  15      & IB02, I09a \\ 
NGC 4945          &  56.8  &  ---   &  80.4  &  ---   &  1300  &  131     & H08 \\ 
NGC 5194/M51      &  2.76  &  ---   &  1.41  &  ---   &  43.5  &   5.7    & IB02, ITB, GP00 \\ 
NGC 5236/M83      &  26.7  &  ---   &  5.98  &  ---   &   291  &  28      & IB01, IB02 \\  
Circinus          &  18.0  &  26.1  &  8.20  &  ---   &   592  &  56      & H08, Z14 \\ 
NGC 6090/Mrk496   &  ---   &  ---   &  0.38  &  ---   &    43  &  ---     & GP00, Ba06 \\ 
NGC 6946          &  21.3  &  ---   &  4.78  &  ---   &   249  &  22      & IB01, IB02, GP00 \\ 
\noalign{\smallskip}
\hline
\end{tabular}
\end{center}
\label{grounddat}
Note: In this table only, all fluxes  refer to the same beamwidth 
of $22"$, except in the cases of NGC~4945 and the Circinus galaxy
where beamwidths of $38"$ apply, similar to the {\it SPIRE} aperture.
b. Reference to line fluxes: Ba06 = Bayet et al. (2006); GP00 = 
Gerin $\&$ Phillips (2000); H08 = Hitschfeld et al. (2008); IWB = 
Israel, White $\&$ Baas (1995); ITB = Israel, Tilanus, $\&$ Baas (2006); 
IB01 = Israel $\&$ Baas (2001); IB02 = Israel $\&$ Baas (2002); 
IB03 = Israel $\&$ Baas (2003); I09a = Israel (2009a); I09b = Israel 
(2009b); Z14 = Zhang $\etal$ (2014).
\end{table*}

Carbon monoxide (CO), the molecular species most commonly observed in
galaxies, is widely used as a tracer for molecular hydrogen
($\h2$). Molecular hydrogen accounts for the bulk of the molecular gas
and is a major component of the interstellar medium (ISM) in galaxies,
but is much harder to observe.  When shielding is reduced (low
metallicity) and radiation enhanced (starburst), CO is readily
photodissociated into its constituent parts, carbon (C) and oxygen
(O).  As the ionization potential of neutral carbon, (C$^{\circ}$) is
close to the dissociation energy of CO, and it can only exist in a
narrow range of physical conditions. This is clearly shown by a
variety of ISM models, for instance, those of Meijerink et al. (2007).

The CO and neutral carbon \ci\ submillimeter lines are the major
coolants for most of the dense molecular gas from which stars are
thought to form. Ground-based observations of \ci\ emission can be
performed from a few dry sites at high elevation, at least in the
$\ci$ (1-0) line, i.e., the $\pci$ transition at 492 GHz.  This is
much less practical for observations of the $\pbci$ line at 809 GHz,
hereafter designated as $\ci$ (2-1).  Like the higher $J$ transitions
of CO, the \ci\ lines are best observed from an airborne platform
(such as {\it SOFIA}), or from a platform in space (such as the now defunct
{\it Herschel Space Observatory}).

Until the {\it Herschel} space mission, the number of published
extragalactic \ci\ observations was limited.  Beyond the Local Group,
the \ci\ 492 GHz line was surveyed in two dozen galaxies by Gerin $\&$
Phillips (2000) and by Israel $\&$ Baas (2002), and mapped in a few bright northern galaxies (see Israel $\etal$ 2009b and
references therein). Gerin $\&$ Phillips found that, relative to the
low-$J$ $\co$ lines, $\ci$ (1-0) is weaker in galaxy nuclei but
stronger in disks, particularly outside star-forming regions.  Israel
$\&$ Baas compared the \ci\ (1-0) with the $J$=2--1 $\thirco$ and
$J$=4--3 $\co$ lines. They find that in galaxy centers, the \ci\
(1-0) brightness temperature is close to or slightly lower than that
of the adjacent $\co$ J=4-3 line, but significantly higher than the
$J$=2-1 $\thirco$ brightness temperature.  This is very different from
the situation pertaining to Galactic photon-dominated regions (PDRs),
where the \ci\ brightness temperatures are invariably much lower, both
as predicted by models (Kaufman $\etal$ 1999) and as observed in
actual fact (Plume $\etal$ 1999, Tatematsu $\etal$ 1999). Israel $\&$
Baas also find that the \ci\ (1-0)/$\thirco$ (2-1) temperature ratio
increases with the \ci\ luminosity in the same beam.  They use
radiative transfer models to show that the C$^{\circ}$ abundance
usually exceeds that of the CO. Both groups concluded that the [CI]
emission arises from a warm and moderately dense
($n=10^{3}$-$10^{4}\,\cc$) gas.

With the mission of the {\it Herschel Space Observatory}, more
measurements of galaxies in both [CI] lines have become available.
Some of these have already been published, and others are presented here
for the first time. Table\,\ref{targetlist} lists the sample of
galaxies in which at least one \ci\ line has been measured, which
will be discussed in this paper. Positions, radial velocities, and
distances\footnote{Cosmology-corrected luminosity distances assuming
  $H_{0}$=73.0 km/sec/Mpc, $\omega_{m}$=0.27, $\omega_{vac}$=0.73,
  corrected to the reference frame defined by the 3K microwave
  background} were taken from the {\it \emph{NASA-IPAC Extragalactic Database}
(NED)}, and the logarithmic far-infrared fluxes $FIR$ were taken from
the \emph{{\it IRAS Point Source Catalog (PSC)}}; $L_{FIR}$ is the corresponding
luminosity in solar units. This sample contains 76 galaxies, of which
22 can be classified as LIRGs (Luminous InfraRed Galaxies,
$L_{FIR}\geq10^{11}$ L$_{\odot}$) and 6 as ULIRGs (UltraLuminous
InfraRed Galaxies, $L_{FIR}\geq10^{12}$ L$_{\odot}$).  Several of
the sample galaxies contain an AGN (active galactic nucleus) in
addition to the starburst that usually dominates the far-infrared
continuum and submillimeter line emission. However, in Centaurus A, the
AGN is dominating the emission over any starburst contribution.

The purpose of this paper is fourfold: (a) to present an extensive
set of \ci\ fluxes covering a wide range of (starbursting) galaxy
centers, ranging from relatively normal galaxies to LIRGs and ULIRGs;
(b) to determine the behavior of the observed \ci\ lines with respect
to each other, $\co$, and $\thirco$; (c) to investigate whether the
derived \ci\ and $\co$ line ratios can be used to characterize the
molecular ISM of the parent galaxies without resorting to the more
elaborate fitting of detailed models (PDR, XDR, CR, etc.) and at the
same time to gain understanding of how the molecular gas properties
define the model results; and (d) to specifically address the
question of whether the \ci\ line luminosity reliably traces molecular
hydrogen column densities so that it can be successfully used to
determine overall molecular mass.

\section{DATA}

\subsection{Herschel-SPIRE}

\nobreak A large sample of (ultra)luminous galaxies was observed as
part of the {\it \emph{HerCULES program}} (cf Rosenberg et al. 2015) with the
Spectral and Photometric Imaging Receiver and Fourier-Transform
Spectrometer ({\it SPIRE-FTS} - Griffin \etal 2010) onboard the {\it
  Herschel Space Observatory}\footnote{{\it Herschel} is an {\it ESA} space
  observatory with science instruments provided by European-led
  Principal Investigator consortia and with important participation
  from {\it NASA}} (Pilbratt et al. 2010) in the single-pointing mode with
sparse image sampling. The observations are summarized in
Table\,\ref{herculesdat}. In addition to these galaxies, we also
extracted a set of more nearby
galaxies from the {\it Herschel} Archives (mostly taken from the SINGS sample). Many of these were
observed with {\it SPIRE} in spectral mapping mode, but for our
purpose we only extracted the central resolution element. These data
are summarized in Table\,\ref{archivedat}. The {\it FTS} has two detector
arrays (the SLW: wavelength range 303-671 $\mu$m corresponding to a
frequency range 447-989 GHz, and the SSW: wavelength range 194-313
$\mu$m) corresponding to a frequency range 959-1544 GHz). All $\co$
lines in the $J$=4-3 to $J$=13-12 transitions were covered, as well as
the two submillimeter \ci\ lines. In this paper, we only discuss the
\ci\ lines, and the adjacent $J$=4-3 and $J$=7-6 $\co$ lines that were
all observed with the SLW.  The $\co$ (4-3) lines occurred close to
the spectrum edge, and their fluxes could not be recovered in some
cases, either because the noise increase at the edge made it
impossible to reliably fit the line or because the line was
red-shifted out of the spectrum altogether, typically at
$V_{LSR}\geq5500\,\kms$. This is also the case for a smaller number
of \ci\ (1-0) lines in galaxies with $V_{LSR}\geq11000\,\kms$.

The {\it SPIRE} spectral resolution of 1.21 GHz is insufficient to
resolve the lines in the SLW spectrum that contains the transitions
discussed in this paper.  Fluxes were first extracted using FTFitter
(https://www.uleth.ca/phy/naylor/index.php?page=ftfitter), a program
specifically created to extract line fluxes from {\it SPIRE-FTS}
spectra.  This is an interactive data language (IDL) based graphical
user interface that allows the user to fit lines, choose line
profiles, fix any line parameter, and extract the flux.  We defined a
polynomial baseline to fit the continuum, derived the flux from the
baseline-subtracted spectrum and fitted it with an unresolved line
profile.  The beam SLW FWHM values are given in the on-line {\it
  Herschel-SPIRE} manual: $40"$ for $\co$ (4-3) and an almost
identical $35"$ for $\co$ (7-6), \ci\ (1-0), and \ci\ (2-1). In
Tables\,\ref{herculesdat} and \ref{archivedat} we present the \ci\,
$\co$ (4-3) and $\co$ (7-6) line fluxes extracted from the {\it SPIRE}
spectra. For a more detailed description of the data extraction and
reduction method used, we refer to Rosenberg $\etal$ (2015).

\begin{figure*}[t]
\begin{minipage}[]{2.54cm}
\resizebox{2.7cm}{!}{\rotatebox{0}{\includegraphics*{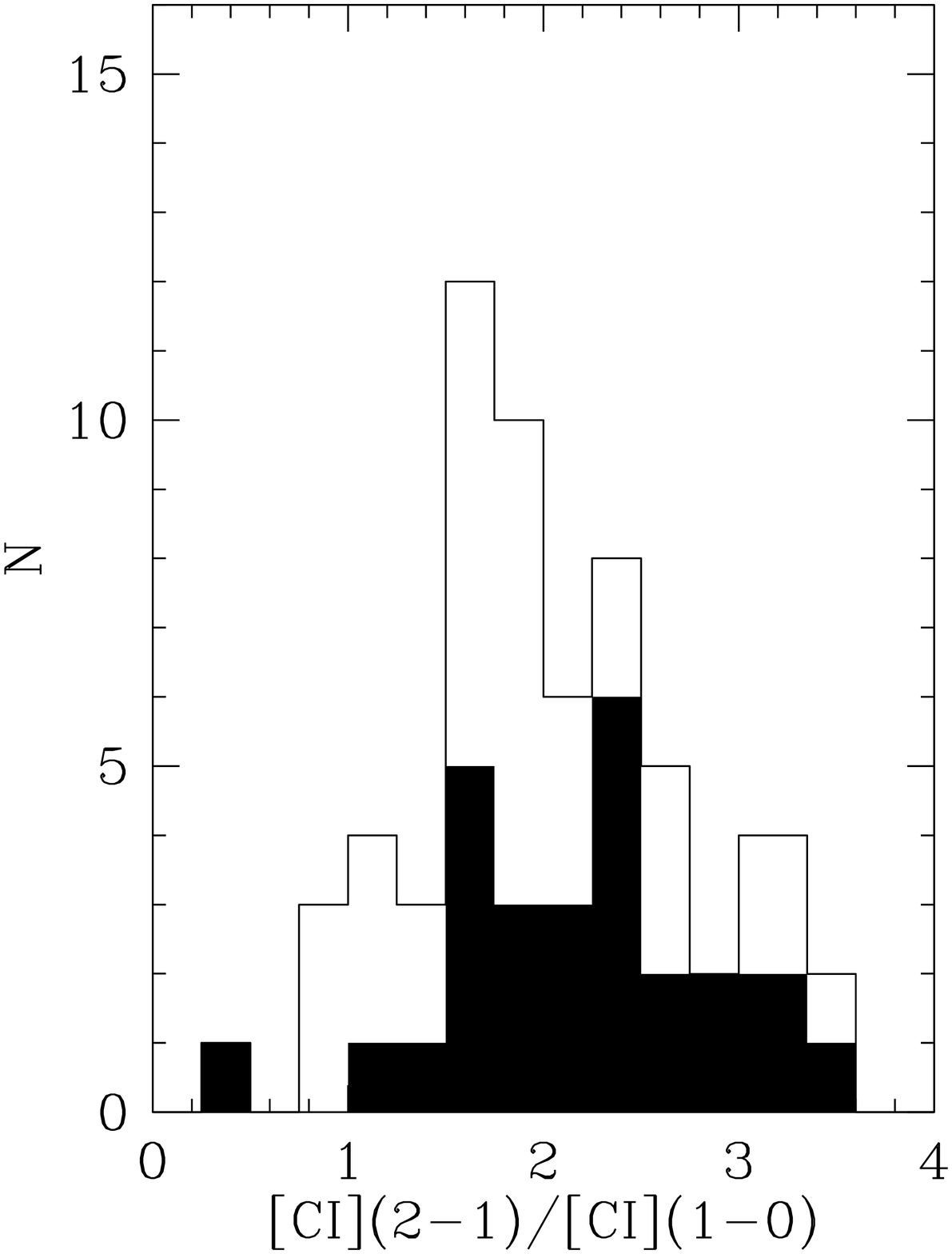}}}
\end{minipage}
\begin{minipage}[]{2.54cm}
\resizebox{2.7cm}{!}{\rotatebox{0}{\includegraphics*{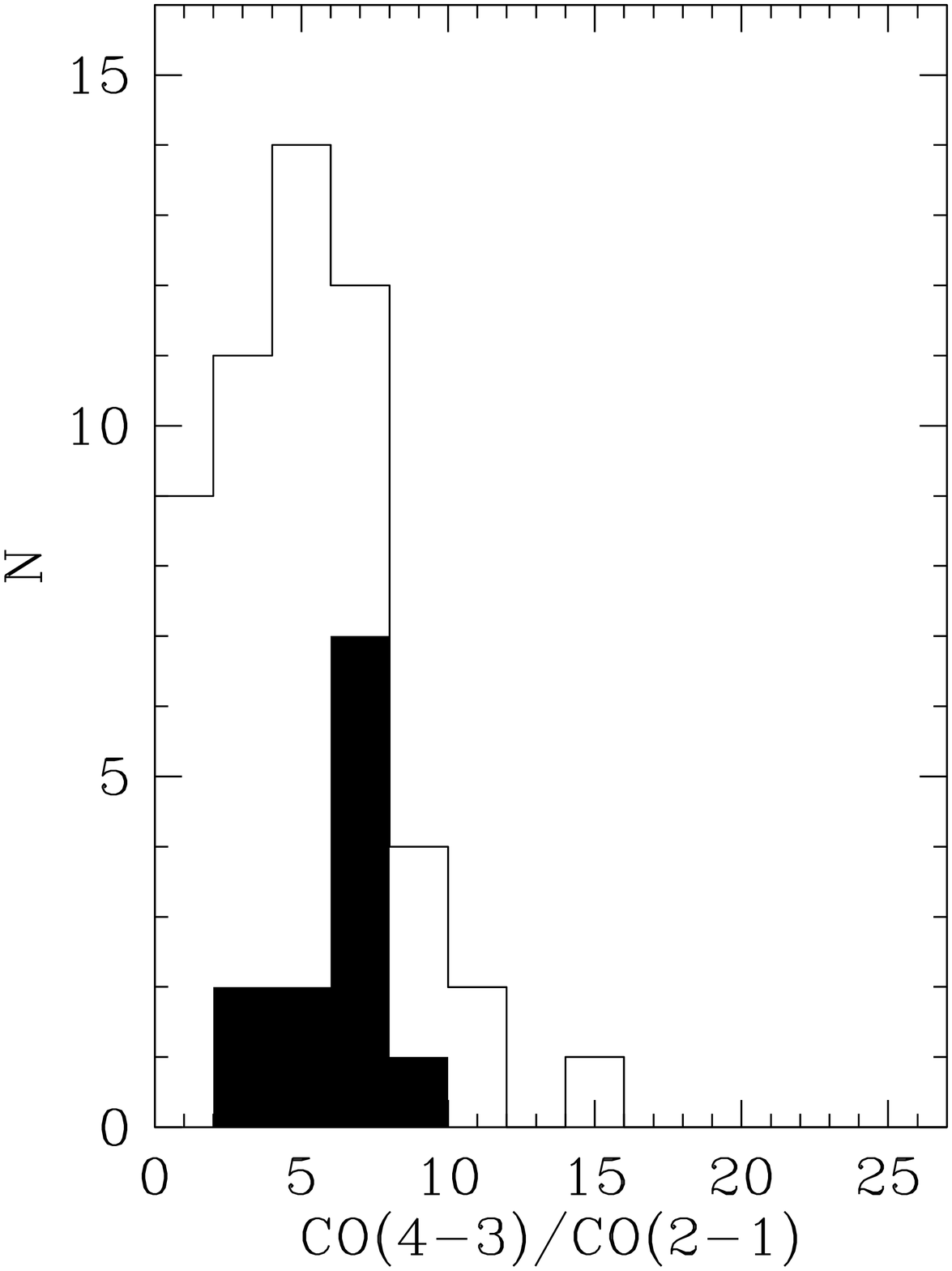}}}
\end{minipage}
\begin{minipage}[]{2.54cm}
\resizebox{2.7cm}{!}{\rotatebox{0}{\includegraphics*{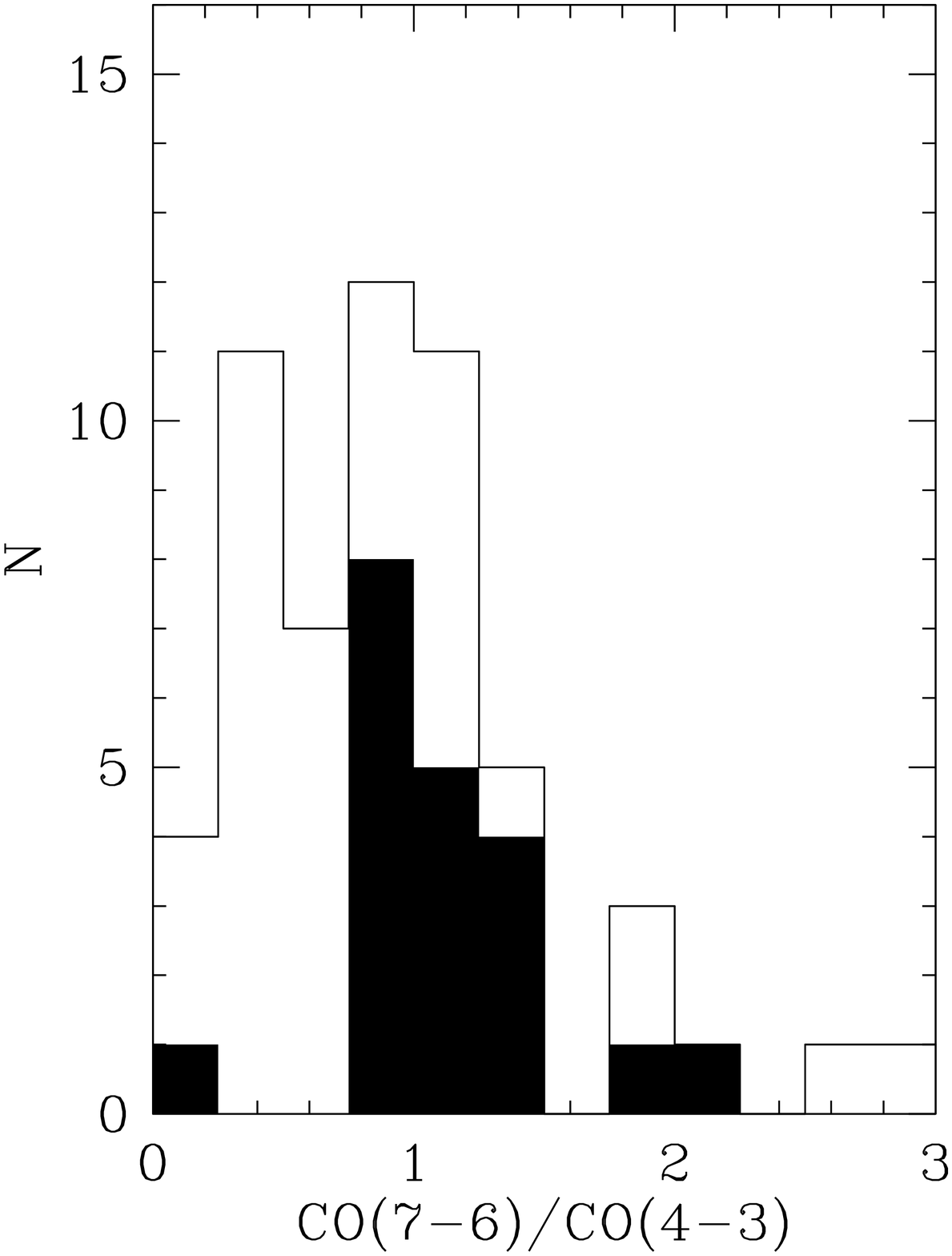}}}
\end{minipage}
\begin{minipage}[]{2.54cm}
\resizebox{2.7cm}{!}{\rotatebox{0}{\includegraphics*{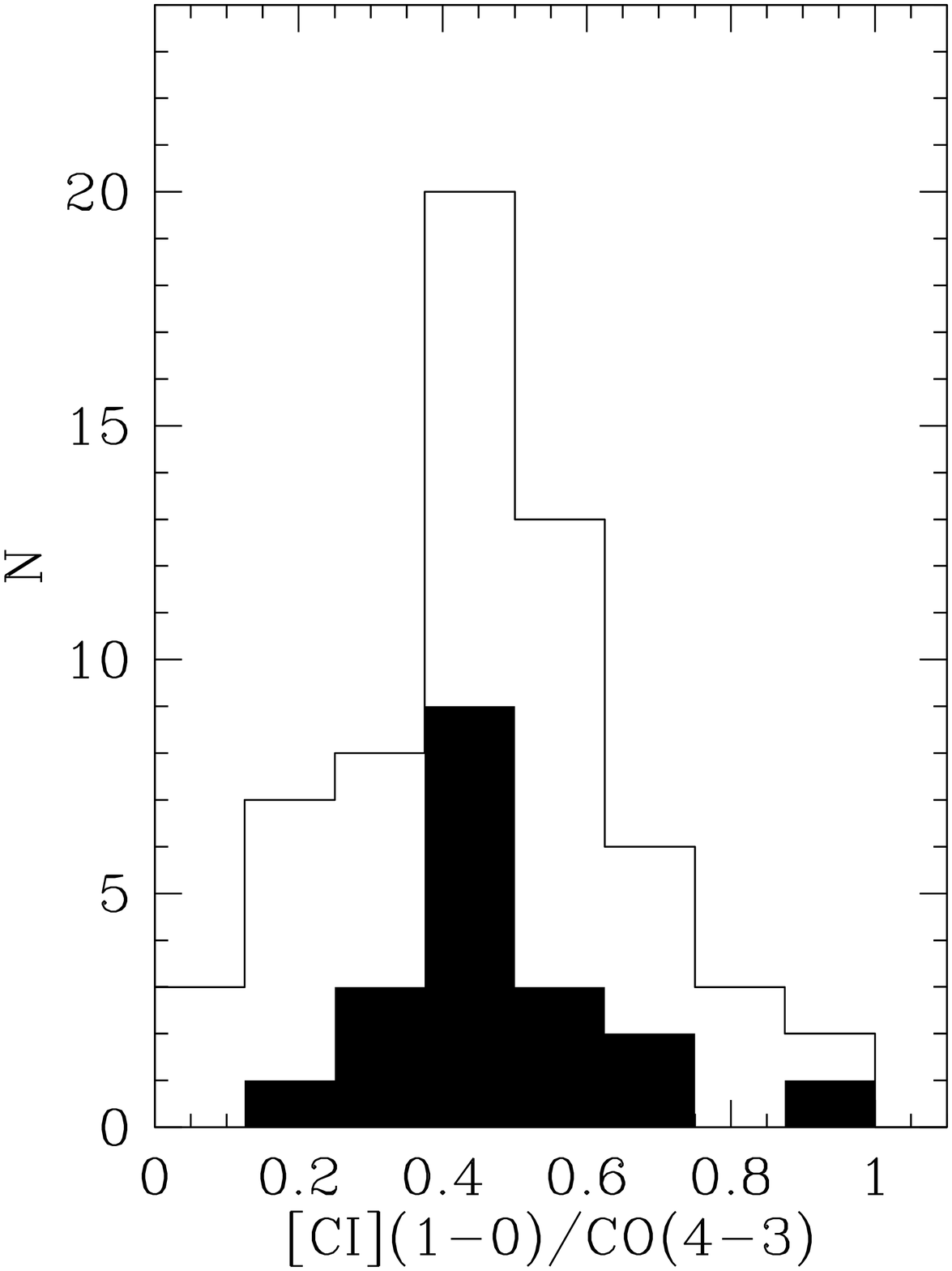}}}
\end{minipage}
\begin{minipage}[]{2.54cm}
\resizebox{2.7cm}{!}{\rotatebox{0}{\includegraphics*{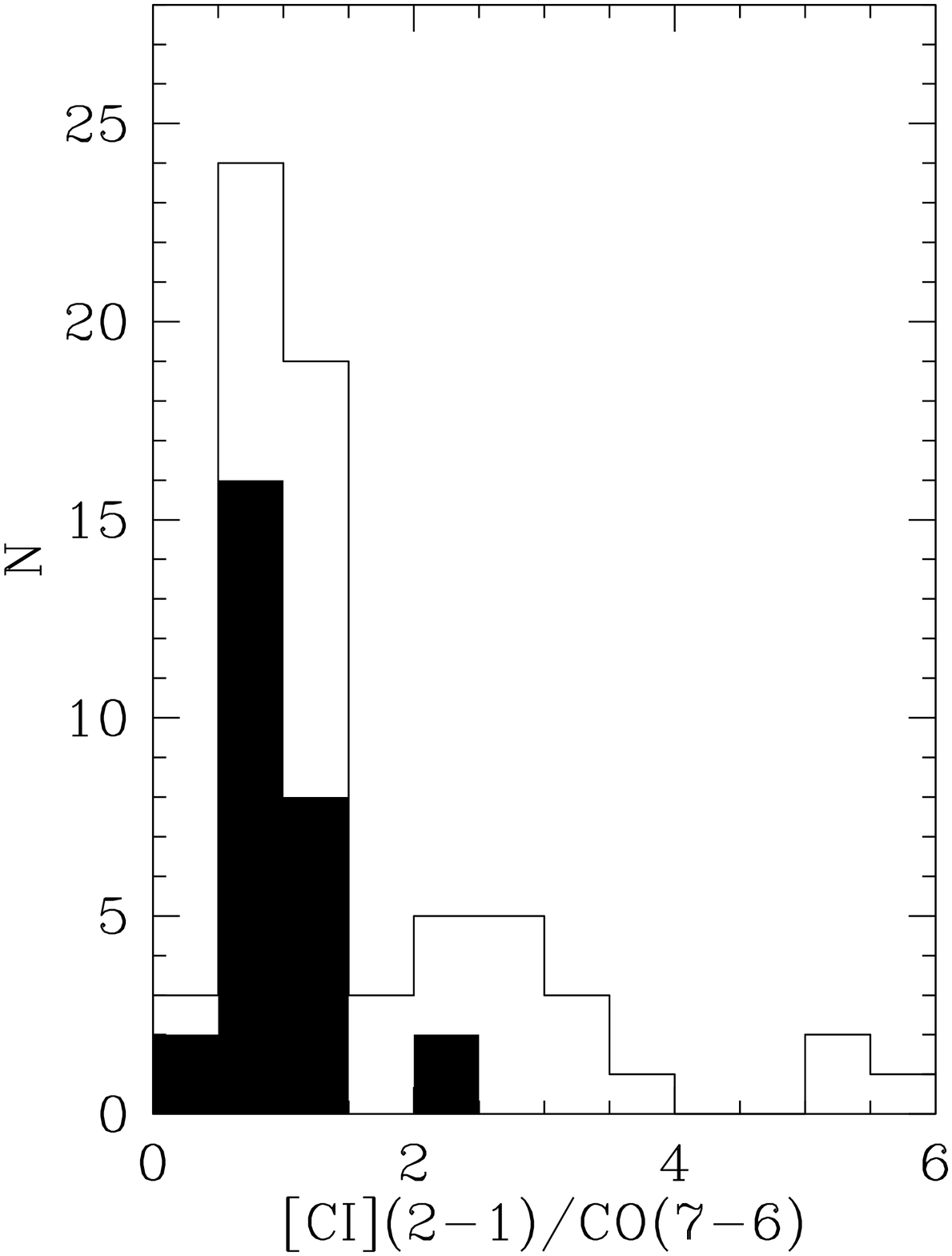}}}
\end{minipage}
\begin{minipage}[]{2.54cm}
\resizebox{2.7cm}{!}{\rotatebox{0}{\includegraphics*{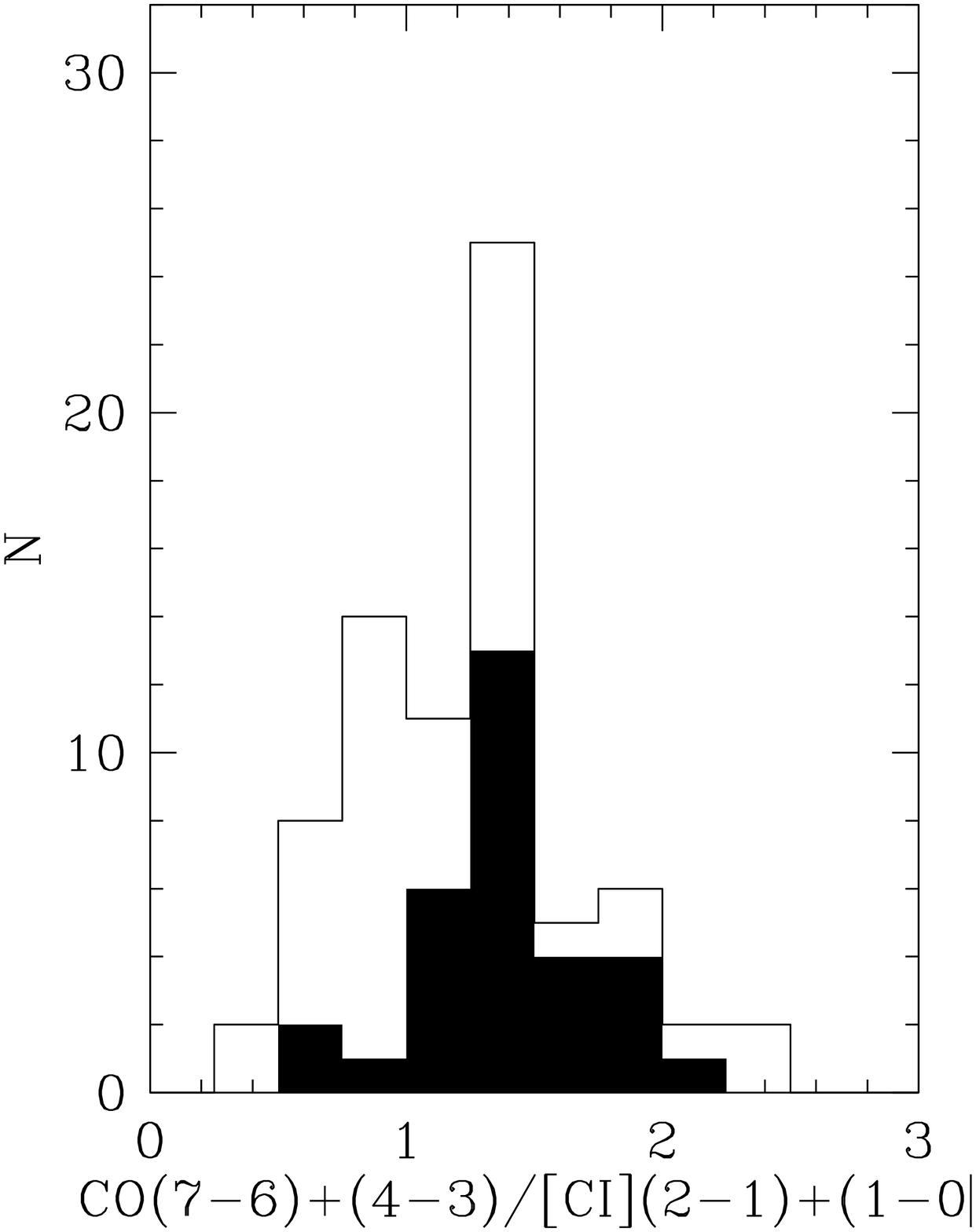}}}
\end{minipage}
\begin{minipage}[]{2.54cm}
\resizebox{2.7cm}{!}{\rotatebox{0}{\includegraphics*{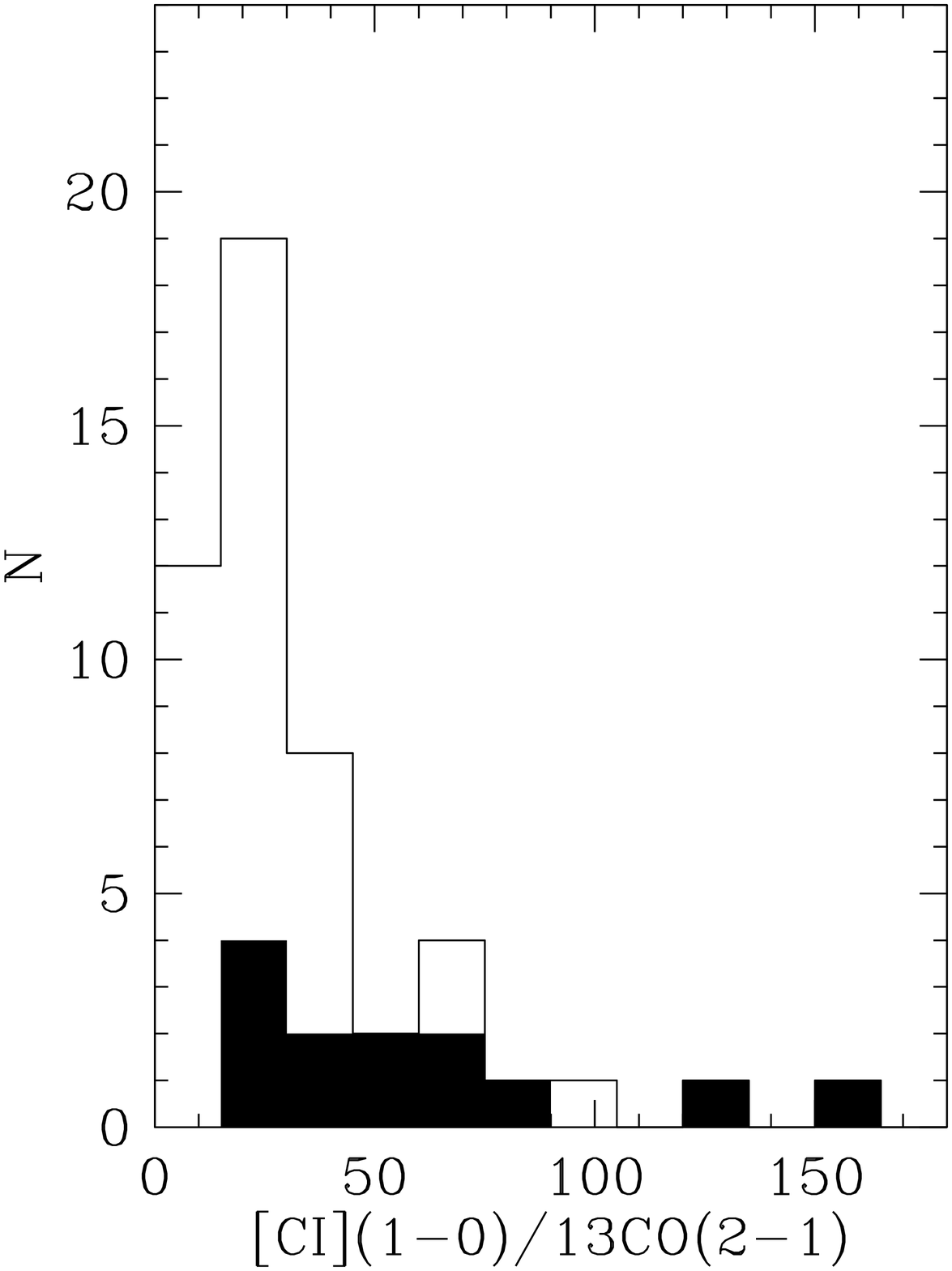}}}
\end{minipage}
\caption[] {Distribution of the various $\ci$ and CO line flux ratios over
  the total sample of galaxies. The subsample of luminous galaxies
  (LIRGS and ULIRGS) is indicated by the shaded area.  }
\label{ratbinfig}
\end{figure*}

\begin{table*}
\caption[]{Average line ratios}
\begin{center}
\begin{tabular}{lcccccc}
\noalign{\smallskip}
\hline
\noalign{\smallskip}
Ratio  & \multicolumn{2}{c}{Complete}  & \multicolumn{2}{c}{Total}  & (U)LIRG  &  Starburst \\
       & \multicolumn{2}{c}{subsample} & \multicolumn{2}{c}{sample} & \multicolumn{2}{c}{Subsample}\\
                 & n  & average        & n  & average \\
\noalign{\smallskip}
\hline
\noalign{\smallskip}  
$\ci$(2-1)/[$\ci$(1-0)         & 35 & 2.08$\pm$0.11 & 63 & 2.12$\pm$0.11 & 2.17$\pm$0.13 & 1.86$\pm$ 0.12\\
CO(4-3)/CO(2-1)                & 35 & 4.22$\pm$0.41 & 55 & 5.08$\pm$0.42 & 5.80$\pm$0.73 & 4.27$\pm$ 0.39\\
CO(7-6)/CO(4-3)                & 35 & 0.84$\pm$0.09 & 57 & 0.97$\pm$0.10 & 1.11$\pm$0.10 & 0.90$\pm$ 0.15 \\
$\ci$(1-0)/CO(4-3)             & 35 & 0.48$\pm$0.04 & 62 & 0.46$\pm$0.03 & 0.47$\pm$0.04 & 0.46$\pm$ 0.03\\
$\ci$(2-1)/CO(7-6)             & 35 & 2.03$\pm$0.30 & 66 & 1.67$\pm$0.19 & 0.92$\pm$0.08 & 2.19$\pm$ 0.29\\
$\ci$(1-0)/$\thirco$(2-1)      & 35 & 32.6$\pm$5.5  & 49 & 33.2$\pm$4.5  & 59.1$\pm$12.3 & 23.9$\pm$ 2.1\\ 
CO(sum)/$\ci$(sum)             & 35 & 1.25$\pm0.09$ & 60 & 1.25$\pm$0.06 & 1.39$\pm$0.10 & 1.16$\pm$ 0.09\\
\noalign{\smallskip}
\hline
\end{tabular}
\end{center}
\label{ratcomp}
\end{table*}

\begin{figure*}[t]
\begin{minipage}[]{4.5cm}
\resizebox{4.7cm}{!}{\rotatebox{0}{\includegraphics*{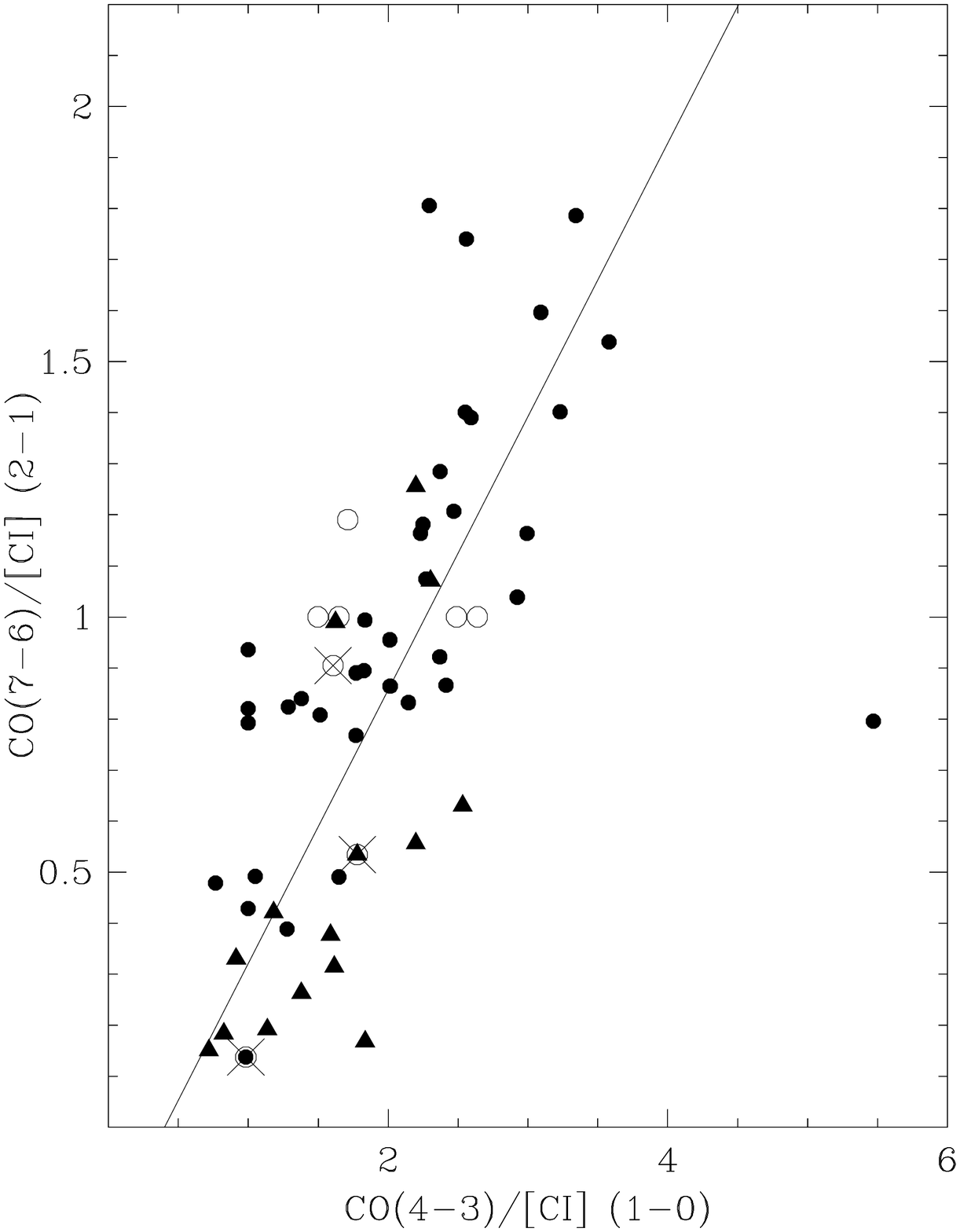}}}
\end{minipage}
\begin{minipage}[]{4.5cm}
\resizebox{4.7cm}{!}{\rotatebox{0}{\includegraphics*{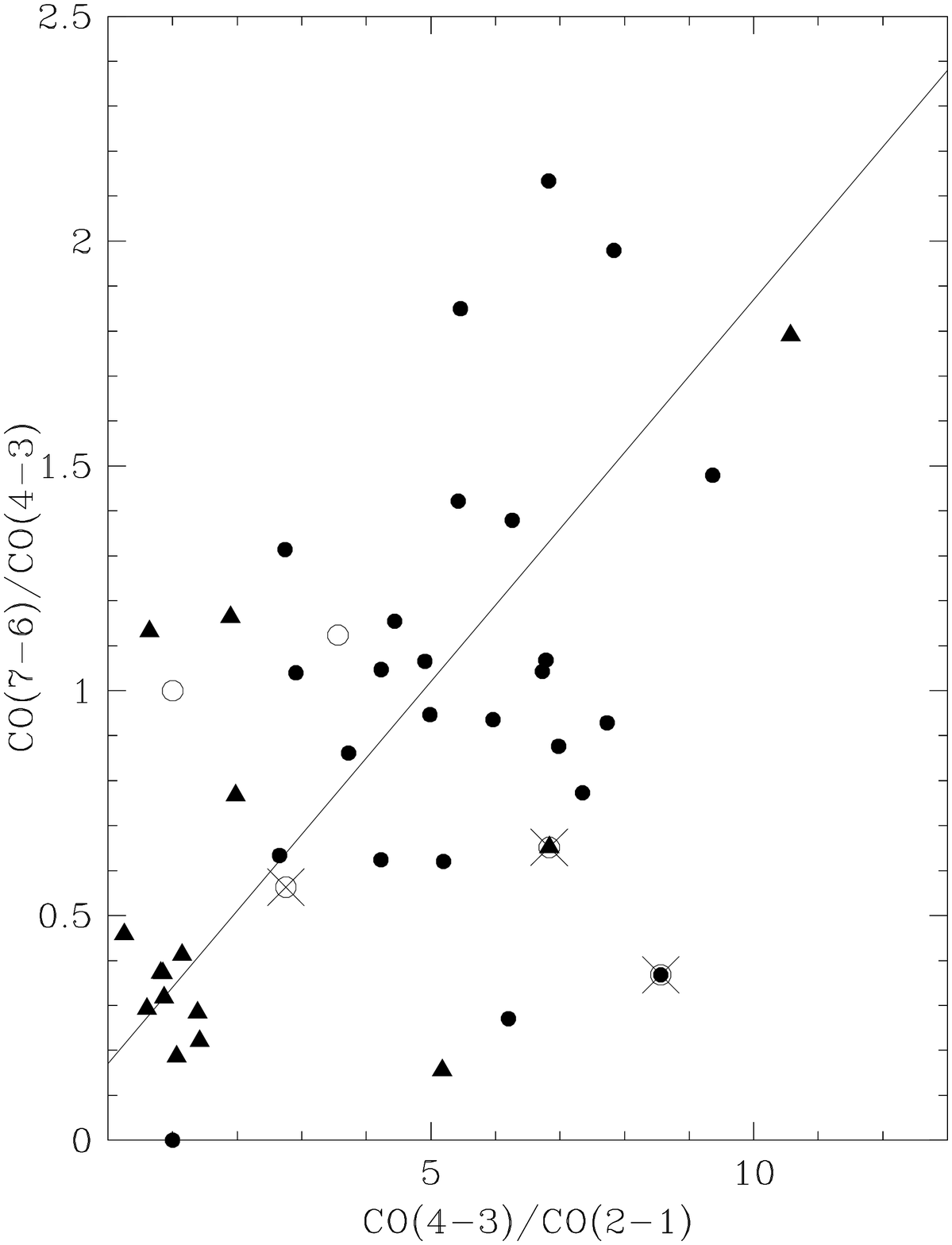}}}
\end{minipage}
\begin{minipage}[]{4.5cm}
\resizebox{4.7cm}{!}{\rotatebox{0}{\includegraphics*{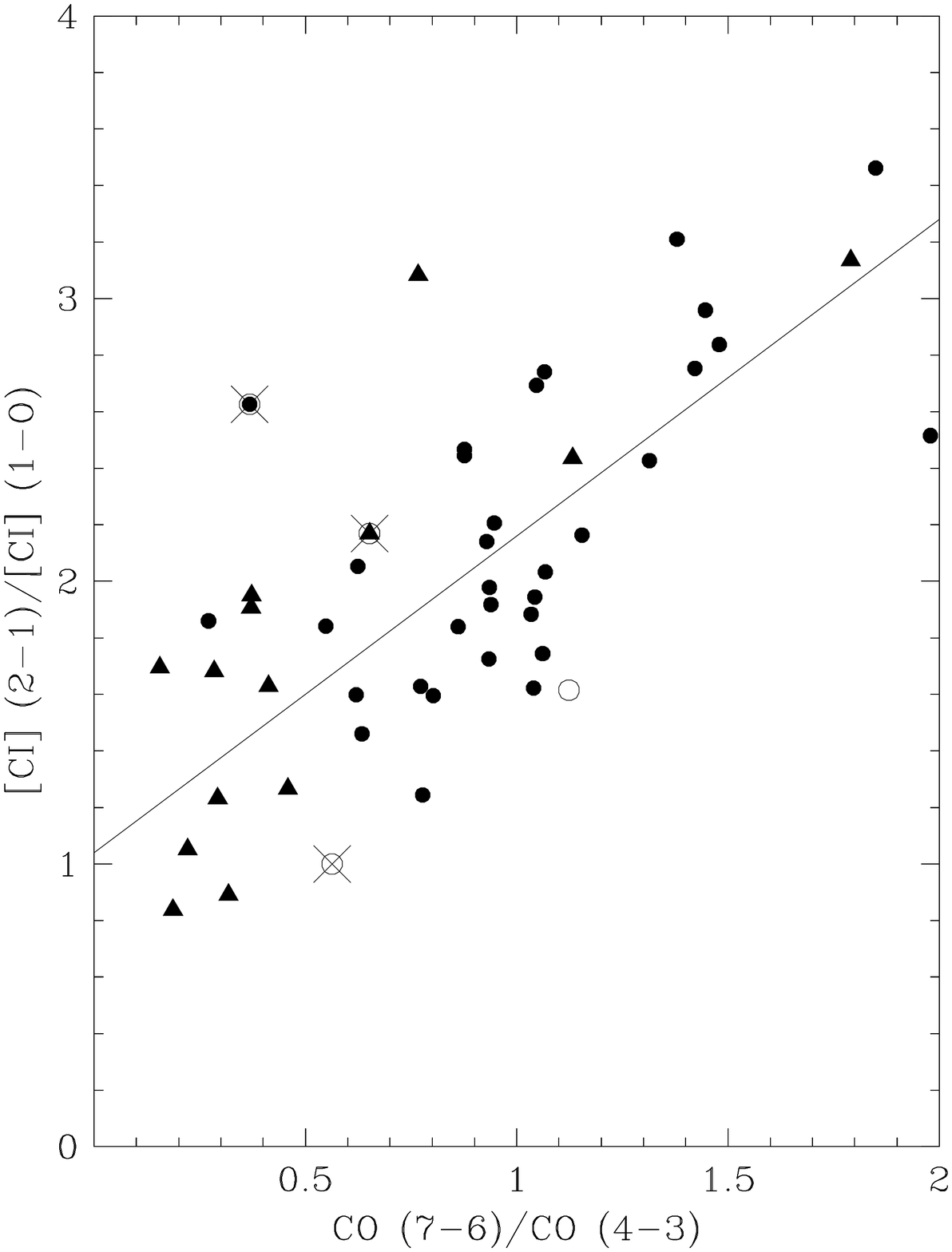}}}
\end{minipage}
\begin{minipage}[]{4.5cm}
\resizebox{4.7cm}{!}{\rotatebox{0}{\includegraphics*{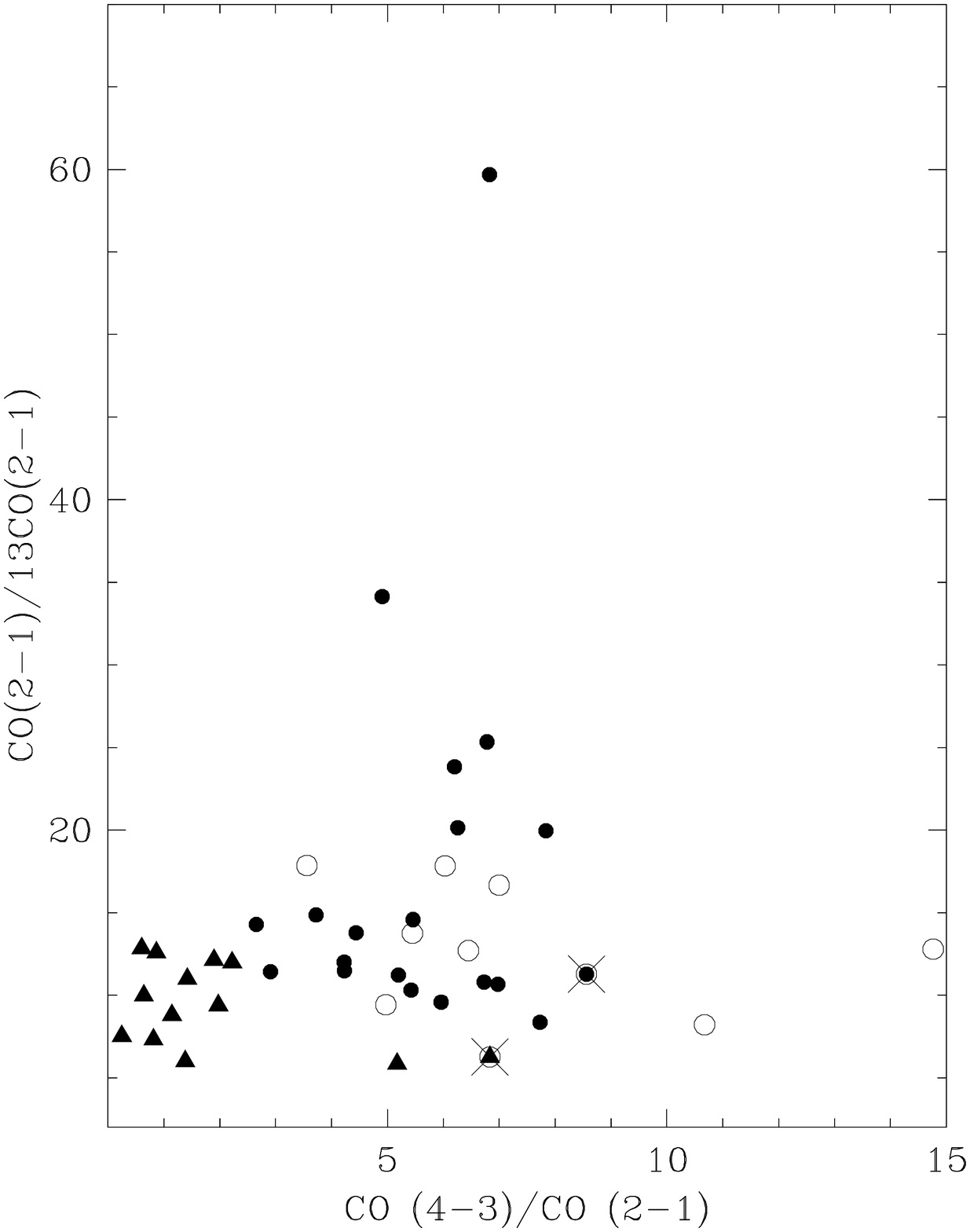}}}
\end{minipage}
\caption[]{Line flux ratio diagrams for the sample galaxies. Ratios derived
  from {\it Herschel-SPIRE} data are marked by filled circles (single
  pointings) and filled triangles (map central peak), those derived
  from ground-based observations are denoted by open circles. Points
  corresponding to Centaurus~A, Perseus~A, and the Milky Way are
  indicated by an additional cross. a. (left): \ci\ lines compared with
  adjacent CO lines. The linear regression correlation marked in the
  panel has a slope of 0.55, and a correlation coefficient
  $r^{2}$=0.66. b. (second from left): CO transition line ratios are
  likewise correlated with a slope of 0.17 and $r^{2}$=0.37.
  c. (second from right) \ci\ line ratio versus the ratio of their
  neigbouring CO lines, slope 1.12 and $r^{2}$=0.67. d. (right):
  $J$=2-1 CO isotopic ratio as a function of the CO (4-3)/CO (2-1)
  ratio; no obvious correlation although the dispersion increases with CO ratio.
 }
\label{ratratfig}
\end{figure*}

\begin{figure*}[t]
\begin{minipage}[]{17.5cm}
\resizebox{6.15cm}{!}{\rotatebox{0}{\includegraphics*{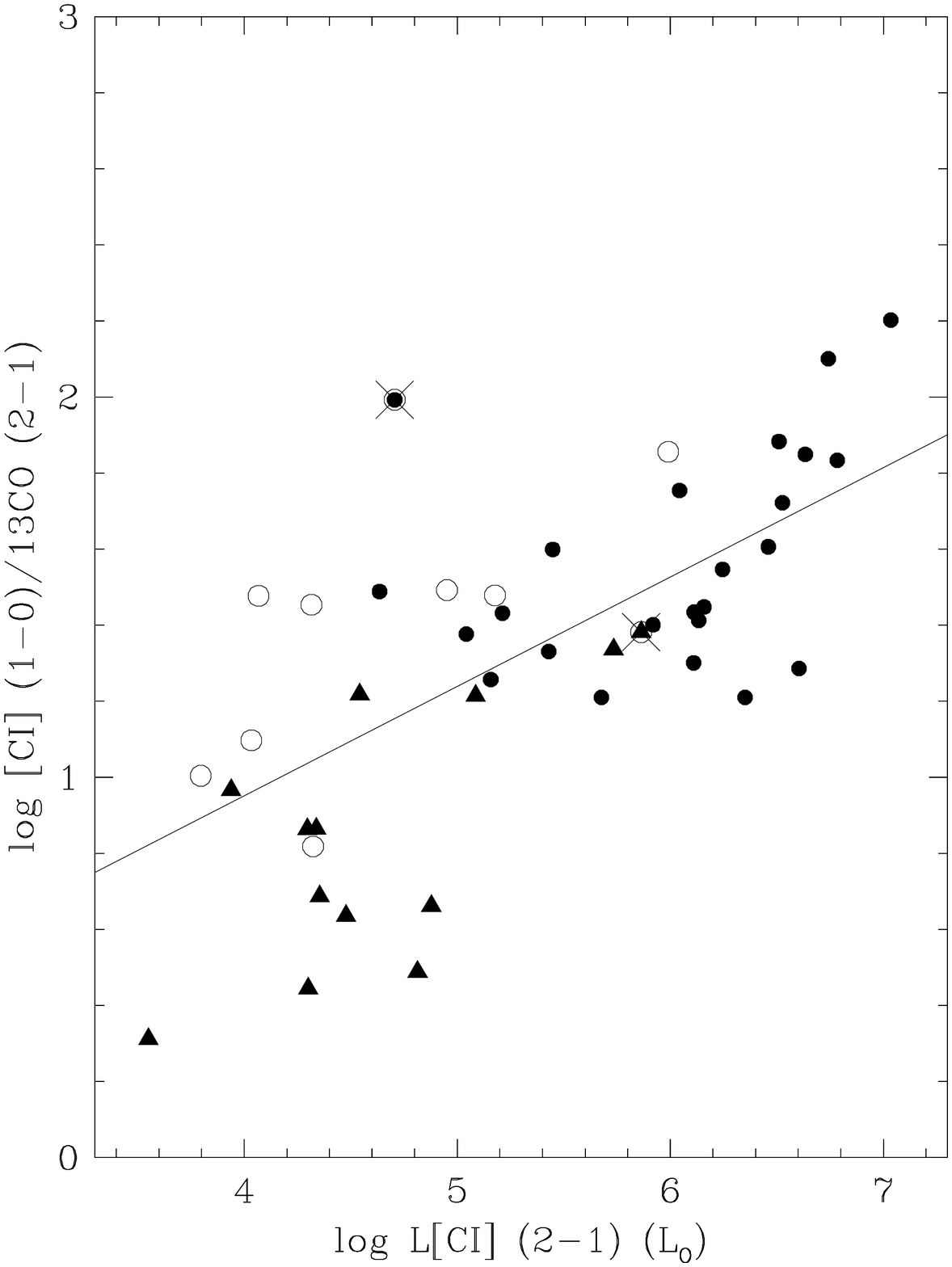}}}
\resizebox{6.15cm}{!}{\rotatebox{0}{\includegraphics*{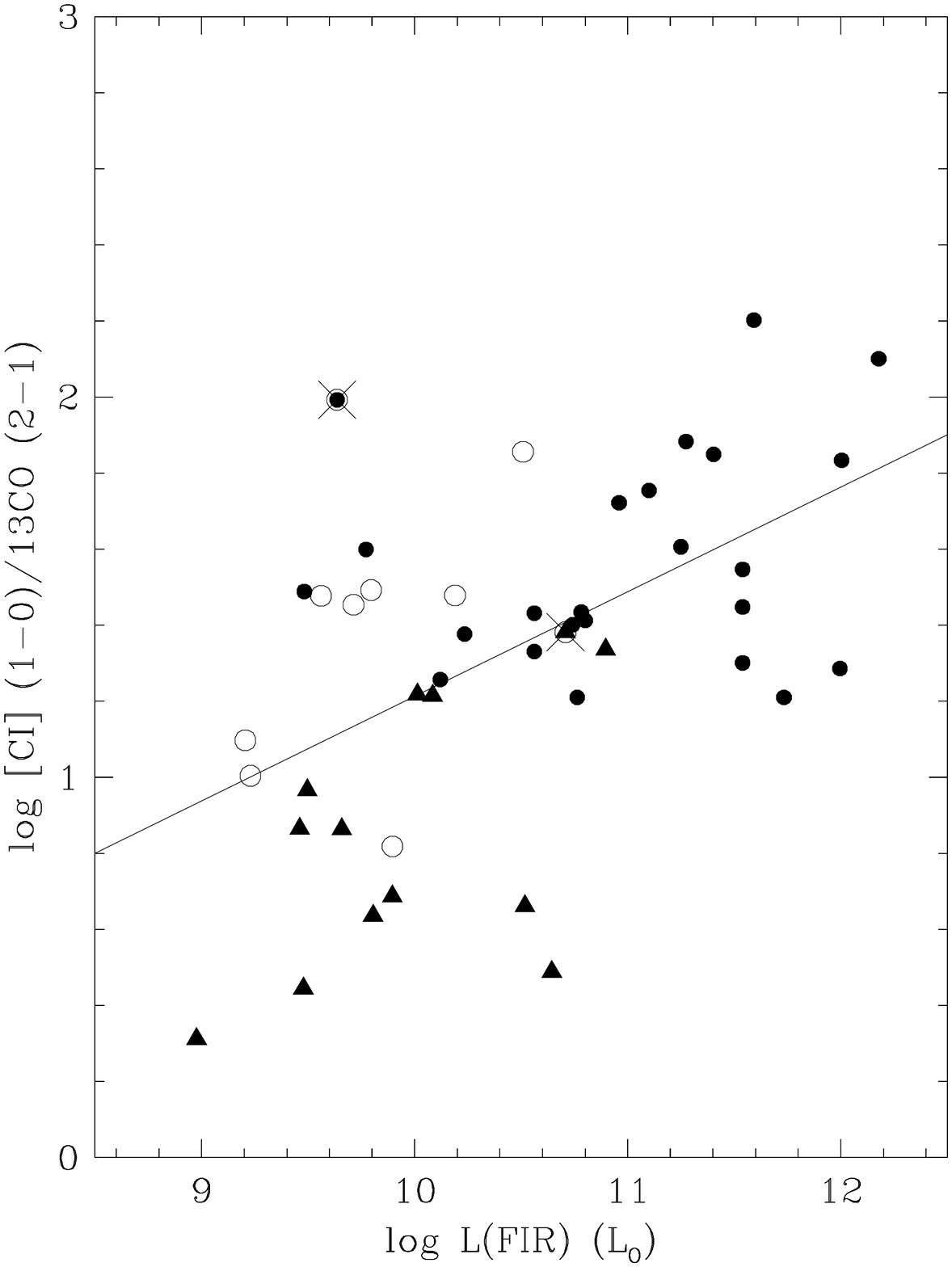}}}
\resizebox{6.15cm}{!}{\rotatebox{0}{\includegraphics*{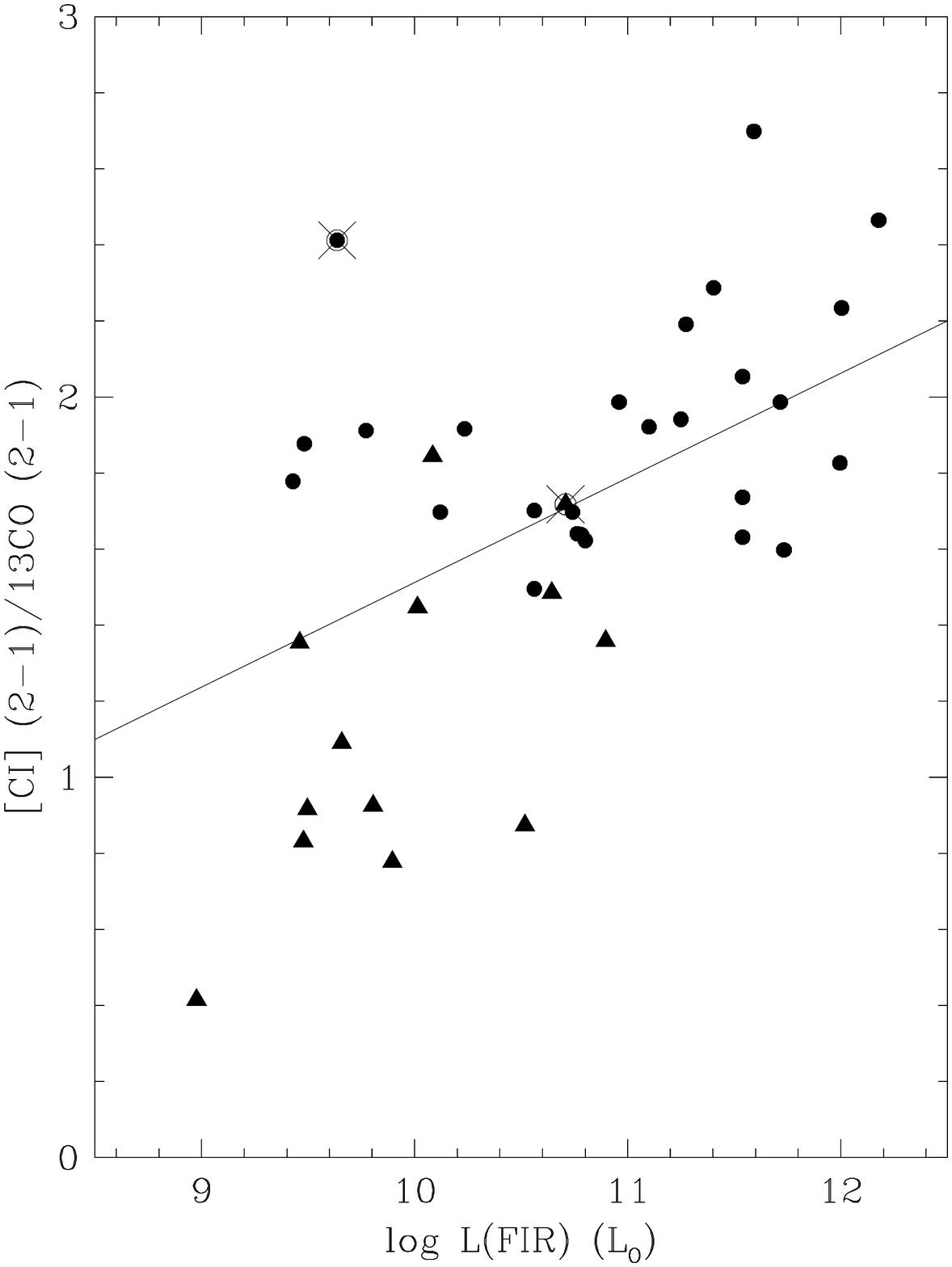}}}
\end{minipage}
\caption[]{Flux ratio of the \ci\ $J$=1-0 to $\thirco$ J=2-1 line in the
  sample galaxies as a function of the \ci\ $J$=1-0 luminosity (left),
  and the total far-infrared luminosity (center); the panel on the
  right depicts the ratio of the $\ci$ $J$=(2-1) line to the $\thirco$
  line, again as a function of far-infrared luminosity. As in the
  previous figure, ratios derived from {\it Herschel-SPIRE} data are
  indicated by filled circles and filled squares, those derived from
  ground-based observations by open circles.  Centaurus~A
  and Perseus~A are indicated by an additional cross. Linear regression
  correlation slopes and coefficients are from left to right:
  0.27/0.42, 0.26/0.28, 0.28/0.32}
\label{ratlumfig}
\end{figure*}

\subsection{JCMT 15m}

\nobreak The 15m James Clerk Maxwell Telescope ({\it
  JCMT})\footnote{The James Clerk Maxwell Telescope is operated by
  the Joint Astronomy Centre on behalf of the Science and Technology
  Facilities Council of the United Kingdom, the National Research
  Council of Canada, and (until 31 March 2013) the Netherlands
  Organization for Scientific Research.} on top of Mauna Kea (Hawaii)
was used between 2003 and 2005 to measure the $J$=2-1 transitions of
$\co$ and $\thirco$ at 230 and 220 GHz, respectively, toward a large
number of galaxy centers. At the observing frequencies, the beam size
was 22$''$ and the main-beam efficiency was 0.7. We used the facility
receiver RxA3 and the Digital Autocorrelating Spectrometer (DAS).  All
observations were taken in beam-switching mode with a throw of 3$'$ in
azimuth.  Spectra were binned to various resolutions; we applied
linear baseline corrections only and scaled the spectra to main-beam
brightness temperatures. Line parameters were determined by Gaussian
fitting and by adding channel intensities over the relevant range. A
subset of the data has already been published (references given at the
bottom of each table).

\subsection{Other data}

In the past few years, {\it Herschel-SPIRE} results have been
published for a variety of galaxies. Both \ci\ line fluxes and those
of the $\co$ (4-3) and $\co$ (7-6) lines in these galaxies are given
in Table\,\ref{spiredat}. The $\ci$ (1-0) 492 GHz and adjacent $\co$
(4-3) 461 GHz line fluxes of galaxies measured from the ground are
given in Table\,\ref{grounddat}. In Tables\,\ref{herculesdat},
\ref{spiredat}, and \ref{grounddat} we have also included $J$=2-1
$\co$ and $\thirco$ line fluxes, when available. For the distant
high-luminosity sources, most of these derive from {\it JCMT} or {\it
  IRAM} 30m observations. In general, these objects have (CO) sizes
that are smaller than the observing beam (see Papadopoulos $\etal$
2012). For the nearer and more extended galaxies, we used {\it JCMT}
or {\it SEST} observations with a beam size of $22"$. Exceptions are
the observations of Arp~299 taken from Aalto $\etal$ (1995: {\it NRAO}
12m, $28"$) and Sliwa et al. (2012: integrated over a combined {\it
  SMA/JCMT} map covering the source - see Rosenberg $\etal$
2014b). The data on M~82 were taken from Ward $\etal$ (2006) and cover
the inner lobes. The data on NGC~4945 and the Circinus galaxy were
taken from Hitschfeld $\etal$ (2008: {\it Nanten2}, $38"$) and Zhang
$\etal$ (2014; {\it APEX}, $14"$, corrected to $22"$). For the nearby
extended galaxies, the \ci\ and $\co$ fluxes closely represent the
emission from the bright central concentrations of gas and dust quite
well, but not at all the more extended disk emission.  In contrast, in
the much more distant luminous galaxies, the emission samples all of
the galaxy. Finally we note that several galaxies, especially the
brighter ones, such as NGC~253, NGC~1068, and the Circinus galaxy, are
found in both the {\it SPIRE} sample and the ground-based sample.

\subsection{Errors}

Most of the data presented in this paper were obtained with the {\it
  SPIRE} instrument. The error budget is discussed in some detail by Rosenberg $\etal$ (2015). The absolute flux values in
Tables\,\ref{herculesdat}, \ref{archivedat}, and \ref{spiredat} have
an absolute calibration uncertainty of $6\%$. To this must be added
another systematic uncertainty of $10\%$ associated with baseline
definition and flux extraction. Sources unresolved by the {\it SPIRE}
instrument thus have an absolute flux uncertainty $\Delta
F\,=\,15\%$. In the case of resolved sources, the correction procedure
used adds another $15\%$ to this: $\Delta F\,=\,30\%$.  Since systematic
errors dominate, the \emph{\emph{{\ line ratios are much more accurate}},} notably
the ratios between the lines that are close in frequency, i.e., the
$\ci$(1-0) and CO(4-3) lines and the $\ci$(2-1) and CO(7-6) line.
Although hard to quantify, their uncertainty should be on the order
of five per cent or less in almost all cases listed.

The ground-based {\it JCMT} $J$=2-1 $\co$ and $\thirco$ line fluxes
have typical uncertainties of $\Delta F\,=\,20\%$ associated with
antenna, receiver, and atmospheric calibration. Thus, the $\co/\thirco$
line ratios that refer to very similar beams have an uncertainty of
about $30\%$. The ratios of {\it SPIRE} CO(4-3) and $\ci$(1-0) to {\it
  JCMT} J=2-1 $\co$ and $\thirco$ have uncertainties ranging from
$25\%$ for compact sources to $35\&$ for extended objects. In a number
of cases, a mismatch in beamsize may introduce an additional
systematic error.  This is further discussed in Section 3.1.

\section{Results}

\subsection{Sample properties and line ratios}

For the sample presented in this paper, we attempted to identify all
galaxies that have good signal-to-noise measurements of at least one
$\ci$ line and two supporting $\co$ or $\thirco$ lines.  The final
selection includes 76 galaxies with distances ranging from the Local
Group to 250 Mpc, with nine galaxies having distances of more than 100
Mpc. The set of line fluxes is not complete, however.  There are \ci\
(1-0) and (2-1) line fluxes for 64 galaxies each. The sample for the
$\co$ (2-1), (4-3), and (7-6) lines consists of 61, 58, and 62
galaxies, respectively.  Finally, $\thirco$ (2-1) fluxes are
available for 46 galaxies. The set of $\co$ and $\ci$ line fluxes is
complete for a total of 35 galaxies, and we will refer to these galaxies
as the complete subsample.  In our subsequent analysis, we give
preference to the {\it SPIRE} data whenever a galaxy was observed both
from the ground and from space.

Because the {\it SPIRE} measurements were obtained with very similar
beam sizes, we can construct \ci\ (2-1)/(1-0), $\co$ (7-6)/(4-3), $\co$
(4-3)/\ci\ (1-0), and $\co$ (7-6)/\ci\ (2-1) ratios that can be
directly compared to one another.  The ground-based measures of the
$J$=2-1 $\co$ and $\thirco$ emission are likewise determined in almost
identical beams, but these are generally smaller than the {\it SPIRE}
beam. This complicates the comparison of line ratios containing either
of these two transitions: the difference in beam sizes may cause a
flux bias where the ground-based emission is underestimated by a
factor of about three.  However, in a number of cases, the $J$=2-1
$\co$ emission was mapped, which allowed us to convolve the
ground-based data to the {\it SPIRE} resolution. In the tables, these
cases are indicated by an asterisk in the last column.  A comparison of
the full resolution and the convolved $\co$ (2-1) fluxes shows that
the correction varies from galaxy to galaxy, but is usually less than
a factor of two (see also Kamenetzky $\etal$ 2014, and Ueda $\etal$
2014).

In Fig.\,\ref{ratbinfig} we show the distribution of the various line
ratios derived from the sample, and Table\,\ref{ratcomp} summarizes
the averages.  Because not all galaxies are measured in all lines, the
question arises whether or not a further bias is introduced by
comparing line ratios derived from different subsamples. As is clear
from Table\,\ref{ratcomp}, the line ratios of the total
(inhomogeneous) sample and those of the complete subsample are within
the errors identical with the marginal exception of the
$\co$(4-3)/(2-1) ratio.

The $\co$ (7-6)/(4-3) flux ratio ranges from 0.3 (low excitation) to 3
(high excitation), with a peak at a ratio of unity. (If we had
expressed the CO line ratio not in flux but in brightness temperature
$T_{mb}$, we would have found the peak to be at a ratio
$T_{mb}$(7-6)/$T_{mb}$(4-3) = 0.2.) The distributions of the other line
ratios are more clearly peaked, suggesting that the measured lines
stem from physical conditions that do not vary greatly over the
sample galaxies.  We note that the \ci\ (1-0) line is usually weaker
by a factor of about two than the adjacent $\co$ (4-3) line. The \ci\
(2-1)/(1-0) ratio is more broadly distributed between values of 1 and
3. The starburst galaxies have a ratio peaking at 1.9, and the more
luminous (U)LIRGs peak at about 2.2.  This is very close to the mean
\ci\ (2-1)/(1-0) ratio of 2.4$\pm$0.3 measured by Walter $\etal$
(2011) for nine galaxies at redshifts $z$ = 2.2 - 6.4.  In the
high-density, low-temperature case, a ratio of two corresponds to a
kinetic temperature $T_{kin}$=25 K; conversely, in the low-density,
high-temperature case the parent gas should have a density of
$n\approx135\,\cc$.

The ratio of the \ci\ (1-0) and $\thirco$ fluxes is of interest
because both lines are expected to be optically thin, in contrast to
the $\co$ lines that are optically thick. The rightmost panel in
Fig.\,\ref{ratbinfig} shows a distribution with a peak at 25 and a
long tail stretching out to NGC~6240, with a ratio of 158. The major
peak already corresponds to a brightness-temperature ratio over 2,
which is much higher than the typical values of 0.3-0.5 found in Milky
Way star-forming complexes ({\it cf} IB02).

In Fig.\,\ref{ratbinfig}, we show both the total sample and the
luminous-galaxy ($L\geq10^{11}$ M$_{\odot}$) subsample. Statistically,
the separation is not `clean', because the low-luminosity galaxies
tend to be at much closer distances.  If they experience an intense
but relatively small-scale circumnuclear starburst, our ratios may
nevertheless locally sample the same physical conditions that occur in
LIRGs globally. As Table\,\ref{ratcomp} also shows, the LIRGs have
clearly higher CO(7-6)/$\ci$(2-1) and $\ci$/$\thirco$ ratios. In their
$\ci$(2-1)/(1-0), CO(7-6)/(4-3), CO(4-3)/CO(2-1), and
CO(4-3)/$\ci$(1-0) ratios, the LIRGs and less luminous starburst
galaxies do not differ much, although the LIRGS tend to have
marginally higher ratios.

Almost all of the fifty galaxies detected in both $J$=2-1 CO isotopes
have an isotopic ratio below 20, with a mean of about 12. Six galaxies
have elevated ratios between 20 and 40, and three galaxies (the
mergers NGC~3256, Mrk~231, NGC~6240) have very high isotopic ratios
between 50 and 70. We note, however, that galaxies with low isotopic
ratios are much easier to detect in $\thirco$ emission. Thus, the
statistics are biased against very high isotopic ratios.

\subsection{Relation between \ci\ and $\co$} 

In Fig.\,\ref{ratratfig}a we compare the \ci-to-adjacent-CO line
ratios for all detected galaxies. As we have already seen in
Fig\,\ref{ratbinfig}, \emph{{\it the CO(4-3) line is stronger than the
    [CI] (1-0) line}} with very few exceptions. Only Centaurus A,
NGC~5135, NGC~3521, NGC~4254, and NGC~4826 have stronger \ci\ (1-0)
lines.  The situation for NGC~4945 and the Circinus galaxy is unclear.
The {\it Nanten2} measurements by Hitschfeld $\etal$ (2008) clearly
show the \ci\ line to be the stronger (Table\,\ref{grounddat}, but
both the {\it Herschel-SPIRE,} and the {\it APEX} measurements clearly
show the $\co$ (4-3) line to be the strongest. The difference in beam
sizes cannot explain this contradictory result.  A similar discrepancy
occurs for NGC~4826, where the ground-based observations show the
$\ci$ (1-0) line to be weaker than the $\co$ (4-3) line. In this case,
however, the difference in observing beam size ($22"$ versus $38"$)
may be relevant.

As far as the $\co$ (7-6)/$\ci$(1-0) line ratio is concerned,
Centaurus A has the lowest ratio in the whole sample, although ratios
that are almost as low are also found for NGC~3521, NGC~4254, NGC~5055, and
NGC~7331. The exceptional nature of Centaurus A has already been noted by
Israel $\etal$ (2014), who drew attention to the high flux of {\it
  \emph{both}} \ci\ lines with respect to the spectrally adjacent $\co$
lines.  The Centaurus~A center is almost unique in having both a $\ci$
(1-0) 492 GHz line flux exceeding that of the $\co$ (4-3) 461 GHz
line and a $\ci$ (2-1) 809 GHz line that is more than seven times stronger
than the $\co$ (7-6) 806 GHz line. In contrast, the CO-to-C ratios of
the other radio galaxy, Perseus~A, do not stand out in any way. As
Fig.\,\ref{ratratfig}a also shows, the CO(4-3)/$\ci$(1-0) and
CO(7-6)/$]ci$(2-1) line ratios are well correlated, with NGC~4536 as
an outlier.

There are relatively good correlations between the CO(7-6)/CO(4-3) and
the CO(4-3)/CO(2-1) ratios (Fig.\,\ref{ratratfig}b) and between the
\ci\ (2-1)/(1-0) and the CO(7-6)/(4-3) ratios
(Fig.\,\ref{ratratfig}c).  As these figures suggest, there is also a
reasonable correlation between the \ci\ (2-1)/(1-0) and the
CO(4-3)/CO(2-1) ratios, although this suffers from larger scatter than
the relation shown in Fig.\,\ref{ratratfig}c.  Because the CO transition
ratios provide a rough indication of the degree of excitation of the
molecular gas, it thus follows that {\it \emph{higher} \ci\ \emph{ratios go with
  more highly excited gas}}. The models by Meijerink $\&$ Spaans
(2005), Meijerink, Spaans, $\&$ Israel (2007), and Kazandjian $\etal$
(2012) show that the higher $\co$ ratios in Figs. 2b and 2c are
inconsistent with heating by UV photons (PDRs) alone and should be
associated with either X-rays or mechanical heating from turbulence
and shocks.  Although the correlation between the \ci\ line ratios and
the $\co$ line ratios is quite evident in Fig.\,\ref{ratratfig}c, it
has a significant dispersion.  By itself, the Centaurus A \ci\
(2-1)/(1-0) ratio is not exceptional, but it is relatively high for
its location in the diagram, which corresponds to a low degree of
excitation as indicated by the $\co$ (7-6)/(4-3) ratio.

Finally, the $J$=2-1 $\thirco$/$\co$ isotopic ratios do not appear to
be clearly correlated with the $\co$ (4-3)/(2-1) (or either the \ci\
(2-1)/(1-0) line ratio or the far-infrared luminosity for that
matter). In Fig.\,\ref{ratratfig}d, over forty galaxies have
$J$=2-1 isotopic ratios below 20. The isotopic ratio does not
noticeably change with increasing $\co$ (4-3)/(2-1) ratio. Six
galaxies (Arp~193, Arp~220, Arp~299A, NGC~3256, NGC~4038, and
CGCG~049-057) have ratios between 20 and 40, and only Markarian~231,
UGC~05101, and NGC~6240 have very high isotopic ratios between 50 and
70, signifying both low CO optical depths and high intrinsic
$^{12}$C/$^{13}$C abundance ratios (see also Mart\'in $\etal$ 2010,
Henkel $\etal$ 2014). {\it \emph{However, there does not appear to be a
  preferred range in $\ci$ or CO line ratios for the highest isotopic
  ratios}.}

\subsection{Relation between \ci\ and $\thirco$}

A direct comparison of \ci\ and $\thirco$ line intensities is of
interest because both are thought to be optically thin (in contrast
to the $\co$ transitions measured, which are optically thick).  These
comparisons were made for small samples of galaxies by Gerin $\&$
Phillips (2000) and by Israel $\&$ Baas (2002; hereafter IB02). A
major and unexpected result of the latter study was the discovery that
in galaxy centers the \ci\ (1-0)/$\thirco$ (2-1) ratio increases with
the (carbon) line luminosity.  In the IB02 sample, most galaxy centers
exhibited a $\ci$/$\thirco$ brightness temperature ratio exceeding the
value 0.3-0.5 that is the rule for (UV) photon-dominated regions
(PDRs) in the Milky Way galaxy.  This finding was supported by
observations of NGC~4945 and the Circinus galaxy (Hitschfeld $\etal$
(2008). The present, much larger sample expanded by both the new
(Tables\,\ref{herculesdat}, \ref{archivedat}) and published
(Table\,\ref{spiredat}) {\it SPIRE} results confirm the positive
correlation (ratio increasing with luminosity) found by IB02.  In
Fig.\,\ref{ratlumfig}a we show the \ci\ (1-0)/$\thirco$ (2-1) ratio as
a function of the \ci\ (1-0) luminosity. This Figure corresponds to
the left-hand panel of Fig. 2 in IB02, with the difference that we now
use metric flux ratios and luminosities instead of values based on
brightness temperatures.  The new data extend the previously found
correlation up to ULIRGs, so that it now covers over three decades in
luminosity.  The correlation is robust because it remains present whether
the \ci\ (1-0)/$\thirco$ (2-1) ratio is plotted as a function of the
$\ci$ line, the $\thirco$ (2-1) line, or even the {\it IRAS}-derived
FIR continuum luminosity.  The emission of the sample galaxies is
mostly due to starbursts, dominating any AGN that may be present.
Notable exceptions are Centaurus~A (NGC~5128) and Perseus~A (NGC~1275),
which are dominated by an AGN rather than a starburst.{ Perseus~A
has `normal' ratios, but Centaurus~A is an outlier with relatively
strong \ci\ emission for its modest far-infrared luminosity.  The
significant increase in $\ci$ (1-0) line fluxes with increasing
luminosity is thus well-established and shared by the $\ci$ (2-1) line
(Fig.\,\ref{ratlumfig}c). {\it \emph{Both $J$=1-0 and the $J$=2-1 \ci\ to
  $^{13}$CO (2-1) line ratios increase with (FIR) luminosity}.}

\section{Line ratios and LVG models}

\subsection{Diagnostic panels}

\begin{figure}[t]
\begin{minipage}[]{9cm}
\resizebox{8.7cm}{!}{\rotatebox{270}{\includegraphics*{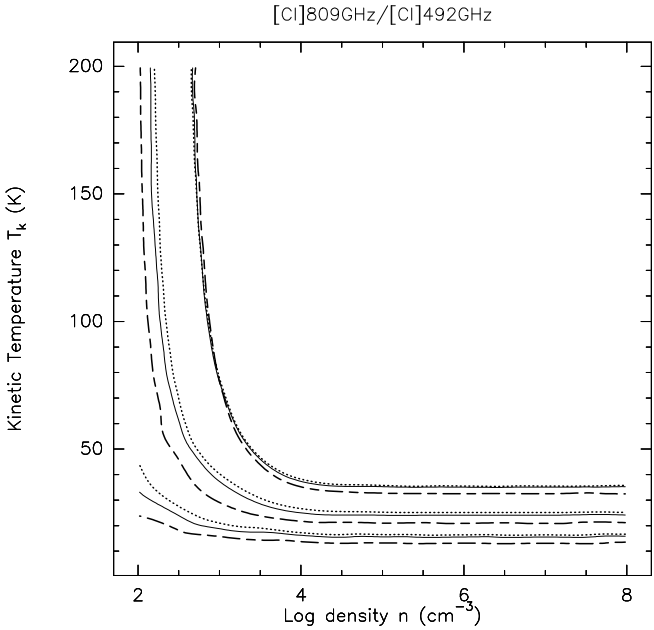}}}
\end{minipage}
\caption[]{LVG results for the $\ci$ line flux ratios observed in the
  galaxy sample.  The panel contains lines of constant model flux
  ratio, as a function of $\h2$ gas kinetic temperature $T_{\rm k}$
  and volume density $n({\rm H}_{2})$, for distinct C column densities
  $N_{\rm C}$/dV. Curves are depicted in groups of constant line ratio
  $\ci$(1-0)/$\ci$(2-1) = 1 (left, bottom), 2 (middle), and 3 (right,
  top).  Within each group of ratios, curves are drawn for column
  densities $N({\rm C})$/dV=$0.3\times10^{17}\,\cm2/\kms$ (dotted
  line), $1\times10^{17}\,\cm2/\kms$ (continuous solid line), and
  $3\times10^{17}\,\cm2/\kms$ (dashed line), respectively.  }
\label{CILVGfig}
\end{figure}

\begin{figure*}[t]
\begin{minipage}[]{6cm}
\resizebox{5.7cm}{!}{\rotatebox{0}{\includegraphics*{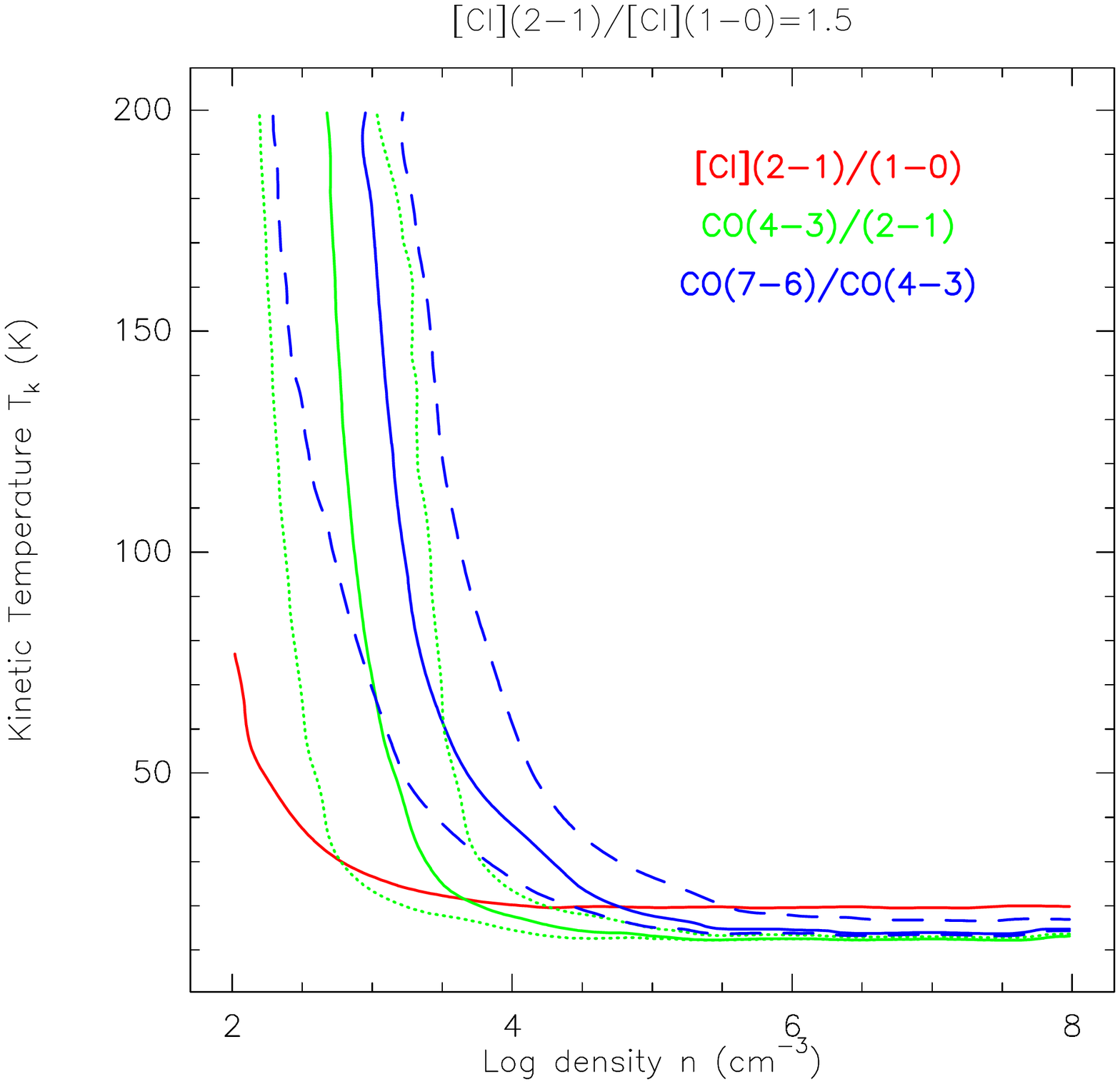}}}
\end{minipage}
\begin{minipage}[]{6cm}
\resizebox{5.7cm}{!}{\rotatebox{0}{\includegraphics*{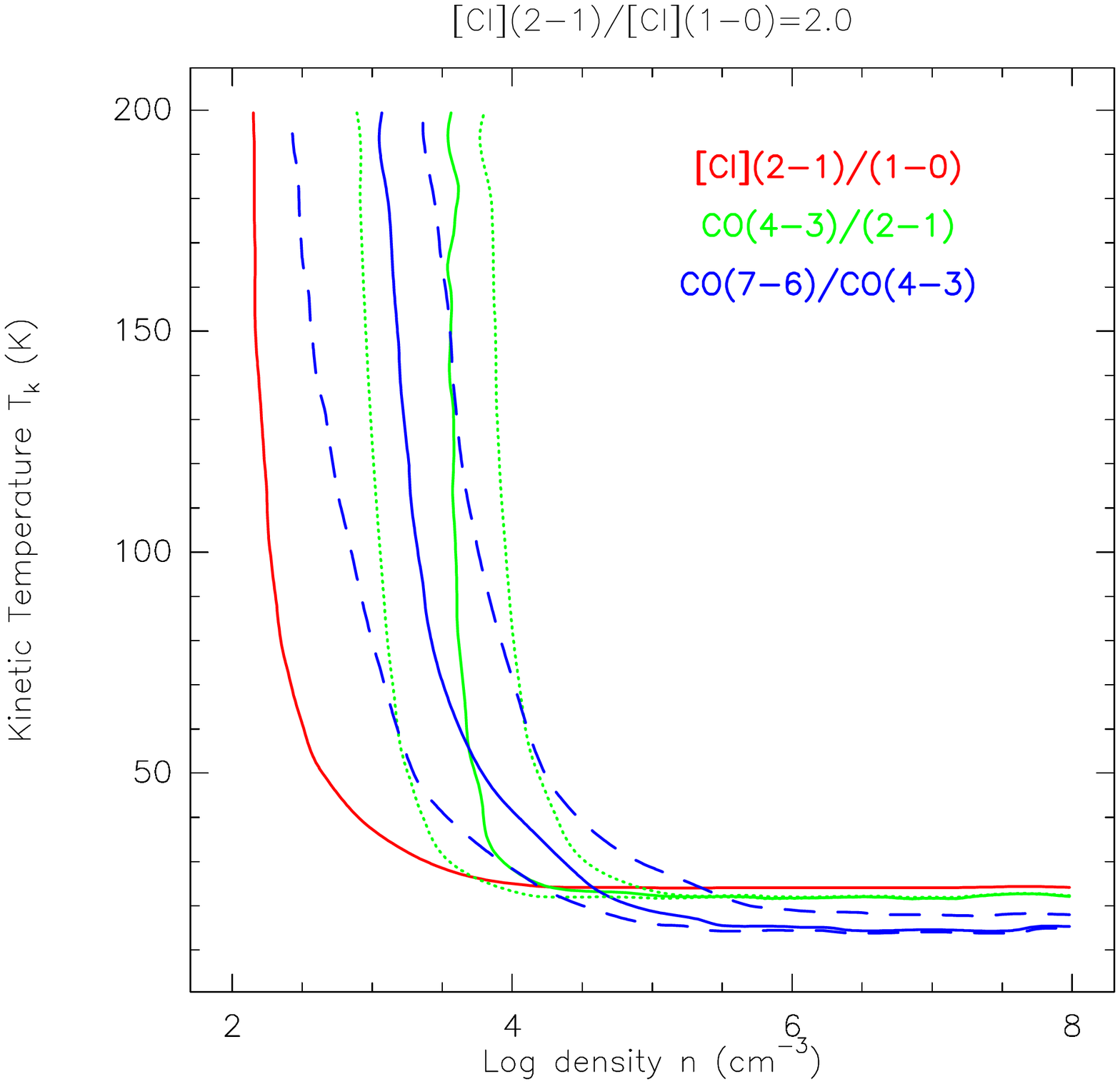}}}
\end{minipage}
\begin{minipage}[]{6cm}
\resizebox{5.7cm}{!}{\rotatebox{0}{\includegraphics*{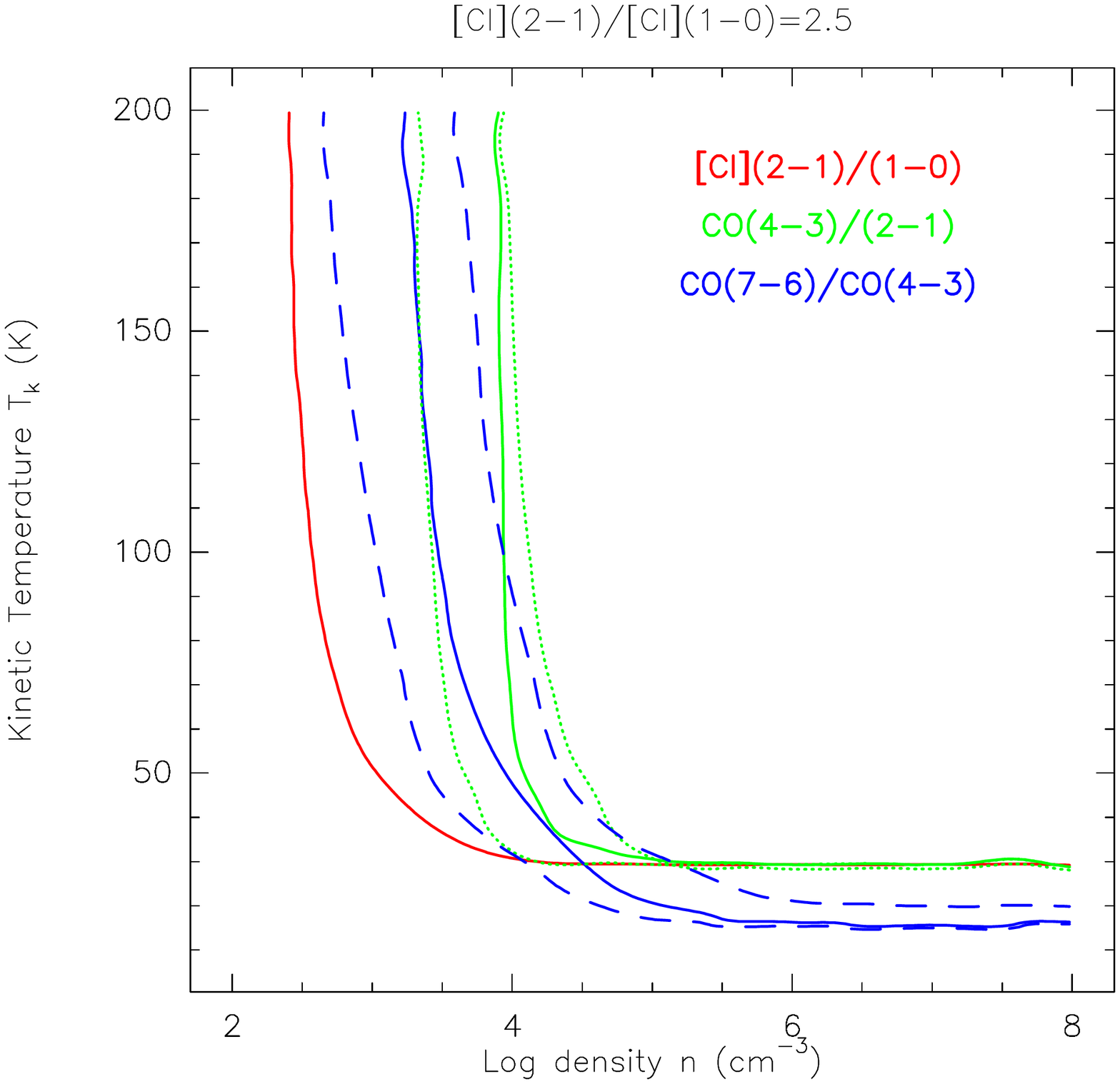}}}
\end{minipage}
\caption[]{LVG results for the $\ci$ and $\co$ line flux ratios observed in
  the galaxy sample.  The panels contain lines of constant flux ratio,
  as a function of $\h2$ gas kinetic temperature $T_{\rm k}$ and
  volume density $n({\rm H}_{2})$, for distinct C and CO column
  densities $N$/dV. From left to right: a. intersection of curves of
  constant line ratio $\ci$ (2-1)/(1-0)=1.5 ($\approx$ 20 K) and
  corresponding line ratios $\co$ (4-3)/$\co$ (2-1)=4.5 and $\co$
  (7-6)/$\co$ (4-3)=0.9; b. intersection of curves of constant line
  ratio $\ci$ (2-1)/(1-0)=2.0 ($\approx$ 25 K) and corresponding line
  ratios $\co$ (4-3)/$\co$ (2-1)=6 and $\co$ (7-6)/$\co$ (4-3)=1.05;
  c. intersection of curves of constant line ratio $\ci$
  (2-1)/(1-0)=2.5 ($\approx$ 30 K) and corresponding line ratios $\co$
  (4-3)/$\co$ (2-1)=7 and $\co$ (7-6)/$\co$ (4-3)=1.35. In all three
  panels, CO curves are given for column densities $N({\rm
    CO})$/d$V=1\times10^{17}\,\cm2/\kms$, (upper curve),
  $3\times10^{17}\,\cm2/\kms$, and $1\times10^{18}\,\cm2/\kms$ (lower
  curve). Since the single $\ci$ (2-1)/(1-0) curve represents a column
  density $N({\rm CO})$/d$V=1\times10^{17}\,\cm2/\kms$, the CO curves
  effectively correspond to [C]/[CO] abundances of 1, 0.3, and 0.1,
  respectively.  }
\label{COLVGfig}
\end{figure*}

To extract physical information from the gas in which the
line emission originates, we evaluated the observed line ratios
using the large-velocity gradient (LVG) radiative transfer code {\it
  RADEX} described by Jansen (1995), Jansen $\etal$ (1994), and
Hogerheijde $\&$ van der Tak (2000) \footnote{see also
  http://www.strw.leidenuniv.nl/home/michiel/ratran/}. LVG models have
the advantage of not requiring any prior specification of a
particular chemical or heating model. The LVG code provides model line
intensities as a function of three input parameters: gas kinetic
temperature ($T_{\rm k}$), molecular hydrogen density ($n({\rm
  H}_{2})$), and the CO (or C) gradient, i.e. the column density per
unit velocity ($N({\rm CO})$/d$V$ resp. $N({\rm C})$/d$V$). In our
analysis, we explored, the parameter range $T_{\rm k}$=10--200 K,
$n({\rm H}_{2})$=10$^{2}$-10$^{8}$ $\cc$,
$N({\co})$/d$V$=(0.1-300)$\times$10$^{17}$ $\cm2/\kms$,
$N({\thirco})$/d$V$=(0.1-300)$\times$10$^{15}$ $\cm2/\kms$, and
$N({\rm C})$/d$V$=(0.1-30)$\times$10$^{17}$ $\cm2/\kms$.  We use
the diagnostic diagrams to constrain the kinetic temperature and gas
density, as well as the C$^{\circ}$/$\co$ and the
C$^{\circ}$/$\thirco$ abundance.

We do not attempt to fit all the sample galaxies individually.
Rather, we consider a narrow range of cases in order to identify
representative physical parameters and trends. This is possible
because, as noted, the $\co$ (7-6)/(4-3) and the $\co$
(4-3)/(2-1) ratios of the sample galaxies increase more or less
linearly with the $\ci$ (2-1)/$\ci$ (1-0) ratio, and the $\co$
(7-6)/$\ci$ (2-1) increases linearly with the $\co$ (4-3)/$\ci$ (1-0)
ratio.  Figure\,\ref{ratratfig}, which illustrates this, also shows
that regardless of the form and degree of correlation, most of the
observed line ratios discussed in this paper cover only a limited
range (see also Fig.\,\ref{ratbinfig}).

The complete analysis only involves galaxies for which all six lines have been
measured
(three $\co$ lines, two $\ci$ lines, and one $\thirco$ line). Because almost all galaxies observed from the ground lack
measurements of the high-frequency $\ci$ (2-1) and $\co$ (7-6) lines,
the analysis is almost exclusively based on the {\it Herschel-SPIRE}
sample. About half of this sample refers to very luminous galaxies,
mostly LIRGs. ULIRGS tend to be at greater distances, with the
consequence that the $J$=4-3 $\co$ line and often also the $\ci$ (1-0)
line are red-shifted out of the {\it SPIRE} band, or suffer from
excessive noise at the band edge. There is only one ULIRG (Arp~220)
for which all six necessary line fluxes are available.

\subsection{$\ci$ line ratios: setting the scene}

The full range of observed $\ci$ (2-1)/(1-0) ratios is rather limited, and
almost fully contained between the values of 1.5 and 3. In
Fig.\,\ref{CILVGfig} we show the result for the modeled $\ci$
(2-1)/(1-0) ratio at discrete values of 1, 2, and 3 over the full
range of densities and temperatures defined above, as a function of
three values for $N(CI$)/d$V$ column densities covering an order of
magnitude. In Fig.\,\ref{CILVGfig}, the line ratios exhibit a characteristic
double degeneracy for temperature and density.  For any given line
ratio, the emitting gas is not uniquely located in the diagram, but
may be at a relatively well-determined low density with an
unconstrained temperature (vertical branch), or at an unconstrained
but relatively high density with a well-determined temperature. The
emitting gas may even be a mixture of such phases which are
indistinguishable observationally.

In the diagrams, higher line ratios are found above lower line ratios
i.e. a higher observed line ratio implies a shift to the right (higher
densities) and to the top (higher temperatures). For any given line
ratio, the curve in the diagram also shifts up and the right when the
column density (abundance) is lowered, down and to the left when the
abundance is increased. As Fig.\,\ref{CILVGfig} shows, increasing the
carbon column density from the reference value $N({\rm
  C})$/d$V$=$10^{17}\,\cm2\kms$ does not shift curves of constant line
ratio by much, and lowering the carbon column density has even less
effect.  This reflects the fact that throughout most of the diagram in
Fig.\,\ref{CILVGfig} the carbon lines are optically thin; optical
depths exceeding unity only occur in the lower left corner of the
panel.

At the highest observed $\ci$ (2-1)/(1-0) ratios of three, gas at
temperatures above $T_{k}$=40 K requires densities to be limited to
$n\leq1000\,\cc$; for ratios of unity, high temperatures are only
allowed at very low densities ($n\leq100\,\cc$). Galaxies that exhibit
this ratio most likely instead contain gas at a well-defined
temperature of 15 K, but with an undetermined density.  It is evident
from Fig.\,\ref{CILVGfig} that the observed $\ci$ line ratios do
constrain the underlying physical conditions, and in particular {\it
  \emph{rule out the presence of significant contributions to the neutral
  carbon emission by gas that is both dense and warm}}.  However,
beyond this, not much more can be concluded from the $\ci$ emission
line ratio.  Further progress requires breaking the $T_{k}/n$
degeneracy by comparing emission from $\ci$ to that of other species,
such as $\co$.

\subsection{$\ci$ and $\co$ line ratios: temperature and density}

The two CO line ratios considered in this paper likewise fall within
well-defined, relatively narrow ranges. All observed $\co$ (7-6)/(4-3) ratios
fall in the range 0.25-2.0, and the $\co$ (4-3)/(2-1) ratios are all
in the range 2 to 10. The scatter in the latter ratio to some extent
reflects that not all $\co$ (2-1) refer to the beam defined by the {\it
  SPIRE} aperture in which the $\co$ (4-3) fluxes were measured.
Although in those galaxies the difference in beam size may be as much
as a factor of three, we have verified that in actual cases, the flux
difference is less than a factor of two because the CO is centrally
concentrated.  In Fig.\,\ref{COLVGfig} we show $T_{k}, n$ diagrams
with modeled $\co$ (7-6)/$\co$ (4-3), $\co$ (4-3)/$\co$ (2-1), and
$\ci$ (2-1)/$\ci$ (1-0) line ratios.

As we have noted before, Fig.\,\ref{ratratfig} shows that the $\ci$ and
$\co$ line ratios are correlated.  We have determined the best-fit
parameters for $\ci$ (2-1)/$\ci$ (1-0) ratios of 1.5, 2, and 2.5 and
the correspondingly increasing $\co$ (4-3)/$\co$ (2-1) and $\co$
(7-6)/$\co$ (4-3) ratios for a range of CO abundances ($N_{CO}$/d$V$).
The three diagrams in Fig.\,\ref{COLVGfig} show curves for
$\ci$ (2-1)/(1-0)=1.5, 2.0, and 2.5 and the mean CO ratios
corresponding to each of these values.  Each of the two CO ratios is
represented by three curves for different CO abundances
($N_{CO}$/d$V$).  As in Fig.\,\ref{CILVGfig}, the CO curves in
Fig.\,\ref{COLVGfig} exhibit the well-known double degeneracy for
temperature and density. However, each of the three sets of degenerate
curves is offset from the others.

If we assume that the three $\co$ lines ($J$=2-1, $J$=4-3, $J$=7-6)
and the two \ci\ ($J$=1-0, $J$=2-1) lines originate in the same gas
phase dominated by a single temperature and a single density, these
can be found from the overlap or intersection of the relevant curves
corresponding to the observed line ratio.  This determination is {\it
  \emph{independent}} of the actual $\ci$/$\co$ abundance (which we
determine later, see Sect. 4.4).  From Fig.\,\ref{COLVGfig} it is
clear that common intersections providing solutions for $T_{k}$ and
$n$ all occur at low temperatures in the range of 20-40 K, and
intermediate densities in the range of $10^{4}$--$10^{5}\,\cc$.  The
assumption that the $\co$ and $\ci$ emitting volumes refer to the same
gas cloud population is supported by the coincidence of the CO/CO and
[CI]/CO intersections.

Emission characterized by ratios $\ci$ (2-1)/(1-0)$\leq1.5 $ -- i.e., with
$\ci$ (1-0) emission relatively strong with respect to both $\ci$
(2-1) and $\co$ (4-3) -- are hard to fit at all. Emission from the center
of the radio galaxy Centaurus A falls into this category. For
C$^{\circ}$ column densities equal to or higher than CO column
densities, i.e. abundances [C]/[CO]$\geq$1.0, marginal fits to the
observed CO line ratios occur only at ratios $\ci$ (2-1)/(1-0)$\geq2$,
at modest kinetic temperatures of 25-30 K but very high densities in
excess of $10^{6}\,\cc$.  The quality of the fits improves with
increasing CO column density, i.e. decreasing [C]/[CO] abundance. As
all fits occur in the medium-density, low-temperature regime, the
$\ci$ line ratios can be translated more or less directly into kinetic
temperatures, with $T_{\rm k}$ smoothly increasing from 20 to 35 K as
$\ci$ (2-1)/(1-0) increases from 1.5 to 3.0.  The fitted density
depends weakly on the column densities assumed.  For instance, at
constant $T_{\rm k}$=30 K and abundance [C]/[CO]=0.3, we find log
$n=5.5\pm0.5$, log $n=4.6\pm0.3$, and log $n=3.9\pm0.4$ $\cc$ for
$N(CO)$/d$V=1\times10^{17}$, $3\times10^{17}$, and $10\times10^{17}$
$\cm2/\kms$, respectively.

When the implied mean temperature and density are relatively low,
multiple gas components are required for good fits.  This applies to
about a third of the sample, characterized by a threshold of
$\ci$(2-1)$\approx$1.65. This corresponds roughly to mean values for
kinetic temperature and density below 20 K and a few times
$1000\,\cc$, respectively.

However, we conclude that the combined $\ci$ and $\co$ line ratios can generally be fit succesfully by a single (molecular) gas component
above this threshold. In that case, \emph{{\it \emph{the observed emission implies
  a gas of mean volume density log} $n({\rm H}_{2})=4.7\pm1.0\,\cc$ \emph{and
  kinetic temperature ranging from} $T_{\rm k}=25$ to 35 K}}. The
limited range of solutions suggests that this is a robust conclusion.
Moreover, with this result we can proceed to constrain the column
densities $N$/d$V$ and even the [C] / [CO] abundances

\subsection{$\ci$ /$\co$ line ratios: the [C] / [CO]  abundance}

\begin{figure*}[t]
\begin{minipage}[]{6cm}
\resizebox{5.7cm}{!}{\rotatebox{270}{\includegraphics*{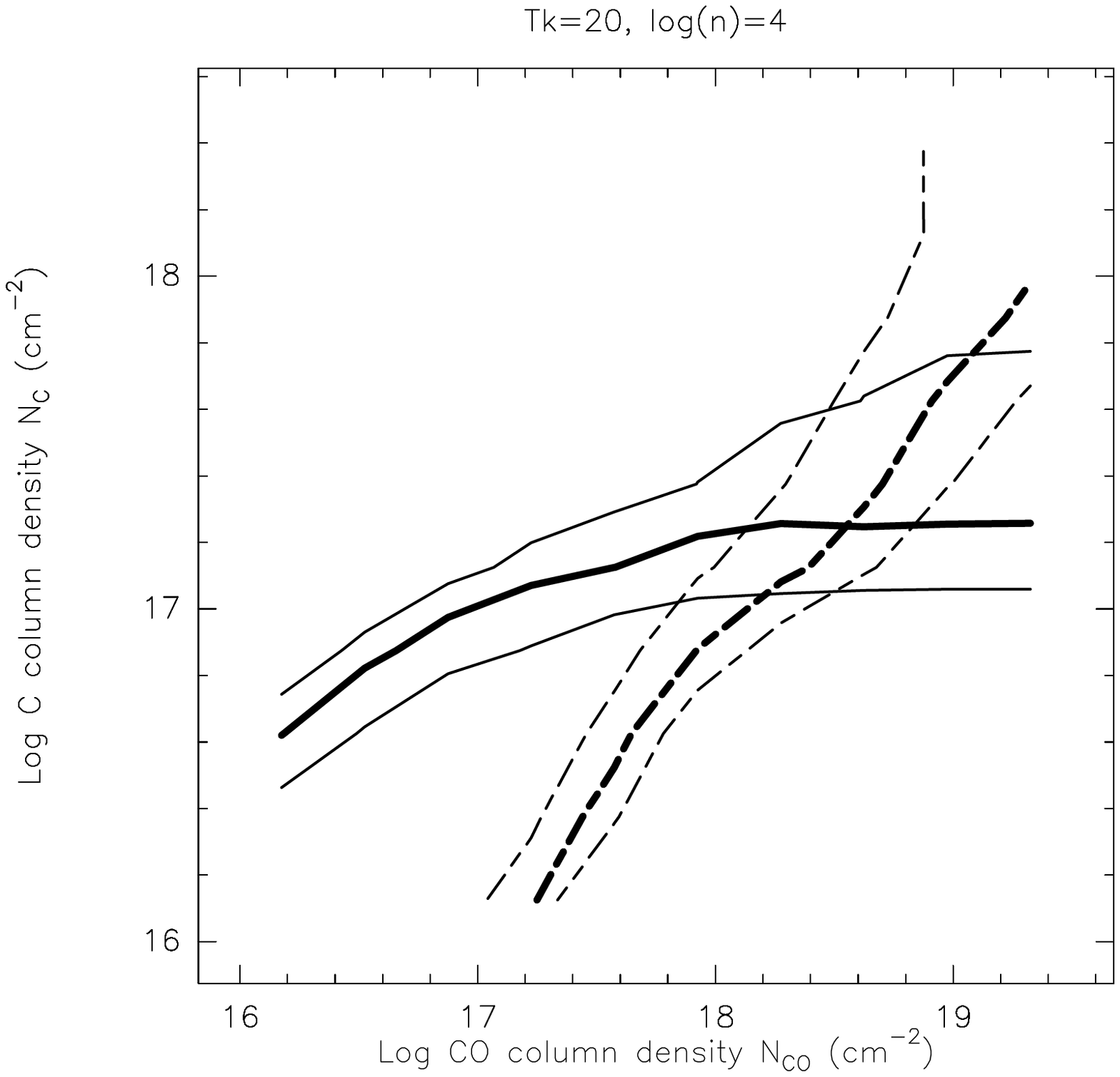}}}
\end{minipage}
\begin{minipage}[]{6cm}
\resizebox{5.7cm}{!}{\rotatebox{270}{\includegraphics*{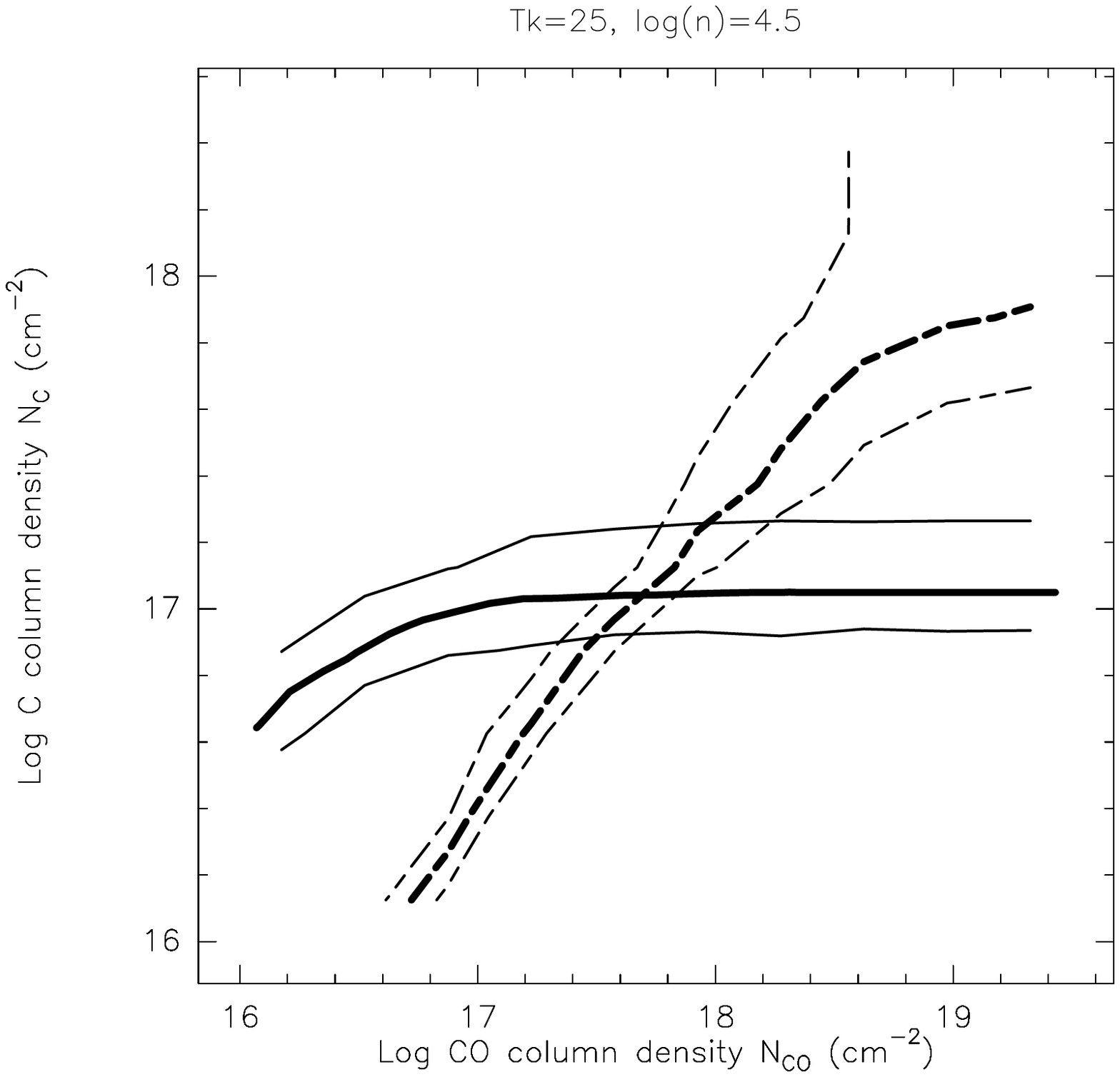}}}
\end{minipage}
\begin{minipage}[]{6cm}
\resizebox{5.7cm}{!}{\rotatebox{270}{\includegraphics*{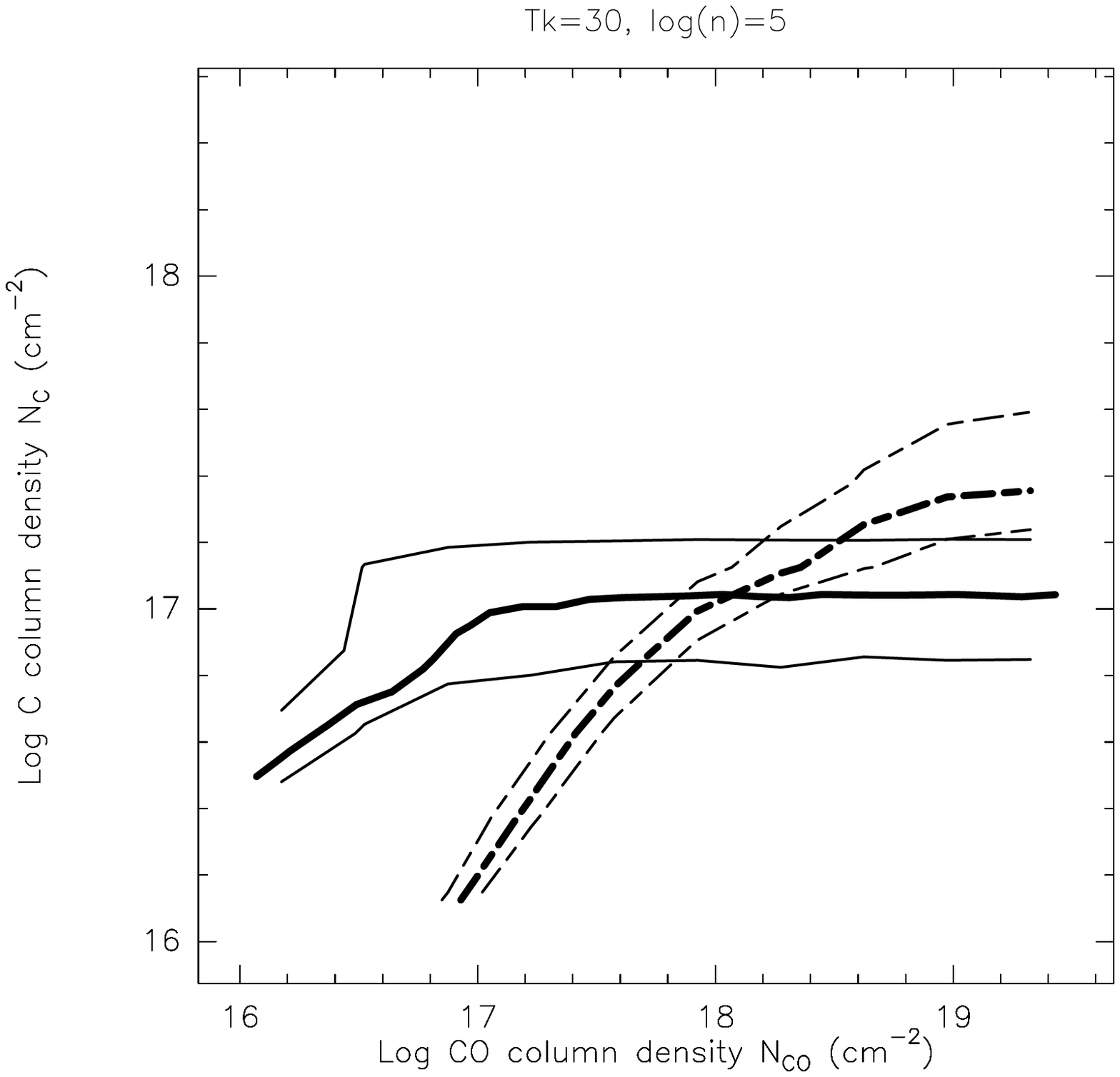}}}
\end{minipage}
\caption[]{LVG results for the $\ci$-to-$\co$ line flux ratios observed in
  the galaxy sample.  The panels contain lines of constant flux ratio
  $\co$ (4-3)/$\ci$ (1-0) (solid lines) and $\co$ (7-6)/$\ci$ (2-1)
  (dashed lines) as a function of carbon and carbon monoxide column
  densities $N_{C}$/d$V$ and $N_{CO}$/d$V$.  From left to right:
  a. model curves at ($T_{k}=20$ K, $n=10^{4}\,\cc$) for ratios $\co$
  (4-3)/$\ci$ (1-0)=1.1, 1.4, 2.0 and $\co$ (7-6)/$\ci$ (2-1)=0.5, 0.8,
  1.1; b. model curves at ($T_{k}=25$ K, $n=3\times10^{4}\,\cc$) for
  ratios $\co$ (4-3)/$\ci$ (1-0)=1.4, 2.0, 2.6 and $\co$ (7-6)/$\ci$
  (2-1)=0.8, 1.1, 1.4; c. model curves at ($T_{k}=30$ K,
  $n=10^{5}\,\cc$) for ratios $\co$ (4-3)/$\ci$ (1-0)=2.0, 2.6, 3.2
  and $\co$ (7-6)/$\ci$ (2-1)=1.1, 1.4, 1.7. The diagram axes are
  column densities for a velocity interval of 1 km/s.  }
\label{NCOLVGfig}
\end{figure*}

In Fig.\,\ref{NCOLVGfig} we plot the ratios of the two $\ci$ lines to
their adjacent $\co$ line, $\co$ (4-3)/$\ci$ (1-0) and $\co$
(7-6)/$\ci$ (2-1) in the $N_{C}$d$V$-$N_{CO}$d$V$ diagram, for a
variety of temperature and density combinations corresponding to
$\ci$(2-1)/(1-0) ratios of 1.5, 2, and 2.5.  The intersections vary only
a little as a function of conditions.  The CO column densities are the
same in all cases within a factor of two, and the C column densities
are all within a factor of three from one another. The implied CO
column densities do not vary much with circumstances, unlike the C
column densities that drop a little with increasing temperature and
density.  Both decrease somewhat with increasing CO/C flux ratio, C
more so than CO. The consistency of the results once more suggests
that our assumption of $\co$ and $\ci$ emission originating in related
gas clouds is correct. Summarizing these results, we find averaged
column density gradients that characterizing the emission {\it log
  $N_{CO}$/d$V=18.0\pm0.3\,\cm2/\kms$ and
  $N_{C}$/d$V=17.1\pm0.5\,\cm2/\kms$}. Individual (inverse) carbon
abundance [CO] / [C] range from 3 to 20, but we find an {\it \emph{averaged
  mean abundance} [CO] / [C]=8$\pm$3}. Thus, the amount of neutral carbon
appears to be a rather small fraction of the amount of carbon monoxide. The
results obtained so far thus indicate that {\it \emph{the observed [CI]
  emission arises from a moderately dense} ($n,=10^{4}$--$10^{5}\,\cc$)
\emph{  and warmish} ($T\approx30$ K) \emph{gas}}, and representing only
$10\%$--$20\%$ of the carbon in the gas phase.  This is consistent with
the results obtained by IB02, except for their conclusion that neutral
carbon and carbon monoxide column densities are similar.  The
abundances found here are also lower than the range of 0.4--3.0 (mean
value 1.2) that we have derived for a total of fifteen nearby galaxy
centers from a more detailed analysis involving two gas phases
(Israel, 2009b and references therein). It may be possible that a more
detailed multiphase analysis of the luminous galaxies in the present
sample would yield higher [C] / [CO] abundances.

\subsection{[CI] /$^{13}$CO line ratios: the [C] / [$^{13}$CO]
  abundance}

\begin{figure*}[t]
\begin{minipage}[]{6cm}
\resizebox{5.7cm}{!}{\rotatebox{270}{\includegraphics*{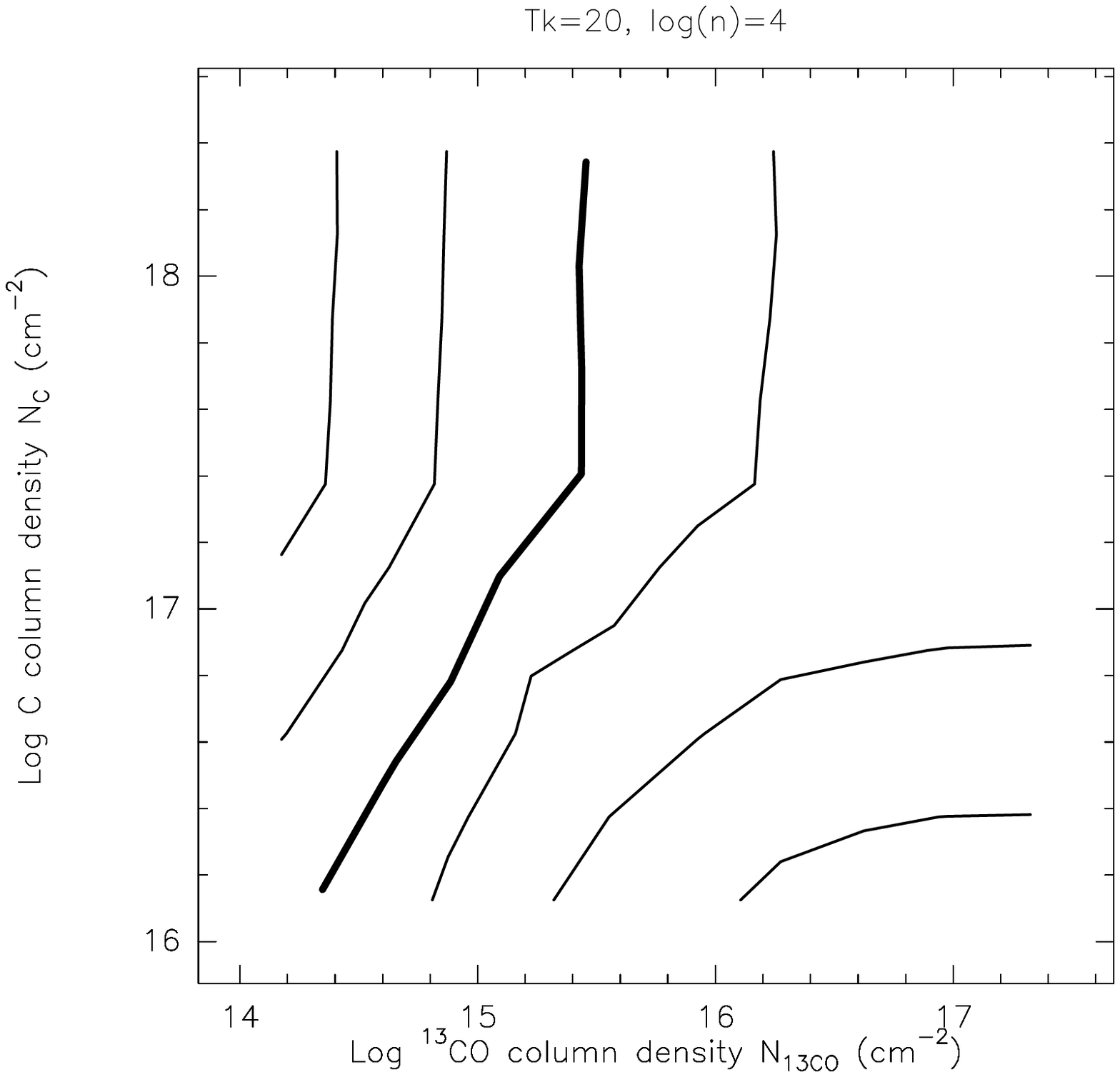}}}
\end{minipage}
\begin{minipage}[]{6cm}
\resizebox{5.7cm}{!}{\rotatebox{270}{\includegraphics*{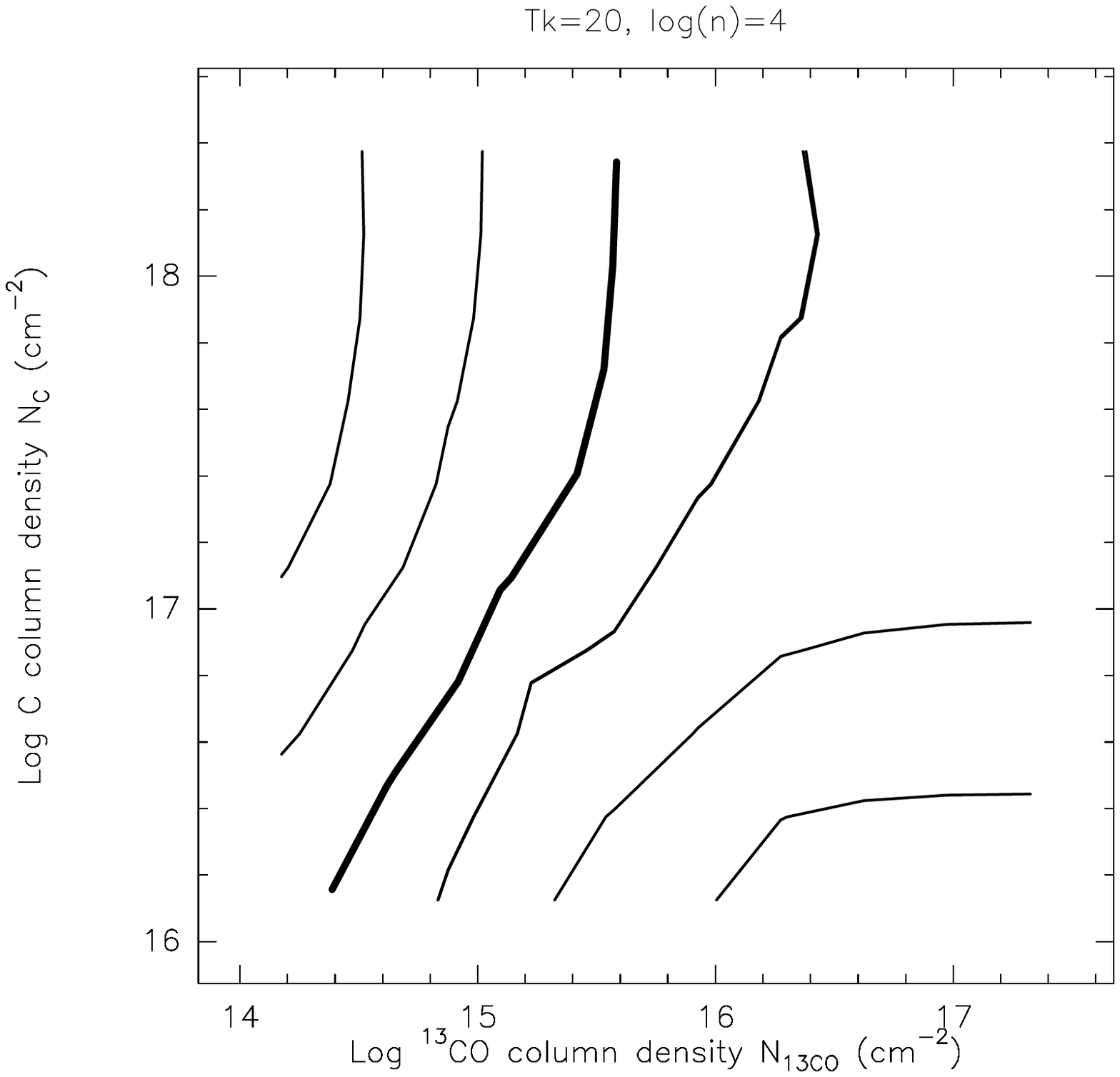}}}
\end{minipage}
\begin{minipage}[]{6cm}
\resizebox{5.7cm}{!}{\rotatebox{270}{\includegraphics*{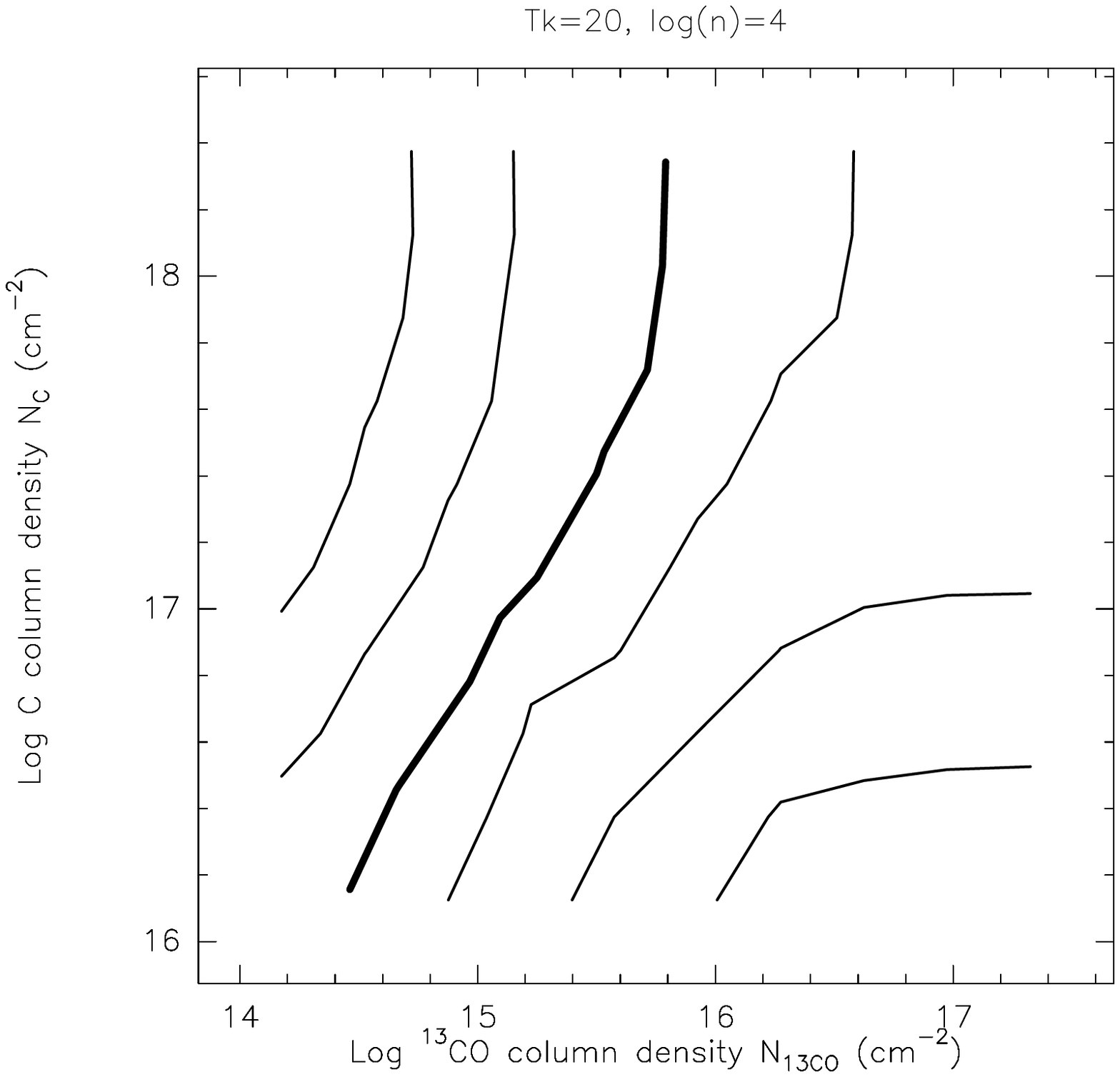}}}
\end{minipage}
\caption[]{LVG results for the $\ci$-to-$\thirco$ line flux ratios observed
  in the galaxy sample.  The panels contain lines of constant flux
  ratio $\ci$ (1-0)/$\thirco$ (2-1) as a function of carbon and
  carbon monoxide column densities $N_{C}$/d$V$ and $N_{13CO}$/d$V$. 
  From left to right: a. model curves at ($T_{k}=20$ K,
  $n=10^{4}\,\cc$) for ratios $\ci$ (2-1)/(1-0)=1, 3, 10, 30 (thick
  line), 100, and 300 increasing from right to left; b. as panel a.,
  but for ($T_{k}=25$ K, $n=3\times10^{4}\,\cc$); c. as panel a., but
  for ($T_{k}=30$ K, $n=10^{5}\,\cc$).  The diagram axes are column
  densities for a velocity interval of 1 km/s.  }
\label{N13COLVGfig}
\end{figure*}

\begin{figure*}[t]
\begin{center}
\begin{minipage}[]{18cm}
\resizebox{5.9cm}{!}{\rotatebox{0}{\includegraphics*{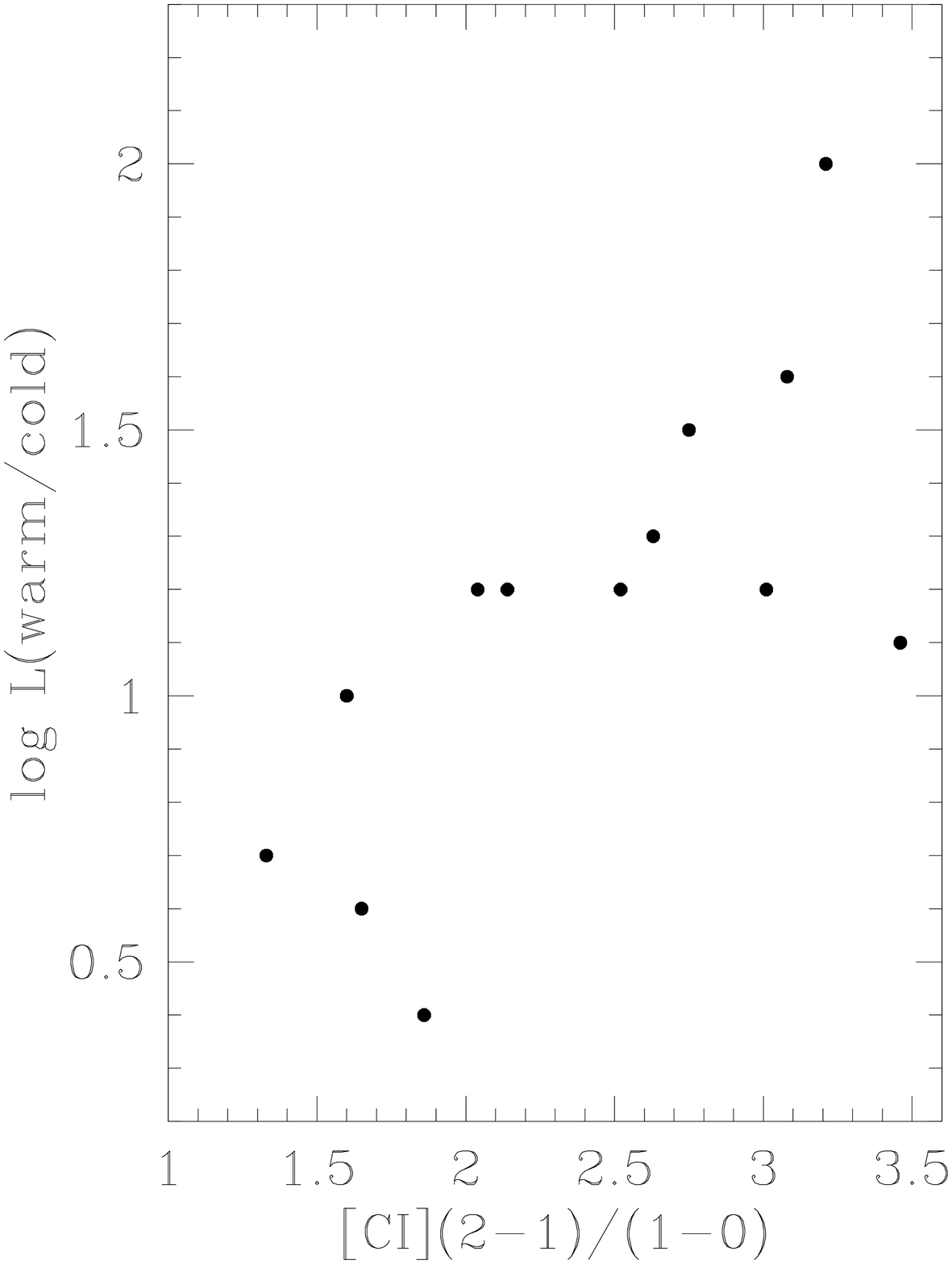}}}
\resizebox{5.9cm}{!}{\rotatebox{0}{\includegraphics*{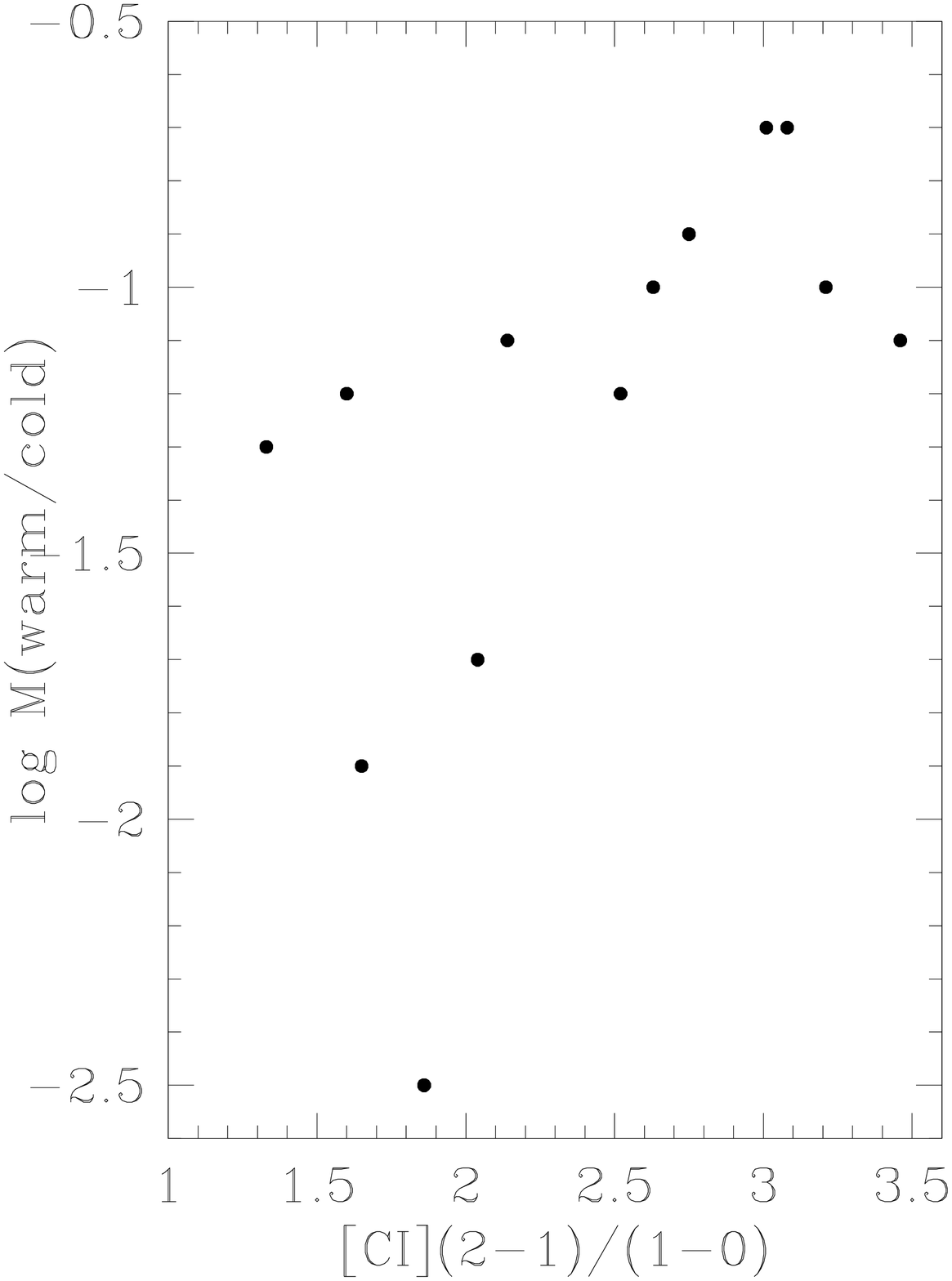}}}
\resizebox{5.9cm}{!}{\rotatebox{0}{\includegraphics*{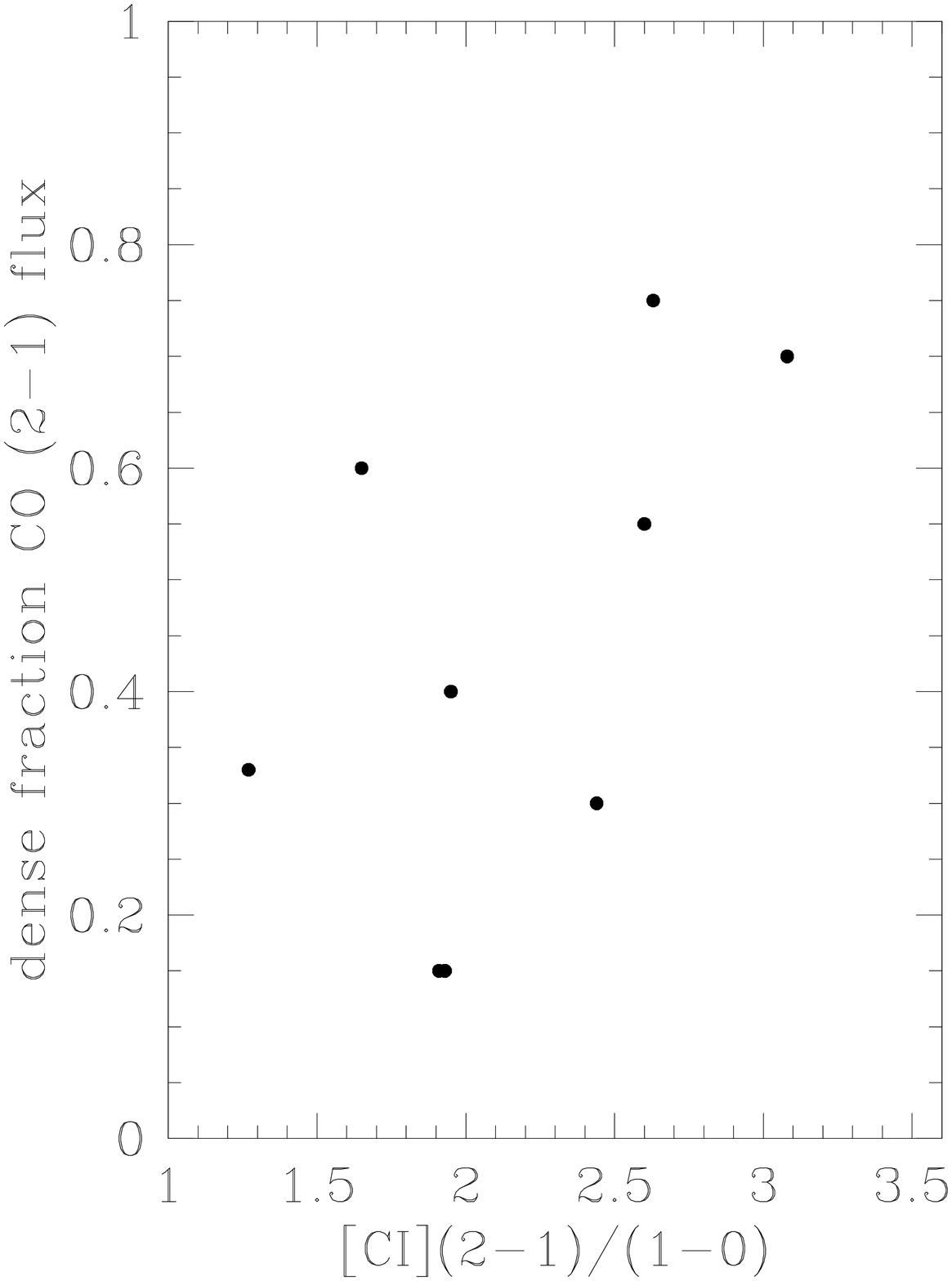}}}
\end{minipage}
\caption[]{Diagnostic two-phase gas ratios as a function of the
  observed $\ci$ (2-1)/(1-0) line flux ratio. The left panel shows the
  behavior of the `warm/cold/ ratio of total CO line luminosities,
  taken from Kamenetzky $\etal$ (2014) versus the $\ci$ line ratio
  from this paper. The center panel shows the corresponding
  `warm/cold' ratio of molecular gas masses. The right panel shows the
  fraction of the $\co$ $J$=2-1 flux that is ascribed to `dense' gas,
  based on the results published by Israel (2009b, and references
  therein). All three quantities increase in favor of `warm' and
  `dense' with increasing $\ci$ line flux ratio.  }
\end{center}
\label{hotcold}
\end{figure*}

In Fig.\,\ref{N13COLVGfig} lines of constant $\ci$ (1-0)/$\thirco$
(2-1) flux in the $N_{C}$/d$V$, $N_{13CO}$/d$V$ diagram. As
Fig.\,\ref{ratbinfig} shows, the sample galaxies have ratios between
10 ($\thirco$ relatively strong) and 100 ($\ci$ that are relatively
strong), with a mean at 30. The flux ratio $\ci$ (2-1/$\ci$ (1-0) that
distinguishes the three panels of Fig.\,\ref{N13COLVGfig} appears to
be uncorrelated with the $\ci$ (1-0)/$\thirco$ (2-1) ratio, so that we
expect the results of any analysis to be very similar for the three
panels. This turns out to be the case. At the neutral carbon column
densities $N_{C}$/d$V$ determined in the previous section,
Fig.\,\ref{N13COLVGfig} shows practically identical values log
$N_{13CO}$/d$V=15.12\pm0.15$, $15.06\pm0.13$, and $15.12\pm0.17$ for
the mean observed line ratio of thirty. At the lowest observed line
ratio of ten, galaxy centers have $\thirco$ column densities higher in
the log by about 0.6: log $N_{13CO}$/d$V\approx15.7$.  Similarly,
galaxies with very high $\ci$ (1-0)/$\thirco$ (2-1 line ratios of a
hundred have $\thirco$ column densities lower to the same degree: log
$N_{13CO}$/d$V\approx14.5$.

We performed the same analysis for
the $\ci$ (2-1)/$\thirco$ (2-1) line ratio (not shown) and find very
similar column densities log $N_{13CO}$/d$V\approx15.1$ for the mean
sample galaxy, but tighter limits of $\Delta$ log $N_{13CO}$/d$V$=+0.4
and -0.2 on the galaxies with the highest and lowest line ratios,
respectively. Although the LVG $\thirco$ column densities are fairly
insensitive to changes in model temperature and density, they do
increase by about a factor of two from galaxies with a low $\co$/$\ci$
flux ratio to those with a high flux ratio in Fig.\,\ref{NCOLVGfig}
and, of course, by a larger factor of 3 to 10 when going from galaxies
with a high $\ci$/$\thirco$ ratio to those with a low ratio in
Fig\,\ref{N13COLVGfig}.  Thus, {\it \emph{the $\thirco$ model column
  densities of the sample galaxies are constrained to the range of}
  $N_{13CO}$/d$V$=$1.25(+1.8,-0.7)\times10^{15}\,\cc/\kms$}. Thus,
Fig.\,\ref{N13COLVGfig} suggests that abundance ratio [$\ci$] /
[$\thirco$] ranges from about 25 (in galaxies with high CO/$\ci$ flux
ratios) to about 250 (in galaxies with low CO/$\ci$ flux ratios).

\section{Discussion}  

\subsection{Importance of dense molecular gas in LIRG ISM}

In the galaxies with the highest $\ci$ (2-1)/$\ci$ (1-0) and $\ci$
(1-0) /$\thirco$ (2-1) flux ratios, the ISM is fully dominated by
dense ($n(\h2)=10^{4}-10^{5}\,\cc$) and modestly warm
($T_{kin}$=20--35 K) gas. Low [C] / [CO] and very low [$\thirco$] /
[$\co$] abundances of 0.1 and 0.01-0.001, respectively, further
characterize this gas. This subsample is about two thirds of the total
sample, and it contains all the (U)LIRGs in addition to fewer less luminous starburst galaxies.  Thus, at least in the
more luminous LIRGs, most of the interstellar gas is in the form of
dense, warmish gas clouds. Diffuse molecular gas clouds of low density
are probably also present, but not as important, contributors to the
line emission from these galaxies.

This is no longer true if we consider galaxies with lower $\ci$
(2-1)/$\ci$1-0) $\leq1.65$ and $\ci$ (1-0) /$\thirco$ (2-1) $\leq30$
ratios.  In these galaxies, the assumption of a single high-pressure
gas phase still yields good fits to the individual observed line
ratios, but the physical parameters derived from these fits become
increasingly inconsistent as we go to ever lower $\ci$ (2-1)/$\ci$
(1-0) and $\ci$ (1-0) /$\thirco$ (2-1) ratios.  For instance, we no
longer can obtain consistent results for density and temperature at
observed ratios $\ci$ (2-1)/$\ci$ (1-0)$\leq$1.5.  Moreover, a
succesful combination of the results of Sects. 4.3 and 4.4 is only
possible if the isotopic ratios are very high [$\co$] /
[$\thirco$]=200-3000 (median value 400).  Such values are\emph{ {\it
    \emph{much higher than the isotopic ratios in the Milky Way,}}}
which are typically 25 to 100 (see, for instance, Giannetti $\etal$
2014, and references therein).  As already noted, high $\co$/$\thirco$
ratios have been found before in starburst galaxies (Mart\'in $\etal$
2010, Henkel $\etal$ 2014), but their values are not nearly as high as
the ones we find here. Even worse, such very high [$\thirco$] /
[$\co$] abundances are actually irreconcilable with observed $J$=2-1
isotopic {\it \emph{flux}} ratios, and the model also predicts
$J$=(1-0) isotopic flux ratios to be higher than those in the $J$=2-1
transition by factors of two to three times, which is contradicted by
observation (see, e.g., Papadopoulos $\etal$ 2012, and Davis 2014).

Thus, our assumption that the $\ci$ and $\co$ line emission is
characterized by a single temperature and a single density breaks down
in galaxies with decreasing $\ci$ (1-0)/$\thirco$ (2-1) and $\ci$
(2-1)/(1-0) flux ratios. Galaxies with far-infrared luminosities log
$L(FIR)\,\geq\,11.3$ can {\it \emph{all}} be fit with a single gas component.
For starburst galaxies with $10.3\,\leq\,L(FIR)\,\leq\,11.3$ this is
true for $60\%$, and for those with $L(FIR)\,\leq\,10.3$ this has
dropped to $45\%$. Although there is no distinct luminosity threshold,
the lower the luminosity of a galaxy, the greater the chance that it
requires modeling with at least {\it \emph{two gas phases}}. Dense gas
($n(\h2)\geq10^{4}\,\cc$) is still in evidence, but the contribution
of gas at lower densities and temperatures, i.e. at lower pressures,
becomes increasingly important and can no longer be ignored.
Assigning part of the observed line emission to a contribution by
low-excitation gas, we might increase the total mass significantly
beyond the mass estimated from the single-phase analysis; for this
reason, we have refrained from attempting to estimate masses in the
preceding sections.

However, such a two-phase LVG analysis has recently been performed and
published by Kamenetzky $\etal$ (2014, hereafter K14). Their sample is
a small subset (15 out of 76 galaxies) of the sample in this paper, which nevertheless covers the full range of far-infrared
luminosities. They also include two (early type) galaxies that we
ignore in the following. In their analysis, K14 likewise relied on
{\it RADEX} to fit {\it \emph{all}} the $\co$ lines up to $J$ = 13-12, and
modeled dust, mid-infrared $\h2$ lines, and the $\ci$, $\cii,$ and
$\nii$ fine-structure lines as well, but they did not include any
$\thirco$ fluxes. To obtain masses, they {\it \emph{assumed}} an
abundance [CO] / [$\h2$]=$2\times10^{-4}$ throughout.  Since K14 used
the same data and the same non-LTE code to analyze the data, their
results should be comparable to ours.  Their $\ci$ LTE
excitation temperatures (their Table 14) indeed lie in the same range
as our non-LTE kinetic temperatures.  K14 distinguish between a `cold'
and a `warm' gas phase contribution to the observed line emission and
derive the parameters for each of these. The cold and warm CO phases
have similar densities, on average $n(\h2)=2000\,\cc$ (range 200-25000
$\cc$) and $n(\h2)=7000\,\cc$ (range 1000-75000 $\cc$), respectively,
but the temperatures are very different, $T_{kin}$=25 K (range 15-250
K) for the cold phase and 725 K (range 250-2500 K) for the warm
phase. The CO column densities they find are very similar to our
values for the cold phase and an order of magnitude lower than what
we find for the warm component. Thus, the more complicated two-phase
modeling by K14 that provides a better fit to (a subset of) the data
in this paper still yields physical results that are not very
different from those derived from our simplified one-phase fits.

The K14 analysis yields molecular gas masses for both `warm' and
`cold' gas phases, as well as their pressures and the luminosity of
the total CO line luminosity. In the above, we concluded that the
contribution from dense gas increasingly dominates the observed line
fluxes for increasing $\ci$ (1-0)/(2-1) ratios. Thus, at the higher
$\ci$ ratios, the `warm' component should also increasingly dominate
the `cold' component in the K14 results.  In Fig. 8, we show both the
CO luminosity ratio and the gas mass ratio of the warm and cold phases
distinguished by K14 (taken from their Table 13), as a function of the
$\ci$ line ratio.  Eight of the galaxies in our sample have also been
mapped and studied in more detail before (Israel 2009b, and references
therein), likewise involving a two-phase LVG analysis based on {\it
  RADEX}. For these eight galaxies, we took the fraction of the total
$\co$ (2-1) flux that the models require to be attributable to the
dense gas phase.  Although the small numbers and the relatively large
dispersions of the data in each panel, as well as differences in
method and lines considered, rule out a fruitful quantitative
comparison between the panels and the results obtained in the above,
all three panels nevertheless show the qualitative behavior that is
expected if the role of dense gas is indeed more dominant at the
higher $\ci$ (2-1)/(1-0) ratios.
 
\subsection{Use of $\ci$ as a tracer of molecular gas}

\begin{figure*}[t]
\begin{center}
\begin{minipage}[]{18cm}
\resizebox{6cm}{!}{\rotatebox{0}{\includegraphics*{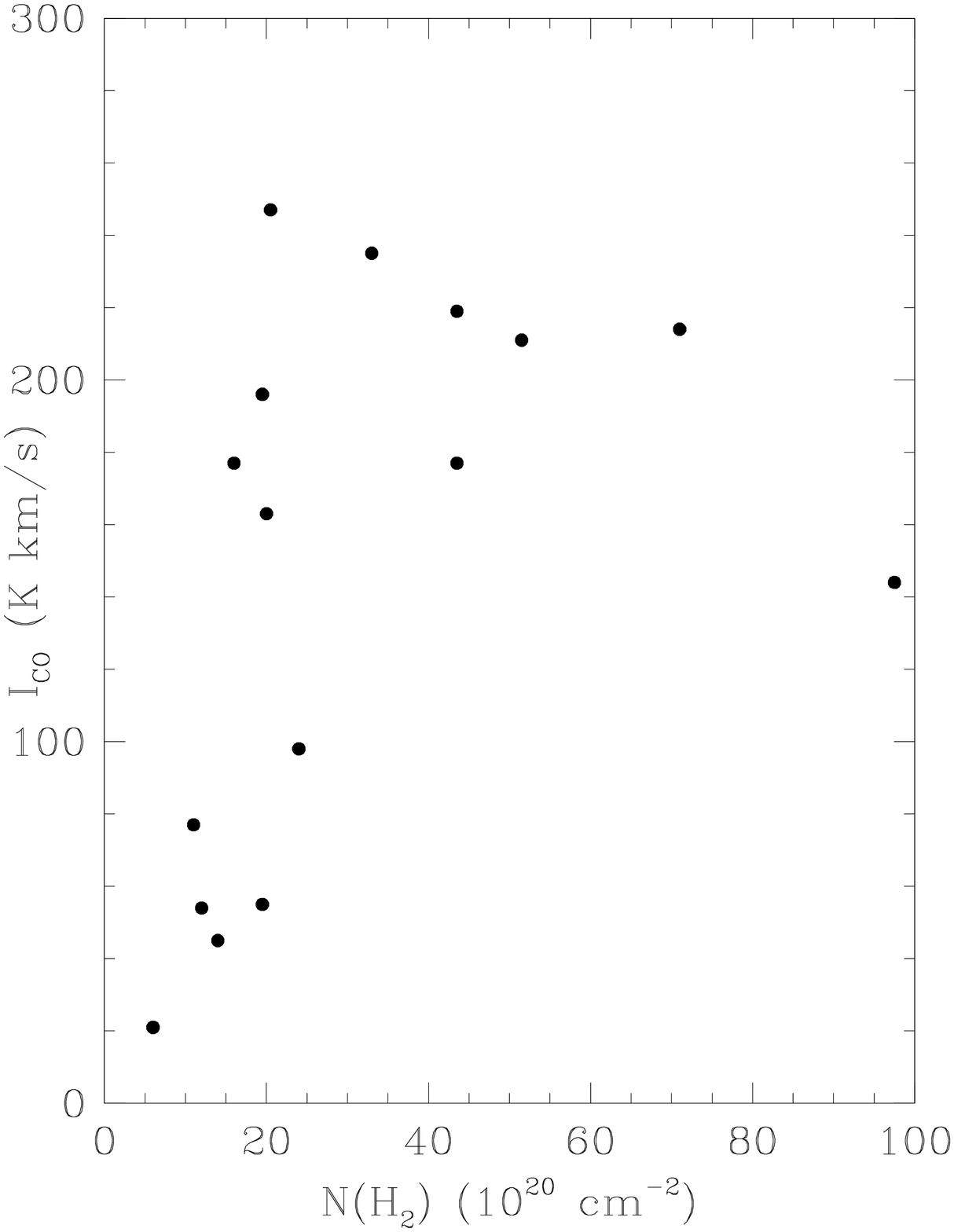}}}
\resizebox{6cm}{!}{\rotatebox{0}{\includegraphics*{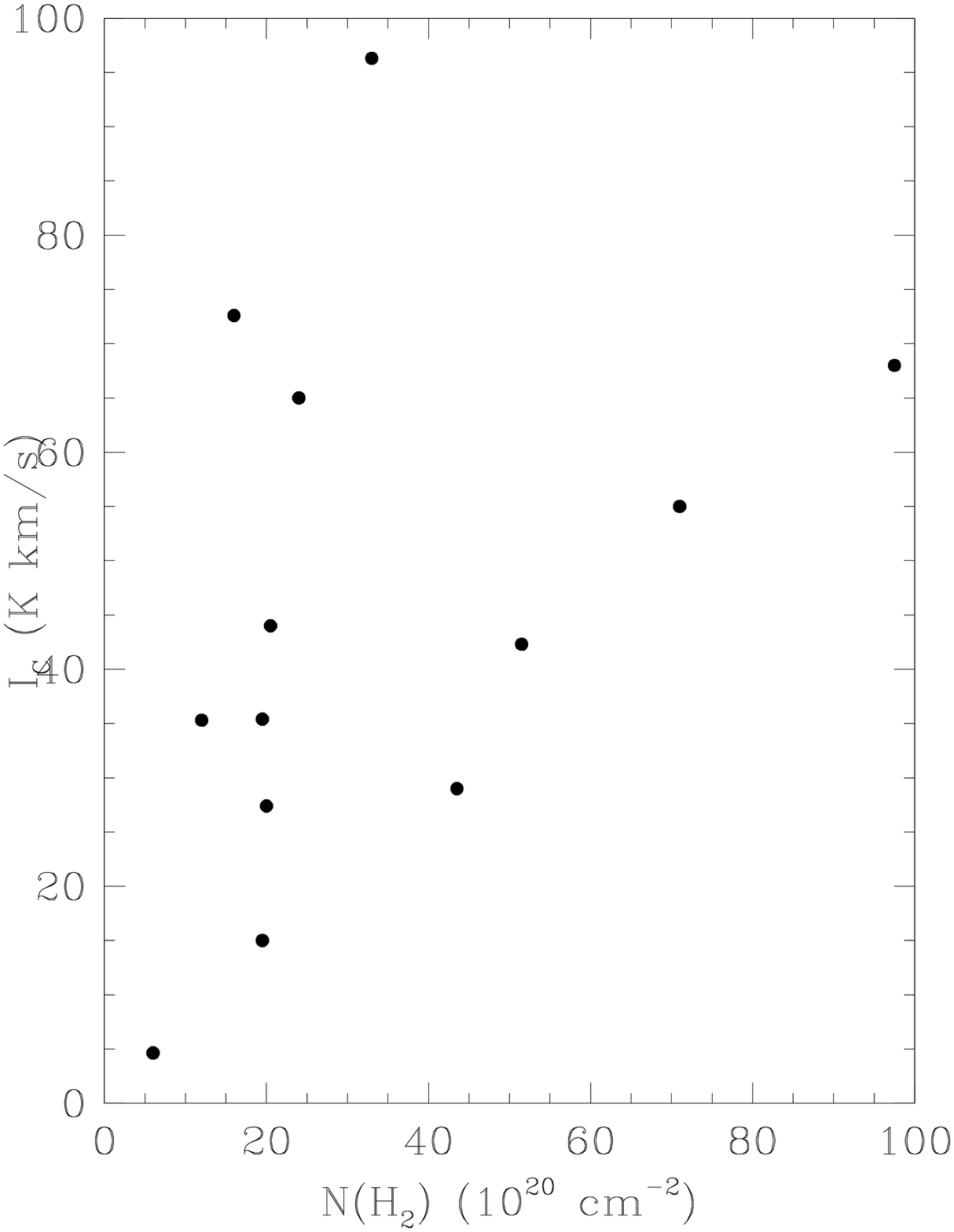}}}
\resizebox{6cm}{!}{\rotatebox{0}{\includegraphics*{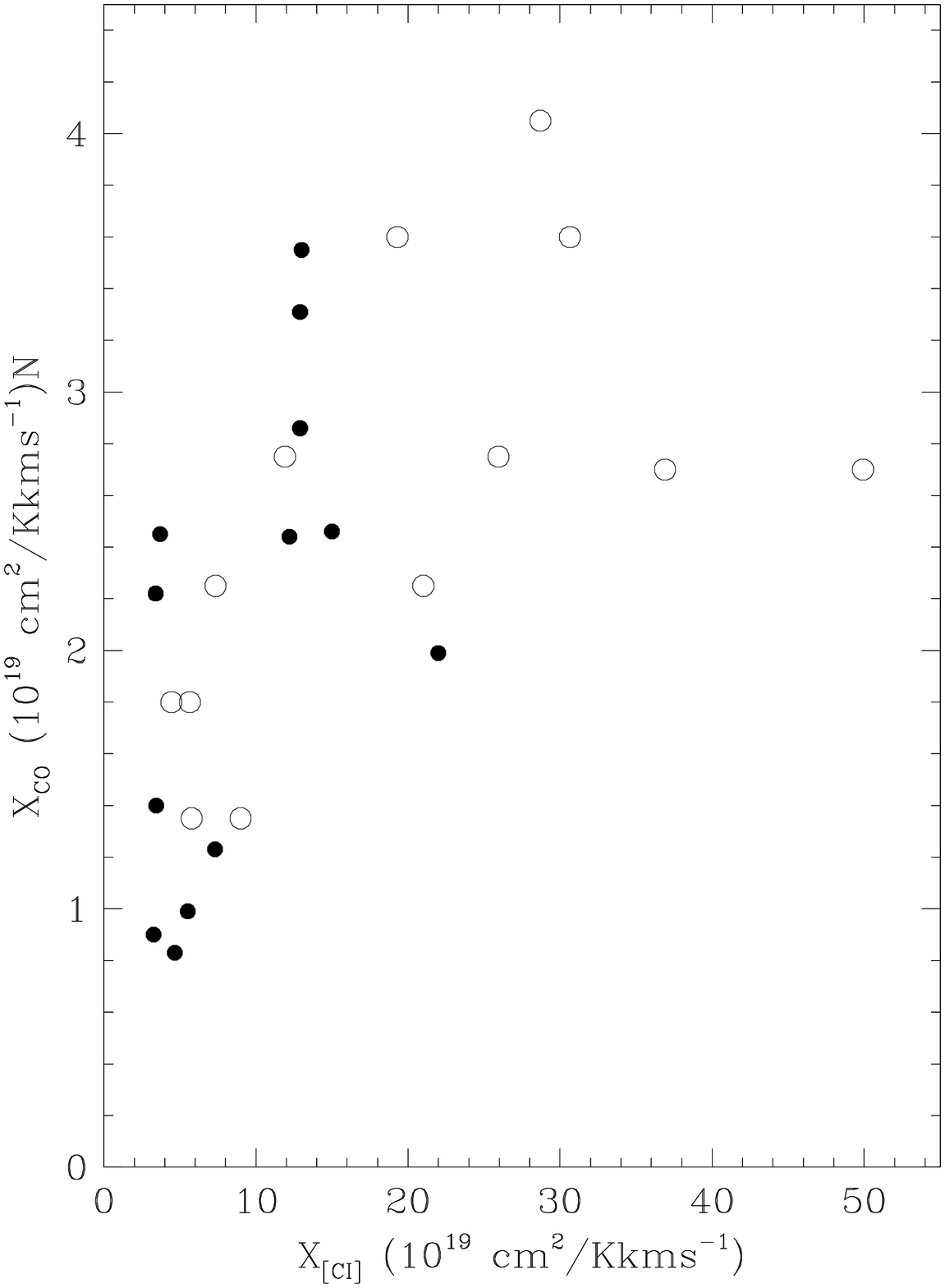}}}
\end{minipage}
\caption[]{Comparison of $N_{\rm \h2}/I_{\rm [CI]}$ and $N_{\rm
    \h2}/I_{\rm CO}$ conversion factors based on the $\ci$ (1-0) and
  $\co (1-0)$ emission from fully modeled galaxy centers (see
  text). The open circles in the rightmost panel represent data
  extracted from the study by Kamenetzky $\etal$ (2014).  }
\end{center}
\label{xfig}
\end{figure*}

Because cold $\h2$ gas is very hard to observe, its surface density
and mass are usually estimated from measurements of dust or tracer
molecules. The ubiquitous CO molecule is such a tracer, even though
the $\co$ lines are optically thick, have metallicity-dependent
abundances, and are subject to various excitation conditions that in
the worst case lead to its dissociation. Nevertheless, the beam-averaged $\co$
(1-0) intensity is often used to derive $\h2$ column densities on the
assumption that variations in the actual ratio of $\co$ intensity to
$\h2$ column density across the (large) linear beam area cancel out,
so that a common factor (called $X_{\rm CO}$) can be used to perform
the conversion. The actual value of $X_{\rm CO}$ to be used and the
ways in which it varies as a function of environmental conditions are
still under debate (see, for instance, the review by Bolatto $\etal$
2013).

It has been suggested by various authors (Papadopoulos $\etal$ 2004,
and more recently Offner $\etal$ 2014, Glover $\etal$ 2015, and
Tomassetti $\etal$ 2014) that $\ci$ {\it \emph{line intensities}} may be a
better tracer of {\it \emph{line-of-sight molecular hydrogen gas column
  densities}} than either $\co$ or $\thirco$, especially the 492 GHz
$\co$ $J$=1-0 transition that is easiest to observe.  Their
conclusions are supported by model simulations of \emph{{\it spatially
  resolved}} Galactic clouds. An important consideration is that they
expect neutral carbon to be widespread and thoroughly mixed with
turbulent molecular gas in filamentary clouds and that it traces
diffuse gas of lower densities than $\co$ and $\thirco$. Tomassetti
$\etal$ (2014) even claim that {\it \emph{spatially unresolved}} $\ci$
measurements may recover the entire $\h2$ mass of a galaxy (but the
kinetic temperature and the neutral carbon abundance need to be known
first).

In the preceding we have found, however, that much or even most of the
$\ci$ emission traces dense $n(\h2)\geq10^{4}\,\cc$ clouds rather
than a diffuse gas, and that conditions may vary considerably as a
function of excitation. This appears to be at odds with one of the
major assumptions underlying the case for $\ci$ emission as a
superior or even useful tracer of molecular gas. It is thus of
interest to compare $\ci$-to-$\h2$ conversion factors to the more
traditional CO-to-$\h2$ conversion factor using actual observations of
{\it \emph{unresolved cloud ensembles}}.  We have done this by determining
the ratios $X_{\rm C}=N_{\h2}/I(\rm [CI]_{492})$ and $X_{\rm
  CO}=N_{\h2}/I(\rm CO_{115})$ for fifteen galaxy centers modeled
using ground-based observations (see Israel, 2009b, and references
therein). For these galaxies, beam-averaged column densities $N_{\h2}$
were derived from the lower $\co$ and $\thirco$ transitions, assuming
a mixture of two gas phases (a warm low-density phase and a colder
high-density phase). Although physically not fully realistic, these
two-phase LVG models are clearly superior to the single-phase LVG
model considered thus far.

The results are shown in Fig. 9.  We find mean values
$X_{\rm CO}=(0.21\pm0.10)\times10^{20}\,\cm2/\kkms$, and $X_{\rm
  C}=(1.0\pm0.6)\times10^{20}\,\cm2/\kkms$.  The {\it \emph{mean}}
CO-to-$\h2$ conversion factor derived from this sample is very close
to the mean values found in the two-phase analysis by K14 and
suggested earlier by Papadopoulos $\etal$ (2012) and Yao $\etal$
(2003).  This mean value of $X_{\rm CO}$ is a full order of magnitude
lower than the value usually assumed for galaxy disks (the `standard'
conversion factor), and a factor of two or more below the value often
applied instead to active galaxy centers. 

The good agreement between the mean $X$ values quoted hides
the fact, however, that neither the individual velocity-integrated $\co$ fluxes
nor the individual $\ci$ fluxes correlate very well with the derived
beam-averaged $\h2$ column densities as shown in the lefthand and
center panels of Fig. 9. In the rightmost one, we also compare
directly the $X_{\rm CO}$ and $X_{\rm C}$ values, derived from the
preceding work. In that panel, we have also included $X$ values
deduced from K14.  Overall, the two are roughly correlated but show
significant variations from galaxy to galaxy over a range that covers an
order of magnitude.

We conclude that in practice the $\ci$ 492 GHz line is not superior to
the $\co$ 115 GHz line as a tracer of molecular ($\h2$) gas. Even
though the data indicate mean conversion factors about an order of
magnitude lower than the `standard' Galactic values, there appear to
be no a priori reliable $\ci$-to-$\h2$ or CO-to-$\h2$ conversion
factors that can be used to find the masses of individual galaxy
centers or luminous galaxies in general. It is therefore quite likely
that the $\cii$ line at 1.9 THz, not studied in this paper, is
superior to both CO and $\ci$ as a molecular gas tracer, and it is
also conceivable that variations in environmental conditions require
analysis of all three lines together to arrive at reliable estimates.

The above considerations reinforce our conviciction that
an attempt to use the results of a simple analysis such as provided
here to derive the total mass of molecular gas in the observed galaxy
centers would be a meaningless, even misleading exercise.

\subsection{Molecular gas in high-redshift galaxies}

Our findings have some relevance for the study of the molecular
content of luminous galaxies in the early universe.  Especially with
millimeter array telescopes, measurements of individual CO and [CI]
lines have become quite feasible, but these are frequently difficult to
interpret. Standard CO-to-$\h2$ conversion factors (although sometimes
applied) are utterly useless, because the physical circumstances in
luminous high-z galaxies are not really known and are certainly different
from the local circumstances on which such factors are based.  In
fact, this is what we actually want to determine.  In any case, the
$X$-factor only (if at all) applies to the $J$=1-0 transition of CO,
which is rarely observed.  By themselves, high CO transitions ($J$>5)
sample a limited fraction of the molecular gas, and far-infrared dust
temperatures often do not usefully constrain the kinetic gas
temperatures. As illustrated in earlier sections of this paper, $\co$
line ratios are highly degenerate for temperature and
density. Emission in one of the $\thirco$ lines, which might be used
to break this degeneracy, are almost always lacking, owing to their
intrinsic weakness (cf. Fig.\,\ref{ratratfig}d). 

The method we have applied to luminous galaxies in the local
universe, however, might also be used to determine parameters for their
early-universe counterparts. If we can measure the three $\co$ lines,
as well as the two $\ci$ lines for these galaxies, we may also
determine the intersection of their ratios, hence their density and
kinetic temperature of the underlying $\h2$ gas, and the C and CO
average column densities. 

Unfortunately, the opacity of the (sub)millimeter sky does not allow
measurement of these lines at all redshifts of interest. However, in
addition to the local universe, {\it \emph{all five lines}} occur in
atmospheric windows for the following redshift ranges: z1=0.15-0.18,
z2=0.9-1.1, z3=1.2-1.4, z4=1.8-2.2, z5=3.1-3.2, and z6=3.7-5.2.  Line
ratios of objects in any of these redshift ranges can be analyzed
immediately using {\it RADEX}-derived diagnostic diagrams of the type
presented in this paper. In principle, such ratios need to be
corrected for the change in cosmic background temperature $T_{bg}$
that increases from 3.2 K at z1 to >13 K at z6. These corrections are
not large. For the gas parameters of the average luminous galaxy in
this paper, the CO(4-3)/CO(2-1) ratio increases up to z6 by no more
than $10\%$, and the CO(7-6)/(4-3) ratio increases by less than $20\%$
The increases in the $\ci$ and in the $\ci$-to-$\co$ ratios are
likewise less than $10\%$ so that even uncorrected use of the diagrams
in Figs.\,\ref{CILVGfig},\,\ref{COLVGfig}, and \ref{NCOLVGfig} produces
useful results.

\section{Conclusions}

We collected flux measurements of the two lines of neutral carbon
$\ci$ at 492 GHz and 809 GHz from 75 galaxies, mostly obtained with
the {\it SPIRE} instrument on the {\it Herschel Space Observatory}, as
well as the fluxes of the $J$=7-6, $J$=4-3, $J$=2-1 $\co$, and $J$=2-1
$\thirco$ lines. In 35 galaxies, all six lines have been measured.

In most galaxies or galaxy centers observed, the ratio of the $\ci$ to
$\thirco$ line flux is much higher than the corresponding ratio in
Galactic sources. The ratio increases with both line and far-infrared
(FIR) continuum luminosity, and this both confirms and expands an
earlier conclusion using data from ground-based facilities alone.  The
flux ratio of the two $\ci$ lines correlates well with both the
$J$=7-6/$J$=4-3 and the $J$=4-3/$J$=2-1 flux ratios of $\co$; higher
$\ci$ and $\co$ ratios trace more highly excited molecular gas.

The observed $\ci$ lines alone do not provide much quantitative
information on the conditions in the interstellar medium (ISM) of the
parent galaxies, because their flux ratio is highly degenerate with
respect to density and temperature. However, when combined with $\co$
line ratios, the $\ci$ lines do significantly constrain the ISM
parameters and provide a useful criterion to classify the parent
galaxies in terms of ISM gas pressure.  The galaxies with the highest
$\ci$ flux ratios and the highest $\ci$/$\thirco$ flux ratios
represent the warmest and most luminous galaxies, classified as
Luminous InfraRed Galaxies (LIRGs). Their emission is reproduced well
by single-gas-phase LVG models.

The interstellar medium (ISM) in the most luminous LIRGs, as traced by
the observed line emission, is fully dominated by dense
($n(\h2)=10^{4}$--$10^{5}\,\cc$) and moderately warm
($T_{kin}\approx35$ K) gas clouds. They appear to have low
[C$^{\circ}$] / [CO] and [$\thirco$] / [$\co$] abundance ratios.  In
the more numerous, less luminous galaxies, a single gas-phase LVG
model no longer produces consistent results. In these galaxies,
emission from gas clouds at lower densities becomes progressively more
important in addition to the emission from dense clouds, and a
multiple-phase analysis is required.

This analysis shows that both $\co$ and $\ci$ velocity-integrated line
fluxes poorly predict molecular hydrogen ($\h2$) column densities in
the high-excitation environments exemplified by (active) galaxy
centers and the ISM of luminous infrared galaxies. In particular,
so-called $X(\ci)$ conversion factors do not outperform $X(\co)$
factors.

The methods and diagnostic diagrams presented in this paper also
provide a means of deriving temperatures, densities, and column
densities of the molecular gas in high-redshift galaxies, from $\co$
and $\ci$ lines alone. We have identified six redshift ranges for
which all five lines required fall in windows of good sky
transparency.

%
%

\end{document}